\documentstyle[aps,harvard,psfig]{revtex}

\citationstyle{dcu}

\begin{document}
\title{Precision Measurements with High Energy Neutrino Beams}
\author{Janet M. Conrad and Michael H. Shaevitz}
\address{Department of Physics, Columbia University, New York, New\\
York 10027}
\author{Tim Bolton}
\address{Department of Physics, Kansas State University, Manhattan,\\
Kansas 66502-2601}
\date{\today}
\maketitle

\begin{abstract}
Neutrino scattering measurements offer a unique tool to probe the
electroweak and strong interactions as described by the Standard Model (SM).
Electroweak measurements are accessible through the comparison of neutrino
neutral- and charged-current scattering. These measurements are
complimentary to other electroweak measurements due to differences in the
radiative corrections both within and outside the SM. Neutrino scattering
measurements also provide a precise method for measuring the F$_2(x,Q^2)$
and $x$F$_3(x,Q^2)$ structure functions. The predicted Q$^2$ evolution can
be used to test perturbative Quantum Chromodynamics as well as to measure
the strong coupling constant, $\alpha _s\,$, and the valence, sea, and gluon
parton distributions. In addition, neutrino charm production, which can be
determined from the observed dimuon events, allows the strange-quark sea to
be investigated along with measurements of the CKM matrix element {\bf $%
|V_{cd}|$ }and the charm quark mass.
\end{abstract}

\newpage

\tableofcontents

\newpage

\section{Introduction}

Neutrino scattering experiments have strongly influenced the development of
particle physics over the past three decades. Neutrinos only couple to other
particles through the weak interaction,
which is well determined in the Standard Model (SM). For this reason,
neutrino scattering measurements can be used as a probe to measure many of
the SM parameters or to 
look beyond the SM for indications of new physics. The effects of the small
interaction cross section for neutrinos has been overcome by  
modern experiments through the use of high intensity beams coupled with
massive detectors which give luminosities in the range of $10^{36}\ $cm$^{-2}
$s$^{-1}$. Data samples in excess of one million events are now available
which allow measurements of strong and electroweak parameters comparable in
precision to other fixed-target and collider determinations.

This Review addresses three broad physics topics accessible to high energy
neutrino deep inelastic scattering experiments: nucleon structure functions
and tests of quantum chromodynamics (QCD) (Sec. \ref{SF Section}), neutrino
induced charm production (Sec. \ref{Dimuon Section}), and electroweak
physics with neutrino beams (Sec. \ref{EW Section}). We begin with a brief
introduction to the relevant kinematics and cross section formulae, and
follow with a summary of experiments whose results we review (Sec. 
\ref{Kinematics Section}). We 
then devote an entire section to the difficult
problem of normalizing data through the determination of the neutrino flux
and total cross section (Sec. \ref{Flux Section}). The main physics sections
follow.

Experiments we review date from roughly 1980 until the present, although in
some cases we reach back further if the results are still of interest. We
also restrict ourselves to high energy measurements; thus we  
emphasize the CERN\ SPS and Fermilab Main Ring and Tevatron programs.
Earlier lower energy experiments have been reviewed elsewhere  
\cite{Barrish,Fisk,Diemoz,Mishra}

\subsection{Kinematic Formalism}
\label{Kinematics Section}

\begin{figure}[ptbp]
\psfig{figure=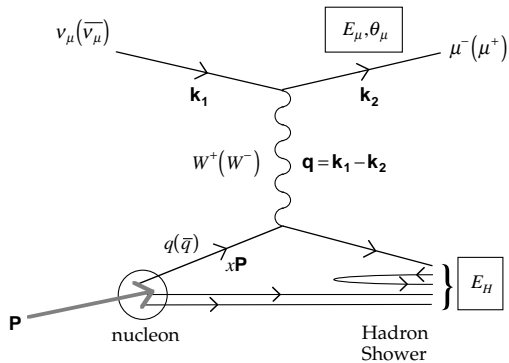,bbllx=0bp,bblly=315bp,bburx=600bp,bbury=645bp,width=3.5in,clip=T}
\caption{The first order Feynman diagram for deep inelastic neutrino
scattering}
\label{CC Feynman}
\end{figure}

The tree-level diagram for charged-current (CC) neutrino-nucleon ($\nu N$)
scattering is shown in Fig. \ref{CC Feynman}. A neutrino (antineutrino) with
incoming four-momentum $k_1$ scatters from a quark or antiquark in the
nucleon via exchange of a $W^{+}$ ($W^{-}$) boson with four-momentum $q$. In
the laboratory frame, the variables that can be measured in this interaction
are the energy and angles of the outgoing muon, $E_\mu $, $\theta _\mu $, $%
\phi _\mu $, and the energy and angles of the outgoing hadrons $E_{had}$, $%
\theta _{had}$, $\phi _{had}$. In practice, experiments are located
sufficiently far from the neutrino production point that the neutrino beam
can be considered, to high accuracy, parallel to the $z$-axis, thus the
difficult-to-measure hadron scattering angles are not necessary to specify
the kinematics. In the laboratory frame, the 4-vectors of the event can then
be written in terms of $E_\mu $, $\theta _\mu $, $\phi _\mu $, $E_{had}$,
and $M$, the proton mass:\footnote{%
We use the conventional units of high energy physics in this Review: $\hbar
=c=1$, and all dimensionful quantities are in GeV unless otherwise indicated.%
} 
\begin{eqnarray}
k_1 &=&(E_\nu ,0,0,E_\nu ), \\
k_2 &=&(E_\mu ,E_\mu \sin \theta _\mu \cos \phi _\mu ,E_\mu \sin \theta _\mu
\sin \phi _\mu ,E_\mu \cos \theta _\mu ),  \nonumber \\
p &=&(M,0,0,0),  \nonumber \\
q &=&k_1-k_2,  \nonumber
\end{eqnarray}
with $E_\nu =E_\mu +E_{had}$. 

Useful invariant quantities which describe the interaction are: 
\begin{eqnarray}
\nu  &=&(p\cdot q)/M~{\rm (energy~transfer),}  \nonumber \\
y &=&p\cdot q/p\cdot k_1~{\rm (inelasticity),} \\
Q^2 &=&-q^2~{\rm (negative~squared~4-momentum),}  \nonumber \\
x &=&Q^2/(2p\cdot q)~{\rm (Bjorken~scaling~variable).}  \nonumber
\end{eqnarray}
In the laboratory, these reduce to:
\begin{eqnarray}
\nu  &=&E_{had}-M,  \nonumber \\
y &=&\frac{(E_{had}-M)}{(E_{had}+E_\mu )}, \\
Q^2 &=&(E_{had}+E_\mu )E_\mu \theta _\mu ^2,  \nonumber \\
x &=&\frac{(E_{had}+E_\mu )E_\mu \theta _\mu ^2}{2M(E_{had}-M)},  \nonumber
\end{eqnarray}
using the small-angle approximation and neglecting the muon mass.

\begin{figure}[ptbp]
\psfig{figure=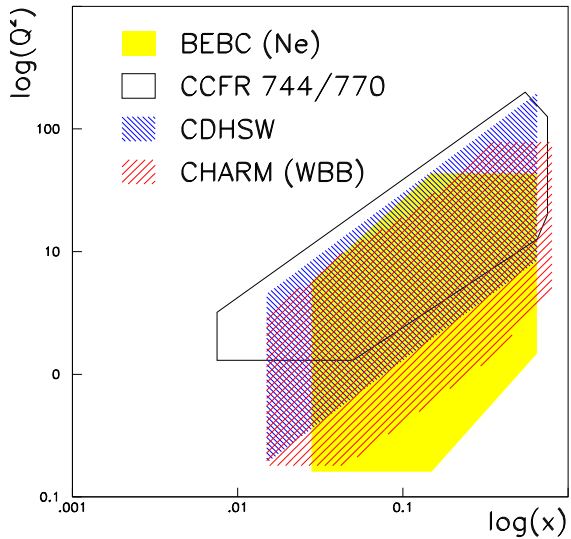,bbllx=0bp,bblly=100bp,bburx=600bp,bbury=645bp,width=3.5in,clip=T}
\caption{The kinematic regions accessible to the high statistics neutrino
experiments discussed in this Review.}
\label{Kinematics Plot}
\end{figure}

In principle, for a given beam of neutrino energy $E_\nu $, the accessible
kinematic region is bounded by: 
\begin{eqnarray}
\frac{m_\mu ^4}{8xME_\nu ^2}\le \nu  &\le &\frac{E_\nu }{(1+2Mx/E_\nu )}%
\rightarrow E_\nu ,  \nonumber \\
\frac{m_\mu ^4}{8xME_\nu ^3}\le y &\le &\frac 1{(1+2Mx/E_\nu )}\rightarrow 1,
\\
\frac{m_\mu ^4}{4E_\nu ^2}\le Q^2 &\le &\frac{2ME_\nu x}{(1+2Mx/E_\nu )}%
\rightarrow 2ME_\nu x,  \nonumber \\
\frac{m_\mu ^2}{2ME_\nu }\le x &\le &1.  \nonumber
\end{eqnarray}
The lower limits are given to leading order in the final state lepton mass
and can always be set to $0$ for the high energies considered; the arrows
indicate the high energy limits of the upper range on the variables, which,
for most purposes, are also sufficient. In practice, experimental
restrictions to select regions of high acceptance and small
corrections    
reduce the available range. The accessible kinematic range of a given
experiment can be described in the plane of any two of the above variables.
The resulting kinematic range in the $x$ and $Q^2$ plane is shown in Fig.~%
\ref{Kinematics Plot} for four high-statistics deep inelastic neutrino
scattering experiments.

In the case of neutral current (NC) scattering, the scattered neutrino is
not reconstructed; all event information must be inferred from the hadron
shower. The important kinematic quantities are identical to the
charged-current case up to small effects due to the muon mass.

\subsection{The Neutrino Scattering Cross Section}

\label{xsec intro}

The unpolarized cross section for neutrino scattering is derived from
placing the constraints of Lorentz invariance and CP invariance on the
product of the leptonic and hadronic tensors. The result is that, for fixed $%
x$ and $Q^2$, only five free parameters describing the hadronic structure
remain. Two are proportional to the outgoing lepton mass, which is small for
the case of $\nu _\mu $ scattering and can be dropped. The remaining three
parameters are functions of $x$ and $Q^2$ and are called ``structure
functions.''

In this Review, we will describe the cross section for scattering from an
isoscalar target in terms of the structure functions $F_2$, $xF_3$ and $R_L$
defined by: 
\begin{eqnarray}
\frac{d^2\sigma ^{\nu (\overline{\nu })N}}{dxdy} &=&\frac{G_F^2ME_\nu }{\pi
\left( 1+Q^2/M_W^2\right) ^2}  \nonumber \\
&&\left[ F_2^{\nu {(\overline{\nu })}N}(x,Q^2)\left( \frac{y^2+(2Mxy/Q)^2}{%
2+2R_L^{\nu {(\overline{\nu })}N}(x,Q^2)}+1-y-\frac{Mxy}{2E_\nu }\right)
\right.   \nonumber \\
&&\left. \pm xF_3^{\nu {(\overline{\nu })}N}y\left( 1-\frac y2\right)
\right] .  \label{eq:sigsf}
\end{eqnarray}
where the $\pm $ is $+(-)$ for $\nu (\overline{\nu })$ scattering. Analogous
functions $F_2^{e(\mu )N}(x,Q^2)$ and $R_L^{e(\mu )N}(x,Q^2)$ appear in the
cross section for deep inelastic charged lepton scattering. $R_L(x,Q^2)$ can
be interpreted as the ratio of the longitudinal to transverse virtual boson
absorption cross-section, and is related to the $2xF_1(x,Q^2)$ structure
function by: 
\begin{eqnarray}
R_L(x,Q^2) &=&\frac{{\sigma _L}}{{\sigma _T}}  \nonumber \\
&=&\frac{F_2(x,Q^2){(1+}4M^2x^2/Q^2{)-}2xF_1(x,Q^2)}{2xF_1(x,Q^2)}{.}
\end{eqnarray}
The function $xF_3(x,Q^2)$ appears only in the cross section for the weak
interaction. It originates 
from the parity-violating term in the product of the leptonic and hadronic
tensors.

\subsection{Neutrino Scattering from Partons}

\label{QPM Section}

The cross section presented in Eq.~\ref{eq:sigsf} makes no assumptions about
the underlying structure of the hadron involved in the interaction. However,
the results of early experiments provided clear evidence that the neutrinos
were interacting with quarks in the nucleon \cite{early}. This section
discusses neutrino-parton scattering at the simplest level. The
interpretation of the cross section within QCD is discussed in more detail
in Sec. \ref{SF Section}.

The ``naive'' quark-parton model (QPM) describes the structure functions in
terms of momentum-weighted parton distribution functions (PDFs). For
neutrino interactions with the partons in the proton, 
\begin{eqnarray}
F_{2,QPM}^{\nu p}(x) &=&2x\left( d(x)+\bar{u}(x)+s(x)+\bar{c}(x)\right) , \\
xF_{3,QPM}^{\nu p}(x) &=&2x\left( d(x)-\bar{u}(x)+s(x)-\bar{c}(x)\right) .
\end{eqnarray}
In these equations $u$, $d$, $s$ and $c$ indicate the PDFs for up, down,
strange and charm quarks, respectively, in the proton. These PDFs describe
the probability that the interacting parton carries a fraction, $x$,  
of the
proton's 4-momentum.\footnote{%
This interpretation of $x$ is an approximation that we employ due to its
intuitive appeal, and because at high energies, this definition of $x$
differs negligibly from its more rigorous interpretation as a
light-cone-variable (See Sec. \ref{Kinematics Section}.).}  
Note that because the neutrino interacts via emission of a $W^{+}$, only the
negatively charged quarks contribute to the interaction.

In a similar way, neutron structure functions can be written in terms of the
proton PDFs by invoking isospin invariance,  
\begin{eqnarray}
F_{2,QPM}^{\nu n}(x) &=&2x\left( u(x)+\bar{d}(x)+s(x)+\bar{c}(x)\right) , \\
xF_{3,QPM}^{\nu n}(x) &=&2x\left( u(x)-\bar{d}(x)+s(x)-\bar{c}(x)\right) .
\end{eqnarray}

When neutron and proton structure functions are combined for an isoscalar
target, one obtains: 
\begin{eqnarray}
F_{2,QPM}^{\nu N}(x) &=&x\left( u(x)+d(x)+2s(x)\right. \\
&&\left. +\bar{u}(x)+\bar{d}(x)+2\bar{c}(x)\right) ,  \nonumber \\
xF_{3,QPM}^{\nu N}(x) &=&x\left( u(x)+d(x)+2s(x)\right. \\
&&\left. -\bar{u}(x)-\bar{d}(x)-2\bar{c}(x)\right) .  \nonumber
\end{eqnarray}

Antineutrino interactions proceed via emission of a $W^{-}$ boson. Therefore
only the positively charged quarks contribute. Following the prescription
for the neutrino case and assuming that $s(x)=\bar{s}(x)$ and $c(x)=\bar{c}%
(x)$, the antineutrino structure functions are: 
\begin{eqnarray}
F_{2,QPM}^{\bar{\nu}N}(x) &=&F_{2,QPM}^{\nu N}(x) \\
xF_{3,QPM}^{\bar{\nu}N}(x) &=&xF_{3,QPM}^{\nu N}(x)-4x\left(
s(x)-c(x)\right) .  \label{xF3QPM}
\end{eqnarray}

In this review $xF_3=\frac 12(xF_3^{\nu N}+xF_3^{\bar{\nu}N})$ will refer to
the average of the neutrino and antineutrino measurements; $\Delta
(xF_3)=xF_3^{\nu N}-xF_3^{\bar{\nu}N}$ 
will indicate the difference between the neutrino and antineutrino structure
functions. We use $F_2$ to refer to either neutrino or antineutrino
scattering because the structure functions are equivalent.

In the QPM, the interaction probability for each PDF depends on the square
of the charge associated with the interaction. In the weak interaction, this
is unity. For charged-lepton scattering mediated by a virtual photon: 
\begin{eqnarray}
F_{2,QPM}^{e(\mu )N}(x) &=&\frac 5{18}x\left[ u(x)+d(x)+\bar{u}(x)+\bar{d}%
(x)\right.  \nonumber \\
&&\left. +\frac 25\left( s(x)+\bar{s}(x)\right) +\frac 85\left( c(x)+\bar{c}%
(x)\right) \right] .
\end{eqnarray}

The structure function definitions of the QPM must be modified to
accommodate strong interactions between the partons and to include mass
effects. Within the framework of QCD, discussed in Sec.~\ref{QCD SF theory}%
, the parton distributions acquire a $Q^2$ dependence. Small transverse
momentum quark-gluon interactions may occur at the time of the scatter. The
probability that the interaction resolves such processes depends on $Q^2$
and modifies the parton distributions, leading to a $Q^2$ dependence known
as ``scaling violations.''  
Thus at Leading Order (LO) in QCD, the structure
functions are: 
\begin{equation}
F_{2,LO}=\sum_{i=u,d..}xq(x,Q^2)+x\overline{q}(x,Q^2),
\label{DIS F_2 definition}
\end{equation}
\begin{equation}
xF_{3,LO}=\sum_{i=u,d..}xq(x,Q^2)-x\overline{q}(x,Q^2).
\label{DIS xf3 definition}
\end{equation}
(In the above equations and subsequent sections, $q$ and $\overline{q}$ are
introduced as the generic notation for the various parton distributions in
the proton.) Parton mass effects further modify the parton distributions, an
effect called ``slow rescaling'' (see Sec. \ref{Dimuon Kinematics}), which
leads to $q(x,Q^2)\rightarrow q(\xi ,Q^2)$, where $\xi $ depends on the mass
of the parton.

To LO, $R_L(x,Q^2)=0$ and the Callan-Gross relation \cite{Callan-Gross}, $%
F_2(x,Q^2)=2xF_1(x,Q^2)$, is valid because spin-$\frac 12$ massless quarks
cannot absorb a longitudinally polarized virtual boson, reverse direction in
the center of momentum frame, and still conserve helicity. A non-zero $%
R_L(x,Q^2)$ is a manifestation of non-perturbative and higher order QCD
effects and is described in Sec.~\ref{QCD SF theory}. In all formulations
that we are aware of, the structure function $R_L$ is the same for neutrino
and antineutrino scattering.

These tree-level expressions can be used to point out some of the
interesting information accessible from $\nu N$ scattering:

\begin{itemize}
\item  $F_2^{\nu N}(x,Q^2)$ and $F_2^{\bar{\nu}N}(x,Q^2)$ measure the sum of
quark and anti-quark PDFs in the nucleon.

\item  The average of $xF_3^{\nu N}(x,Q^2)$ and $xF_3^{\bar{\nu}N}(x,Q^2)$
measure the valence quark PDFs, 
\begin{eqnarray}
xF_3 &=&u(x,Q^2)+d(x,Q^2)-\bar{u}(x,Q^2)-\bar{d}(x,Q^2) \\
&\equiv &u_V(x,Q^2)+d_V(x,Q^2).  \nonumber
\end{eqnarray}

\item  $F_2^{\nu N}(x,Q^2)$ and $F_2^{e(\mu )N}(x,Q^2)$ together confirm the
fractional electric charge assignment to the quarks and provide sensitivity
to the strange quark PDF.

\item  The difference $\Delta (xF_3)=xF_3^{\nu N}(x,Q^2)-xF_3^{\bar{\nu}%
N}(x,Q^2)$ is sensitive to the strange and charm quark content in the
nucleon.
\end{itemize}

Neutrino scattering from $d$ or $s$ quarks can produce a $c$ quark in the
final state which can be identified by either a lifetime or, more easily, a
semileptonic decay tag. This ability to tag final state charm permits a
series of unique strong and electroweak tests including the direct
measurement of $s(x,Q^2)$ and $\bar{s}(x,Q^2)$, extraction of the CKM\
elements $V_{cd}$ and $V_{cs}$, a measurement of the mass of the charm quark 
$m_c$, and a sensitive probe of the dynamics of LO and next-to-leading order
(NLO) QCD. This physics is described in Sec. \ref{Dimuon Section}.

Neutral-current $\nu N$ scattering is described by a closely related set of
structure functions. While these can be measured, the primary interest in
this process lies in inclusive cross-section measurements for electroweak
physics tests of the Standard Model. The detailed description of $\nu N$ NC
scattering will be discussed in Sec. \ref{EW Section}.

\subsection{Overview of Experiments}

\label{Experiments Section}

This section provides an overview of the beams and experimental apparatus
used in deep inelastic neutrino scattering. Table~\ref{Experiment Overview}
summarizes the data samples obtained by the high-precision neutrino
experiments which are discussed in more detail below:

\begin{table}[tbp]
\begin{center}
\begin{tabular}{ccccccc}
Experiment & Beam Energy (GeV) & target & $\nu$ events & $\overline{\nu}$
events & Flux Method & ref \\ \hline
CCFR (E616) & 30-300 & Fe & 150,000 & 23,000 & Secondary Monitoring &  \cite
{MacFar} \\ 
CCFR (E701) & 30-300 & Fe & 35,000 & 7,000 & Sec. Mon., $y$-int. &  \cite
{Auchincloss} \\ 
CCFR (E744/770) & 30-360 & Fe & 1,300,000 & 270,000 & Fixed $\nu_0$ &  \cite
{Seligman} \\ 
CDHSW & 20-212 & Fe & 221,000 & 20,500 & iter. $y$ fit &  \cite{CDHSWtot} \\ 
CHARM & 10-160 & CaCO$_2$ & 116,000 & 6000 & Secondary Monitoring &  \cite
{Charmtot} \\ 
CHARM II & 5-100 & Glass & 750,000 Q.E. & 1,300,000 Q.E. & $%
\sigma^{\nu}_{tot}$ &  \cite{CHARM II Detector} \\ 
BEBC & 15-160 & Ne (D$_2$) & 15,000 (12,000) & 10,000 (11,000) & iter. $y$
fit &  \cite{BEBCratio}
\end{tabular}
\end{center}
\caption{Overview of High Precision Neutrino Experiments. Target refers to
the bulk of the material in the experimental target. Total events vary for
each type of analysis depending on cuts; the numbers here are approximate. }
\label{Experiment Overview}
\end{table}

\subsubsection{CCFR Experiment}

The CCFR collaboration has performed a series of experiments at Fermi
National Accelerator Laboratory (hereafter, ``Fermilab'') using the Lab E
neutrino detector. This paper includes results from data runs E616, E701,
E744 and E770, spanning a period from 1979 through 1988. A new experiment,
E815, is presently taking data.

The E616 and E701 experiments took data using the Fermilab dichromatic
neutrino beam. The primary protons had an energy of 400 GeV. The secondary
beam traversed a 60 m chain of magnets and collimators, which selected pions
and kaons based on energy and charge. The neutrino energy spectrum from this
beam has two relatively narrow peaks from the pion and kaon decays which
gives the beam its dichromatic nature.

The most recent results from the CCFR Collaboration come from the 1985
(E744) and 1987 (E770) fixed-target runs at Fermilab. The CCFR neutrino beam
resulted from decays of pions and kaons produced in interactions of the
800~GeV FNAL proton beam with a beryllium target. The FNAL Quadrupole
Triplet beam line, which had no sign-selecting magnets, was used to
transport the secondaries. This resulted in a wide-band beam of neutrinos
and antineutrinos, with a relatively uniform energy spectrum ranging in
energy up to 600~GeV. This beam resulted in approximately 86.4\% $\nu_\mu $,
11.3\% $\overline{\nu }_\mu $, and 2.3\% $\nu _e$ and $\overline{\nu }_e$
events.

\begin{figure}[ptbp]
\psfig{figure=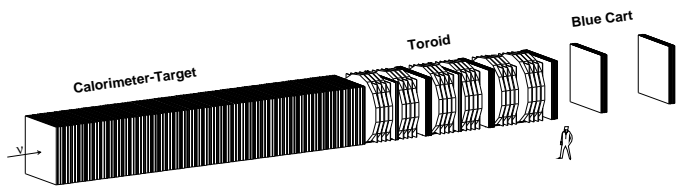,width=3.5in,bbllx=0bp,bblly=230bp,bburx=567bp,bbury=420bp,clip=t}
\caption{Schematic representation of the CCFR detector. The neutrino beam
travels from left to right. The target-calorimeter is on the left and the muon
spectrometer (toroid) is on the right.}
\label{CCFR Detector Pict}
\end{figure}

Neutrino events were observed in the Lab E neutrino detector \cite{detector},
shown in Fig.~\ref{CCFR Detector Pict}. The detector is constructed as a
target calorimeter followed by a toroid muon spectrometer. The 690 ton
calorimeter consists of eighty-four 3 m$\times $3 m$\times $10 cm iron
plates interspersed with scintillators and drift chambers, with a measured
hadronic-energy resolution $\sigma /E_{had}=0.89/\sqrt{E_{had}}$. 
The toroid spectrometer has five sets of drift chambers for muon tracking as
well as hodoscopes for triggering; its momentum resolution is limited by
multiple Coulomb scattering to $\sigma _P/p=0.11$. The detector was
periodically calibrated by a test beam of pions and muons.

NuTeV  \cite{NuTeV}, the next generation of deep inelastic neutrino scattering
experiments at Fermilab, began taking data in May, 1996. NuTeV has modified
the beam line to select the sign of the charged pions and kaons \cite{NuTeV
Beam}. The resulting beam is almost purely $\nu _\mu $ or $\overline{\nu }%
_\mu $, depending on the sign selection. The enhanced purity of the
antineutrino sample will improve a number of QCD studies, especially those
concerned with small-$x$ physics. Separate running will also permit the
first high-statistics simultaneous measurements of the electroweak mixing
parameter $\sin ^2\theta _W$ and neutral-to-charged-current coupling
strength ratio $\rho $. The substantially upgraded Lab E neutrino detector
is continuously calibrated with a test beam of pions, electrons or muons
throughout the running. The new test beam also provides a broader range of
hadron and muon energies than were available to CCFR.

\subsubsection{The CDHSW Experiment}

The CDHSW experiment \cite{CDHexpt2,CDHexpt1} measured the total neutrino
cross section using 100, 160 and 200 GeV narrow band neutrino beams and
performed precision electroweak and structure function measurements with a
wide-band beam during the early 1980's. Primary protons from the CERN SPS
were used to produce secondary pions and kaons. For the wide-band running,
the secondary beam was focussed using a magnetic horn. Depending on the sign
selection set by the magnetic field, the tertiary beam consisted mainly of
neutrinos or antineutrinos. An antineutrino contamination in a neutrino beam
will result in single-muon events with a muon that has positive rather than
negative charge. This contamination is therefore called the ``wrong-sign''
background. The fraction of wrong-sign background events for CDHSW neutrino
running was 2.3\%. For antineutrino running, the wrong-sign background was
17\%  \cite{CDHSWSF}.

This experiment used the upgraded CDHS detector, which consisted of
toroidally-magnetized iron plates sandwiched between planes of scintillator
and drift chambers. The calorimeter was 3.75 m in diameter, 21 m long, and
had a total mass of approximately 1100 tons. The hadronic energy resolution
of the detector varied, depending on the segmentation of the iron, from $%
\sigma /E_{had}\approx 0.58/\sqrt{E_{had}}$ to $0.70/\sqrt{E_{had}}$. The
momentum of the muon was determined by the bend in the magnetic field, with
an average resolution of 9\%.  
The field of the toroid was chosen to focus $\mu ^{+}$ or $\mu ^{-}$  
based on the setting of the magnetic horn.

\subsubsection{The CHARM Experiment}

The CHARM experiment \cite{CHdet} made a high precision measurement of the
total cross section for neutrinos and antineutrinos using the 160 GeV narrow
band neutrino and antineutrino beam at the CERN SPS during the early 1980's.
The CHARM detector was composed of a fine-grained calorimeter followed by a
muon spectrometer. The calorimeter consisted of 78 marble plates (CaCO$_3$)
of dimension 3 m$\times $3 m$\times $ 8 cm. Scintillation counters,
proportional drift tubes, and streamer tubes were located between the
plates. The resulting hadronic energy resolution was $\sigma /E_{had}=0.49/%
\sqrt{E_{had}}$. The vertex position resolution was 3 cm at
$E_{had}=50$ GeV.  
The muon spectrometer consisted of an iron toroid spectrometer surrounded by
a magnetized iron frame and instrumented with proportional chambers. The
calorimeter was calibrated with electron and pion beams.

\subsubsection{The CHARM II Experiment}

The CHARM II experiment \cite{CHARM II Detector} collected data during the
period 1987-1991, using the wide-band neutrino beam described in the above
CDHSW experiment section. This experiment was optimized for detection of
electrons. The detector's target was composed of 48 mm(0.5 radiation length)
thick plates of glass, with a total mass off  690 tons.  
Interspersed between the plates were streamer tubes, and between every five
sets of plates and streamer tubes were scintillation counters. The
electromagnetic energy resolution from the streamer tubes was $\sigma
/E=0.09+0.15/\sqrt{E}$. The muon spectrometer consisted of magnetized iron
toroids instrumented with scintillation counters and drift chambers. The
resolution was $\Delta P/P=13\%$ at $20$ GeV.

\subsubsection{The BEBC Experiment}

The Big European Bubble Chamber (BEBC) experiment \cite{BEBCexpt} ran with
both narrow- and wide-band beams provided by the CERN SPS during a series of
experiments that began in the late 1970's and continued during the 1980's
using a variety of targets. This Review mainly discusses results from the
high statistics BEBC runs with deuterium (WA25) and neon (WA59) targets. The
cryogenic bubble chamber had a 10~m$^3$ fiducial volume surrounded by a 3~T
superconducting magnet and supplemented with a two-plane, 150~m$^2$ external
muon identification system.

\section{Neutrino Flux and Total Cross Section Measurements}

\label{Flux Section}

Accurate measurement of the flux of neutrinos in a beam is necessary for all
of the analyses discussed in this Review. Flux monitoring is a difficult
task for a beam of neutral weakly interacting particles; hence indirect
methods must be employed. Flux determinations can be based on measurements
of the secondary beam. Alternatively, experiments have used fits to the $y$
distribution of the data or integration over the observed low $y$ events.
These methods rely upon the relation between the cross section, flux and
number of events ($N_{\nu ,{\overline{\nu }}}$): 
\begin{equation}
\frac{dN_{\nu ,{\overline{\nu }}}}{dy}=\Phi _{\nu ,\overline{\nu }}(E)\frac{%
d\sigma _{\nu ,\overline{\nu }}}{dy},  \label{eq:Nfluxsig}
\end{equation}
and on assumption that as $y\rightarrow 0$, 
\begin{eqnarray}
\frac 1{E_\nu }\frac{d\sigma ^{\bar{\nu}N}}{dy} &=&\frac 1{E_\nu }\frac{%
d\sigma ^{\nu N}}{dy}  \label{low y limit} \\
&\rightarrow &C,  \nonumber
\end{eqnarray}
where $C$ is a constant independent of energy.  
The Flux measurement errors are among the largest systematic errors
associated with the neutrino precision measurements, therefore an extended
discussion of the subject is presented here.

\subsection{Absolute Flux Determination}

Neutrino beams result mainly from the decay of secondary pions and kaons
which are produced when a proton interacts with a target. Most pion and kaon
decays are two-body, thus the neutrino energy and flux can be determined
using kinematic constraints if the integrated intensity, energy, and spacial
distribution of the secondary beam are known. This method is most successful
when applied to narrow-band beams and was employed by many experiments to
measure structure functions and the total cross section. These experiments
include CCFR E616 \cite{Oltman,Auchincloss}, CDHSW \cite{CDHSWtot}, and CHARM%
 \cite{Charmtot}.

The primary and secondary beams are monitored using a variety of detectors,
including RF cavities, toroids, Cerenkov counters, segmented wire ionization
chambers, and hodoscopes. Redundancy is important to reduce the systematic
errors in the measurement of the secondary beams. Two uncertainties
associated with this technique result from three-body decays of kaons and
pion or kaon decays that occur before the momentum selection magnets \cite
{Auchincloss}. The former are largely from $K\rightarrow \pi \mu \nu $,
which can be corrected using a Monte Carlo simulation of the neutrino flux.
The latter correction can be measured directly by placing a dump in the beam
after the magnet, thus absorbing the sign-selected secondary beam. All
interactions in this mode would result from neutrinos produced in decays
before the magnet. As an additional constraint to reduce errors further, the
CHARM collaboration also monitored the muon flux from the tertiary beam \cite
{Charmtot}.

\subsection{Relative Flux}

The structure function measurements described in Sec.~\ref{SF Section} are
very sensitive to the accurate determination of the relative flux among the
various energy bins and between neutrinos and antineutrinos. Therefore, it
is important to establish methods of normalizing between neutrino energy
bins which have small systematic errors and little dependence on assumptions
of nucleon structure. These methods are crucial for experiments using
wide-band neutrino-beams, where beam monitoring methods cannot provide
sufficient accuracy. In a narrow-band beam, these methods can be used to
test the accuracy of the measurements from secondary beam monitoring.

The ``$y$-intercept technique'' relies on the premise that near zero
hadronic energy transfer, $y=E_{had}/E_\nu \rightarrow 0$, the differential
inelastic cross section divided by energy is independent of energy and is
the same for neutrinos and antineutrinos (Eq. \ref{low y limit}).%
The differential cross section for neutrino scattering, given by Eq.~\ref
{eq:sigsf}, in the $y\rightarrow 0$ limit reduces to: 
\begin{eqnarray}
&&\Bigl[{\frac 1{{E_\nu }}}{\frac{{d\sigma ^\nu }}{{dy}}}\Bigr]_{y=0}=\Bigl[{%
\frac 1{{E_\nu }}}{{\frac{d\sigma ^{\overline{\nu }}}{{dy}}}}\Bigr]_{y=0} 
\nonumber \\
&=&\ \ \ {\frac{G_F^2M}\pi }\int_0^1F_2(x,Q^2\rightarrow 0)dx=Const.
\end{eqnarray}
This is true independent of incident neutrino energy or neutrino type for
scattering on an isoscalar target. Thus, for each bin in energy, the flux of
neutrinos or antineutrinos is given by: 
\begin{equation}
\lim\limits_{y\rightarrow 0}\left( \frac 1{E_\nu }\frac{dN{^{\nu (\overline{%
\nu })}}(E_\nu )}{dy}\right) =Const.\times \Phi ^{\nu (\overline{\nu }%
)}(E_\nu ).
\end{equation}
This method was applied by CCFR (E616) to cross-check the flux obtained from
secondary-beam measurements for the determination of the neutrino total
cross section.

For wide-band neutrino beams, the beam energy and radial distribution are
usually determined through iterating the Monte Carlo simulation of the
neutrino events, successively varying the input flux and structure functions
until the Monte Carlo matches the observed distribution of events. Initial
distributions for the structure functions are taken from previously
published results. These structure functions and initial distributions for
the neutrino flux are used to generate events which are then put through the
Monte Carlo simulation of the detector. The ratio of simulated events to
detected events in each $x$-$y$-$E_\nu $ bin provide the correction factors
for the flux and structure functions for the next iteration. Variations on
this method are used by both the CDHSW and CCFR collaborations.

The CCFR E744/E770 ``fixed-$\nu _0$ method'' provides a specific example of
one iterative technique for extracting the relative flux \cite{Seligman}.
This method exploits the $y$-dependence of the data to obtain the flux. The
relationships of Eq.'s ~\ref{eq:Nfluxsig} and \ref{eq:sigsf} can be extended
to small $\nu $ as a polynomial in terms of $y={\nu /E_\nu }$, 
\begin{equation}
{\frac{dN_{\nu ,{\overline{\nu }}}}{d\nu }}=A_{\nu ,\overline{\nu }}+B_{\nu ,%
\overline{\nu }}\/(\nu /E_\nu )+(C_{\nu ,\overline{\nu }}/2)\/(\nu /E_\nu
)^2,  \label{eq:Nyparam}
\end{equation}
where $A$, $B$ and $C$ are given by 
\begin{eqnarray}
A &=&\frac{G_F^2M}\pi \int_0^1F_2(x,Q^2)\Phi (E_\nu)dx,  \nonumber \\
B &=&-\frac{G_F^2M}\pi \int_0^1(F_2(x,Q^2)\mp xF_3(x,Q^2))\Phi (E_\nu)dx, \\
C &=&B-\frac{G_F^2M}\pi \int_0^1F_2(x,Q^2)\tilde{R}(x,Q^2)\Phi (E_\nu)dx, 
\nonumber
\end{eqnarray}
with 
\begin{equation}
\tilde{R}(x,Q^2)\equiv \frac{{1+2Mx/\nu }}{1+R_L(x,Q^2)}-{\frac{{Mx}}{{\nu }}%
}-1.
\end{equation}

Except for small variations due to the scaling violations of the structure
functions, $A$, $B$ and $C$ are relatively independent of $\nu$ and $E_\nu $%
. Integrating equation \ref{eq:Nyparam}, the number of events with $\nu <\nu
_0$ for a given energy, $N_{\nu <\nu _0}^{Obs}(E_\nu)$, is given by: 
\begin{equation}
N_{\nu <\nu _0}^{Obs}(E_\nu) = \Phi (E_\nu)\int_0^{\nu _0}d\nu A \left[ 1+\frac \nu {%
E_\nu }\frac BA - \frac{\nu^2}{2E_\nu ^2} \left( \frac BA-\frac{\int F_2%
\tilde{R}}{\int F_2}\right) \right].
\end{equation}
From the values of $N_{\nu <\nu _0}^{Obs}(E_\nu)$, $\Phi (E_\nu)$ can be determined
for both neutrinos and antineutrinos up to an overall normalization
constant. The overall normalization is determined from to the world-average
cross-section for $\nu $Fe (Sec. \ref{Nu Cross Section}) \cite{Seligman}, 
\[
\sigma ^{\nu Fe}/E_\nu=(0.677\pm 0.014)\times 10^{-38}{\rm cm}^2/{\rm GeV}. 
\]

The assumption that the integrals over the structure functions are
independent of $\nu $ is only approximately true. Recall that $Q^2=2M\nu x$,
therefore for fixed $\nu $, using $F_2$ as an example, 
\begin{equation}
\int_0^1F_2(x,Q^2)=\int_0^1F_2(x,2M\nu x)\equiv F_2^{int}(\nu )
\label{eq:F2int}
\end{equation}
The systematic error introduced by this residual $\nu $-dependence was
studied by varying the $\nu$ range from which the flux was extracted and
found to be less than 0.5\%  \cite{Seligman}.

\subsection{Electron Neutrino Flux}

\begin{figure}[ptbp]
\psfig{figure=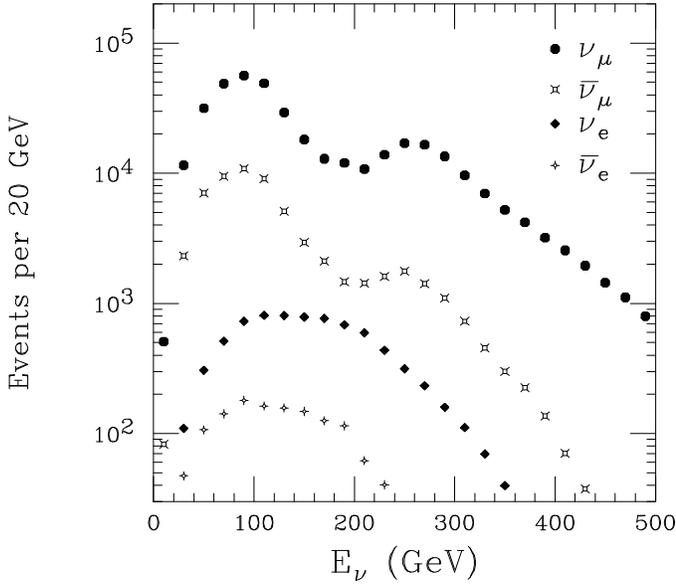,width=3.5in}
\caption{The CCFR 744/770 neutrino and antineutrino event spectrum.}
\label{CCFR nue flux}
\end{figure}

Most neutrino beams are composed mainly of muon-type neutrinos and
antineutrinos, with a small contamination from the electron flavor. Accurate
determination of this contamination is important for the precision
measurement of $\sin ^2\theta _W$ discussed later. The $\nu_e / \overline{\nu%
}_e$ flux is evaluated through detailed Monte Carlo simulations which are
tied to the direct muon-flavor flux measurements. As an example of the
magnitude of the contamination, the CCFR E744/770 Monte Carlo Flux
determination is shown in Fig.~\ref{CCFR nue flux} \cite{King}. As a test of
the accuracy of this method, a recent analysis of the CCFR data to determine
directly the total number of $\nu _e$ interactions has confirmed the Monte
Carlo prediction \cite{Romosan}.

\subsection{Measurements of the Total Cross Section}

\label{Nu Cross Section}

The total neutrino and antineutrino cross sections have been measured in a
large number of experiments. This section reports on those experiments which
have measured the total cross section to an accuracy of at least 10\%. The
CHARM \cite{Charmtot}, CDHSW \cite{CDHSWtot}, BEBC \cite{BEBCtot2,BEBCtot1},
CCFR E616 \cite{MacFar}, and CCFR E701 \cite{Auchincloss} results are from
experiments described in detail above.   
Two older experiments, the 15-ft Bubble Chamber (15-ft BC) at Fermilab \cite
{BC15tot1,BC15tot2}, and FNAL E310 (WHPFOR), a counter experiment with an
iron target \cite{Mann}, also have results with small errors and are
discussed below. All results are presented for interactions with an
isoscalar target, and a correction has been applied to the iron data to
compensate for the neutron excess.

\begin{figure}[ptbp]
\psfig{figure=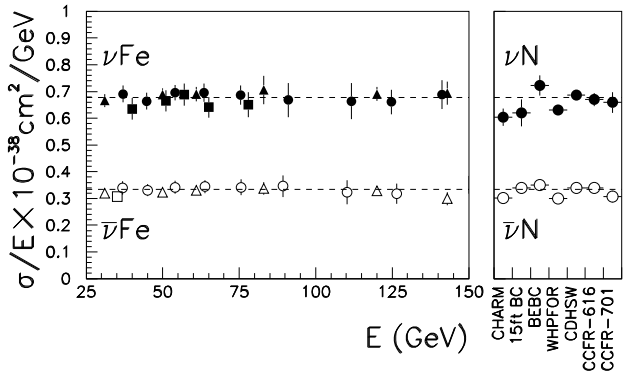,bbllx=33bp,bblly=350bp,bburx=380bp,bbury=600bp,width=3.5in,clip=T}
\caption{Neutrino (solid) and antineutrino (open) measurements of $%
\sigma_{tot}/E_\nu$. Left plot: $\sigma_{tot}/E_\nu$ versus 
$E_\nu$ for iron targets. Circles --
CCFR E616, Squares-- CCFR E701, triangles -- CDHSW. Right plot: average $%
\sigma_{tot}/E$ for a variety of targets (isoscalar corrected).  Dashed lines
indicate average for iron data (see text). (The error bars
include both statistical and systematic errors.)}
\label{Cross Section Plot}
\end{figure}

A linear dependence of the total cross section on energy is expected if
point-like scattering between neutrinos and quarks dominates the scattering
mechanism. This behavior is observed in the neutrino-iron total cross
section measurements from CCFR E616 \cite{MacFar}, CCFR E701 \cite{Auchincloss}
and CDHSW \cite{CDHSWtot}, as shown in Fig.~\ref{Cross Section Plot}. The
ratio $\sigma _{tot}/E_\nu $ as a function of $E_\nu $ is in agreement with
a straight line with no slope for both neutrino and antineutrino
interactions.

The total cross section for neutrino scattering on iron is used to provide
the overall normalization of the flux for the CCFR E744/E770 experiment.
Table~\ref{Cross Section Table} summarizes the isoscalar corrected
measurements for neutrino-iron experiments. The world averages for these
experiments are indicated by the dashed lines in Fig.~\ref{Cross Section
Plot}(a). The data from all experiments are in good agreement with these
averages. The E744/E770 data also yield a measurement of $\sigma ^{\overline{%
\nu }}/\sigma ^\nu $=$0.509\pm 0.010$ \cite{Seligman} which is listed with
the other measurements in Table~\ref{Cross Section Table}.

\begin{table}[tbp]
\begin{center}
\begin{tabular}{ccccc}
Experiment & $\sigma^{\nu}/E$ & $\sigma^{\overline{\nu}}/E$ & $\sigma^{%
\overline{\nu}}/\sigma^{\nu}$ & ref. \\ \hline
CCFR (E616) & $0.669 \pm 0.024$ & $0.340 \pm 0.020$ & $0.499 \pm 0.025$ & 
 \cite{MacFar} \\ 
CDHSW & $0.686 \pm 0.020$ & $0.339 \pm 0.022$ & $0.495 \pm 0.010$ &  \cite
{CDHSWtot} \\ 
CCFR (E701) & $0.659 \pm 0.039$ & $0.307 \pm 0.020$ & $0.467 \pm 0.028$ & 
 \cite{Auchincloss} \\ 
CCFR (E744/770) &  &  & $0.509 \pm 0.010$ &  \cite{Seligman} \\ \hline
World Ave. & $0.677 \pm 0.014$ & $0.334 \pm 0.008$ & $0.500 \pm 0.007$ & 
\end{tabular}
\end{center}
\caption{Measurements of the neutrino cross section on iron}
\label{Cross Section Table}
\end{table}

\begin{figure}[ptbp]
\psfig{figure=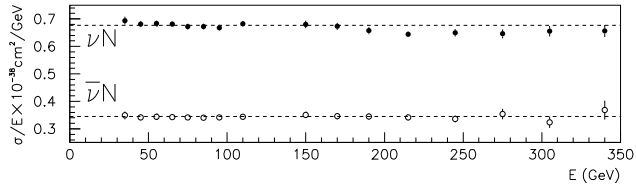,bbllx=0bp,bblly=500bp,bburx=540bp,bbury=700bp,width=3.5in,clip=T}
\caption{$\sigma_\nu/E_\nu$ for $\nu_\mu$ and $\overline{\nu}_\mu$ from CCFR
E744/E770. Normalization is based on the world averages given in 
Table~\protect\ref{Cross Section Table}}
\label{CCFR sigma/E}
\end{figure}

The world-average $\bar{\nu}/\nu $ cross-section ratio, in combination with
the world-average neutrino cross section, sets the overall normalization for
CCFR E744/E770 data. Given this normalization, CCFR has measured the energy
dependence of $\sigma _{tot}/E_\nu $ up to energies of 400 GeV. The result
is shown in Fig.~\ref{CCFR sigma/E}, where systematic and statistical errors
have been added in quadrature. The data have been fit to a model which
allows a slope in $\sigma _{tot}/E_\nu $, denoted $\varepsilon _{\nu (\bar{%
\nu})}$ for neutrinos (antineutrinos), 
\begin{equation}
\sigma _{tot}^{\nu (\bar{\nu})}/E_\nu =\sigma ^{\nu (\bar{\nu})}\left(
1+\varepsilon _{\nu (\bar{\nu})}E_\nu \right) ;
\end{equation}
the results are
\begin{eqnarray*}
\varepsilon _\nu  &=&(-2.2\pm 0.7)\%/(100\text{ GeV)\quad(CCFR),} \\
\varepsilon _{\bar{\nu}} &=&(-0.2\pm 1.3)\%/(100\text{ GeV)\quad(CCFR).}
\end{eqnarray*}
A small slope for the neutrino cross section is expected as a consequence of
QCD and heavy quark effects \cite{Hin77}.

Figure~\ref{Cross Section Plot}(b) shows the average total cross section
measurement for neutrino-iron targets as well as neutrino-nucleon scattering
from various isoscalar targets by CHARM \cite{Charmtot}, the 15-ft BC \cite
{BC15tot1,BC15tot2}, and BEBC \cite{BEBCtot2,BEBCtot1}. The approximate
average beam energy of these lower energy measurements is $25~$GeV. The
results are in good agreement with the average for the high energy
neutrino-iron cross section indicated by the dashed line.

\section{Measurements of Nucleon Structure from Inclusive Scattering}

\label{SF Section}

Precision deep inelastic neutrino scattering experiments provide
opportunities to test QCD evolution and extract the QCD parameter $\Lambda $%
, which sets the scale of the strong interaction. In the kinematic regions
where the structure of the nucleon can be interpreted in terms of quarks,
neutrino scattering possesses high sensitivity to many individual parton
distributions. Probing the nucleon with neutrinos complements charged-lepton
scattering experiments, and comparisons between neutrino and charged-lepton
deep inelastic experiments provide tests of the universality of the
structure functions. Global analyses which include the neutrino data and
other cross section measurements yield the best parameterizations of the
individual parton distributions.

This section reviews the contributions of neutrino deep inelastic scattering
experiments to precision measurements of nucleon structure. A brief review
of the theory of the structure functions is provided. Experimental
techniques and issues which must be considered in the analyses are reviewed.
Results of the high precision neutrino measurements are compared between
experiments and to charged-lepton scattering results. Finally, the QCD
analysis of the structure functions is presented.

\subsection{Theoretical Framework}

This section describes the theoretical framework in which the structure
function measurements will be interpreted. The QCD expectation is described
along with a discussion of non-perturbative and nuclear effects.

\subsubsection{The Structure Functions within QCD}

\label{QCD SF theory}

The general form for the differential cross section depends upon three
structure functions, $F_2$, $R_L$ and $xF_3$, as shown in Eq.~\ref{eq:sigsf}%
. In the QPM, the neutrino structure functions can be written as functions
of sums and differences of the momentum-weighted quark probability
densities, as described in Sec.~\ref{QPM Section}.  
QCD modifies this parton model interpretation in order to account for the
interactions between the partons.

Within QCD, partons with higher fractional momentum may contribute to
interactions at any lower $x$ through radiation or gluon splitting. The
probability that the interaction resolves such a splitting depends on $Q^2$,
as described by the Dokshitser-Gribov-Lipatov-Altarelli-Parisi (DGLAP)
equations \cite{DGLAP1,DGLAP2,DGLAP3,DGLAP4}: 
\begin{eqnarray}
\frac{dq^{NS}(x,Q^2)}{d\ln Q^2} &=&\frac{\alpha _S(Q^2)}{2\pi }\int_x^1\frac{%
dy}yq^{NS}(y,Q^2)P_{qq}(x/y),  \nonumber \\
\frac{dq^S(x,Q^2)}{d\ln Q^2} &=&\frac{\alpha _S(Q^2)}{2\pi }\int_x^1
\frac{dy}y \left[ q^S(y,Q^2)P_{qq}^S(x/y) \right. \\
 &+&\left. G(y,Q^2)P_{qG}(x/y) \right],  \nonumber \\
\frac{dG(x,Q^2)}{d\ln Q^2} &=&\frac{\alpha _S(Q^2)}{2\pi }\int_x^1
\frac{dy}y \left[ q^S(y,Q^2)P_{Gq}(x/y) \right. \nonumber \\
 &+&\left. G(y,Q^2)P_{GG}(x/y) \right].  \nonumber
\end{eqnarray}
The splitting functions, $P_{ij}(x/y)$, give the probability that parton $j$
with momentum $y$ will be resolved as parton $i$ with momentum $x<y$ . The
symbols $q^{NS}=\sum\limits_i(q_i-\overline{q}_i)$ and $q^S=\sum%
\limits_i(q_i+\overline{q}_i)$ refer to the non-singlet and singlet quark
distributions, respectively. The probability of finding a gluon in the
nucleon carrying a fractional momentum $x$ is represented by $G(x,Q^2)$.

From the DGLAP equations, it can be seen that the change in the structure
functions with $Q^2$ as a function of $x$ depends on $\alpha _s$. At next to
leading order, $\alpha _S$ is given by: 
\begin{equation}
\alpha _S(Q^2)=\frac{4\pi }{\beta _0\ln (Q^2/\Lambda ^2)}\left( 1-\frac{
\beta _1}{\beta _0}\frac{\ln [\ln (Q^2/\Lambda ^2)]}{\ln (Q^2/\Lambda ^2)}
\right) ,
\end{equation}
where $\beta _0=11-2n_f/3$ , $\beta _1=102-38n_f/3$, and $n_f$ is the number
of quark flavors participating in the interaction at this $Q^2$. The QCD
coupling constant $\Lambda $ must be introduced when the renormalization
technique is applied to remove the divergences within QCD. For the
discussion here, the modified minimal subtraction scheme ($\overline{ MS}$)
is used.

The structure function $R_L=\sigma _L/\sigma _T$ is zero in the simple
parton model. However, QCD effects such as quark-gluon bremsstrahlung 
and $q%
\overline{q}$ pair-production introduce transverse momentum leading to a
small, non-zero value of $R_L$ which has a predicted $x$ and $Q^2$
dependence. Within QCD \cite{RL Theory}:
\begin{eqnarray}
&&R_{L,QCD}(x,Q^2)= \\
&&\frac{\frac{\alpha _s}\pi \int_x^1\frac{dy}y\left[ \sum\limits_{i=q,%
\overline{q}}q_i(x)\sigma ^L\left( \frac xy\right)
+\sum\limits_{i=q}G(x)\sigma ^L\left( \frac xy\right) \right] }{%
\sum\limits_{i=q,\overline{q}}q_i(x)+\frac{\alpha _s}\pi \int_x^1\frac{dy}y%
\left[ \sum\limits_{i=q,\overline{q}}q_i(x)\sigma ^L\left( \frac xy\right)
+\sum\limits_{i=q}G(x)\sigma ^L\left( \frac xy\right) \right] .}  \nonumber
\end{eqnarray}
Given well-measured quark distributions and $\alpha _S$ from $xF_3$ and $F_2$%
, a precise measurement of $R_L$ can be a sensitive probe of the gluon
distribution.

\subsubsection{Nonperturbative QCD Effects}

\label{nonpert}

Non-perturbative QCD processes which contribute to the structure-function
measurements are collectively termed higher-twist effects. These effects
occur at small $Q^2$ where the impulse approximation of scattering from
massless non-interacting quarks is no longer valid. Examples include target
mass effects, diquark scattering, and other multiparton effects. Because
neutrino experiments use heavy targets in order to obtain high interaction
rates, nuclear effects must also be considered.

\paragraph{Target-Mass and Higher-Twist Effects.}

The target-mass correction \cite{Targmass} to the structure functions
accounts for the mass of the nucleon $M$ by rescaling the apparent
fractional momentum of the quark: $x\rightarrow \xi =2x/(1+k)$, where $%
k=\left( 1+4x^2M^2/Q^2)\right) ^{\frac 12}$. The structure functions are
then: 
\begin{eqnarray}
F_2^{TMC} &=&\frac{x^2}{k^2}\frac{F_2^{QCD}}{\xi ^2}+\frac{6M^2}{Q^2}\frac{%
x^3}{k^4}I_1+\frac{12M^4}{Q^4}\frac{x^4}{k^5}I_2, \\
xF_3^{TMC} &=&\frac{x^2}{k^2}\frac{xF_3^{QCD}}{\xi ^2}+\frac{2M^2}{Q^2}\frac{%
x^3}{k^3}I_3.  \nonumber
\end{eqnarray}
In the above equations, 
\begin{eqnarray}
I_1 &=&\int_\xi ^1duF_2^{QCD}(u,Q^2)/u^2,  \nonumber \\
I_2 &=&\int_\xi ^1du\int_u^1dvF_2^{QCD}(v,Q^2)/v^2, \\
I_3 &=&\int_\xi ^1duxF_3^{QCD}(u,Q^2)/u^2.  \nonumber
\end{eqnarray}
For the QCD analysis presented in Sec. \ref{QCD SF Analysis}, the structure
functions are corrected to remove the target-mass effect.

\begin{figure}[pthb]
\psfig{figure=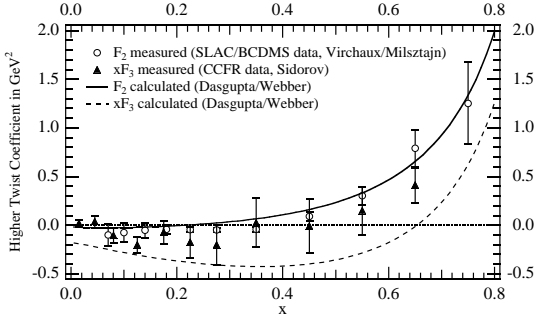,bbllx=0bp,bblly=490bp,bburx=540bp,bbury=800bp,width=3.5in,clip=T}
\caption{Coefficients of the Higher-Twist parameterization.  Solid --
$xF_3$ from Sidorov (1996).  
Open -- $F_2$ from Virchaux and Milsztajn (1992). 
The solid(dashed) line is the $F_2$($xF_3$) result from Dasgupta and
Webber (1996).}
\label{sidfig}
\end{figure}

The remaining higher-twist effects cannot be calculated a priori from PQCD,
but must be measured. Virchaux and Milsztajn \citeyear{VirchMil} compared SLAC
electron scattering and BCDMS muon scattering measurements of $F_2$  to QCD
expectations.  
Deviations were attributed entirely to higher-twist effects and were fit to
the form: 
\begin{equation}
F_2^{Measured}/F_2^{predicted}=(1+C_i/Q^2).
\end{equation}
The constants, $C_i$, calculated for each $x$-bin are shown in Fig.~\ref
{sidfig}.
This determination was done using a value of $\alpha_s$ smaller than
the current world average which may make these higher twist
corrections too large.  

The \citeasnoun{VirchMil} corrections are applicable to a measurement 
of $F_2$ in neutrino
scattering, however $xF_3$ may not have the same higher-twist corrections as $%
F_2$. Figure~\ref{sidfig} also shows the measured 
higher-twist contribution extracted from the 
CCFR $xF_3$ data based on a NLO analysis following the prescription used
for  $F_2$ \cite{Sidorov}.  In addition, Fig.~\ref{sidfig} shows the  
calculation by \citeasnoun{DW} which includes infrared 
renormalon contributions.  
The measured higher-twist contribution to the two structure
functions are consistent; however, the coefficient for the $x=0.65$ bin of $%
xF_3$ is almost a factor of two lower than that for $F_2$. 
A preliminary analysis
of the recent CCFR $F_2$ and $xF_3$ results also indicates 
higher-twist corrections that are smaller than the \citeasnoun{VirchMil}
or \citeasnoun{DW} determinations.  For these reasons, the new CCFR
analysis uses one-half of the \citeasnoun{DW} values, with a systematic
error given by repeating the analysis with no correction and with the
full correction.

\paragraph{Nuclear Effects in Neutrino Scattering}

\label{nu nukes}

Dependence on the mass number $A$ of the nucleus could arise from several
effects:

\begin{itemize}
\item  Fluctuations of the intermediate virtual boson to mesons. This might
suppress the bound nucleon (iron) structure function compared to that from a
free (hydrogen) or nearly free (deuterium) nucleon  \cite{VDM}. The
vector-meson-dominance model ascribes the cause of shadowing to fluctuations
of the boson into mesons leading to strong interactions near the ``surface''
of the nucleus. This model was developed for charged-lepton scattering,
where the photon can fluctuate only into vector mesons.  
However, for neutrinos the $W$ has an axial as well as a vector component.
Vector and axial vector meson dominance effects are expected to mainly
affect the low-$x$ region.   

\item  Gluon recombination which can occur in a large nucleus between
partons of neighboring nucleons \cite{recomb1,recomb2}. This leads to an $A$%
-dependent depletion of low-$x$ partons.  

\item  The EMC Effect, which denotes a suppression of the structure
functions from high-$A$ targets compared to deuterium in the $0.2<x<0.7$
range. Many theoretical explanations have been developed to explain the EMC
effect \cite{Geesaman}, which has been most clearly observed in charged
lepton scattering experiments.   
These include multiquark-clusters \cite{clusters1,clusters2},
dynamical-rescaling \cite{dynresc}, and nuclear-binding effects \cite
{bind1,bind2}. In each of these models, the effect is independent of the
type of boson probe, and thus is expected to appear also in neutrino
scattering using high-$A$ targets.

\item  Fermi motion of nucleons within the nucleus, which is expected to
affect nuclear targets at very high $x$  \cite{BodRich1,BodRich2}.
\end{itemize}

\begin{figure}[ptbp]
\psfig{figure=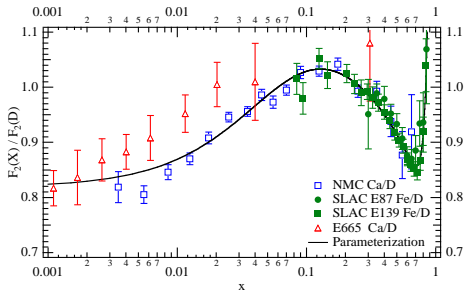,bbllx=10bp,bblly=250bp,bburx=600bp,bbury=550bp,width=3.5in,clip=T}
\caption{Fit to the NMC, E665 and SLAC nuclear data. The resulting fit is
used to correct the charged-lepton deuterium data for comparison with the
CCFR result}
\label{nukes}
\end{figure}

Figure~\ref{nukes} shows the $A$-dependence of structure function data
measured in deep inelastic charged-lepton scattering. If this $A$-dependence
is entirely attributed to effects within the nuclear target, as opposed to
propagator effects, then this figure should also describe the expected
nuclear dependence for neutrino scattering from a high-$A$ target.

\subsection{Experimental Techniques for Measuring $xF_3$ and $F_2$}

Various experimental techniques are used to extract $F_2$ and $xF_3$ from
the measured number of events. This section considers the methods used by
CCFR, CDHSW and CHARM as examples. In all cases, the assumptions concerning
the longitudinal structure function $R_L$ affect the final result and must
be considered in any comparisons.

\subsubsection{The CCFR Measurement of $F_2$ and $xF_3$}

\label{SF CCFR}

The recent CCFR E744/E770 measurements of $F_2$ and $xF_3$ provide an
example of a technique of extracting structure functions in neutrino
experiments with very small systematic errors. The resulting measurements
demonstrate the high precision which recent neutrino experiments have been
able to achieve. As a result of this precision, neutrino experiments now
quantitatively test QCD at next-to-leading order.

The structure functions $F_2$ and $xF_3$ are extracted from $\nu N$ and $%
\bar{\nu}N$ differential cross sections given the fluxes ($\Phi _\nu ,\Phi _{%
\bar{\nu}}$) determined by the method described in Sec. \ref{Flux Section}%
 \cite{Seligman}. The number of events in each $x$ and $Q^2$ bin is related
to the differential cross section by: 
\begin{equation}
N=\rho LN_A\int_{x-bin}\int_{Q^2-bin}\left[ \int_{all~energies}\frac{%
d^2\sigma }{dxdQ^2}\Phi (E)dE\right] ,
\end{equation}
where $\rho $ is the target density, $L$ is the target length and $N_A$ is
Avogadro's 
number. The structure functions are extracted by varying the
Monte Carlo input to match the number of events in each acceptance-corrected
bin.

The CCFR results presented here are from the final analysis of the
E744/E770 data which differ from the preliminary results of %
 \citeasnoun{quintas}. Detailed information on the improvements to the
analysis which lead to differences from the preliminary results can be found
in  the articles by \citeasnoun{Seligman}.

\begin{figure}[ptbp]
\psfig{figure=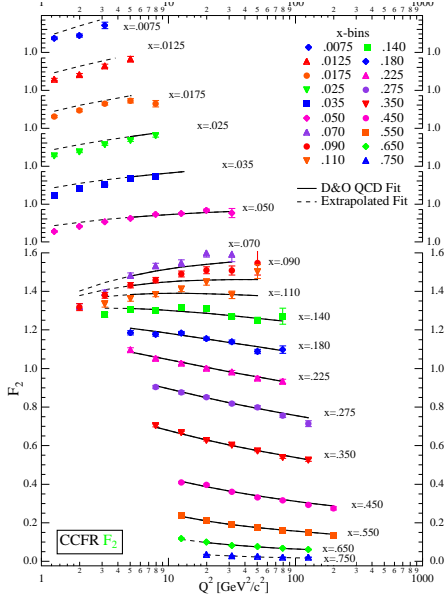,bbllx=0bp,bblly=100bp,bburx=600bp,bbury=700bp,width=3.5in,clip=T}
\caption{CCFR $F_2$ measurement. Errors are statistical only. Lines indicate
the CCFR QCD fit discussed in the text.}
\label{CCFR-F2-plot}
\end{figure}

\begin{figure}[ptbp]
\psfig{figure=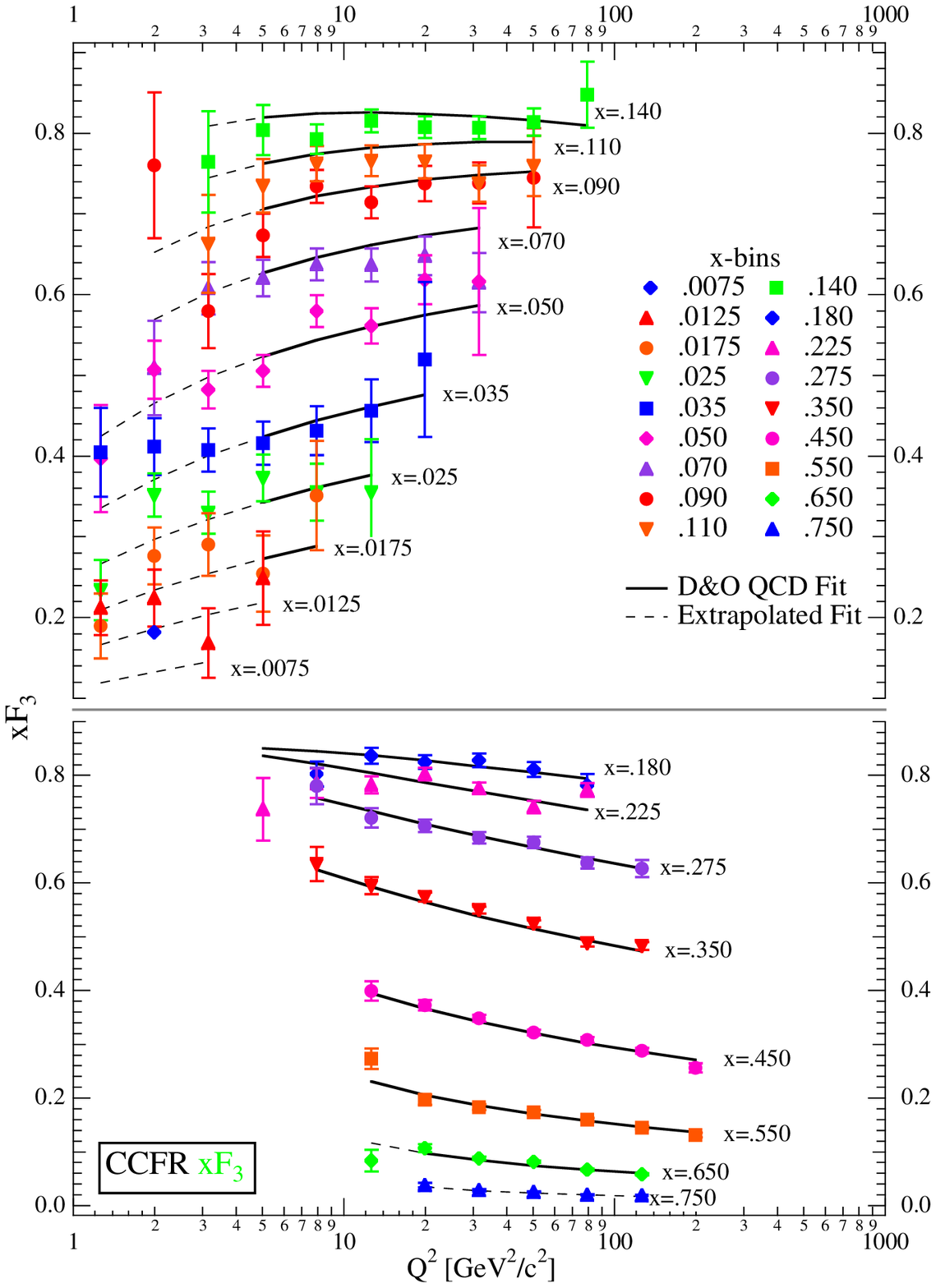,bbllx=0bp,bblly=100bp,bburx=600bp,bbury=700bp,width=3.5in,clip=T}
\caption{CCFR $xF_3$ measurement. Errors are statistical only. Lines
indicate the CCFR QCD fit discussed in text. }
\label{CCFR-xF3-plot}
\end{figure}

Figures~\ref{CCFR-F2-plot} and \ref{CCFR-xF3-plot} show the CCFR
measurements of $F_2$ and $xF_3$ as a function of $Q^2$ in various bins of $x
$. Only statistical errors are shown, and the systematic errors are
approximately 2\%. The solid and dashed curves on the plot indicate the QCD
fits which are discussed in Sec. \ref{QCD SF Analysis}. The precision of the
data is such that subtle deviations from the QCD expectation can be probed.

\subsubsection{$F_2$ and $xF_3$ Extracted from the Cross Section}

The structure functions can be extracted by forming the sums and differences
of the differential cross section given in Eq.~\ref{eq:sigsf}. This
technique was used by the CDHSW and CHARM experiments. To make the
measurement, the number of events in each $x$ and $Q^2$ bin is related to
the differential cross section by 
\begin{equation}
\frac 1{E_\nu }\frac{{d^2\sigma }}{dxdQ^2}=\frac{N(x,Q^2,E_\nu {)}}{N(E_\nu )%
}\frac{\sigma _{tot}/E_\nu }{dxdQ^2},
\end{equation}
where $N(x,Q^2,E_\nu )$ is the number of events in each $\left( x,Q^2,E_\nu
\right) $ bin and $N(E_\nu )$ is the number of events in each $E_\nu $ bin
integrated over $x$ and $Q^2$. The structure functions can then be expressed
in terms of the sums and differences of neutrino and antineutrino cross
sections as 
\begin{eqnarray}
\frac 1{E_\nu }\frac{{d^2\sigma ^\nu }}{dxdQ^2}+\frac 1{E_\nu }\frac{{%
d^2\sigma ^{\overline{\nu }}}}{dxdQ^2} &=&f(y,R_L)F_2+g(y)\Delta xF_3, 
\nonumber \\
\frac 1{E_\nu }\frac{{d^2\sigma ^\nu }}{{dxdQ^2}}-\frac 1{E_\nu }\frac{{%
d^2\sigma ^{\overline{\nu }}}}{{dxdQ^2}} &=&2g(y)xF_3.
\end{eqnarray}
where 
\begin{eqnarray}
f(y,R_L) &=&(1-y+\frac{y^2}{2(1+R_L)})\frac y{Q^2},  \nonumber \\
g(y) &=&y(1-\frac y2)\frac y{Q^2},
\end{eqnarray}
and $F_2$ and $xF_3$ can be extracted by solving the coupled equations.

The difference between $xF_3$ in neutrino and antineutrino scattering, $%
\Delta \left( xF_3\right) $, is related to the difference between the
strange and charm sea distributions, $(s-c)$, as shown in Eq.~\ref{xF3QPM}
and discussed in Sec. \ref{Dimuon Section}.

\subsubsection{Assumptions concerning $R_L$}

The ratio of the longitudinal-to-transverse absorption cross section, $R_L$,
can be measured from the $y$-dependence of deep inelastic scattering data.
Fits to the special function $F$ defined as follows can be used to determine 
$R_L$. 
\begin{eqnarray}
F(x,Q^2,\epsilon )&=&\frac{\pi (1-\epsilon )}{y^2G_F^2ME}\left[ \frac{%
d^2\sigma ^\nu }{dxdy}+\frac{d^2\sigma ^{\overline{\nu }}}{dxdy}\right] 
\nonumber \\
&=& 2xF_1(x,Q^2)\left[ 1+\epsilon R(x,Q^2)\right].
\end{eqnarray}
In this equation, $\epsilon \simeq 2(1-y)/(1+(1-y)^2)$ is the polarization
of virtual $W$ boson. This equation assumes $xF_3^\nu =xF_3^{\overline{\nu }%
} $ and, therefore, a correction for $\Delta \left( xF_3\right)$ must be
applied. The values of $R_L=\sigma _L/\sigma _T$ are extracted from linear
fits to $F$ versus $\epsilon $ at fixed $x$ and $Q^2$ bins.

\begin{figure}[ptbp]
\psfig{figure=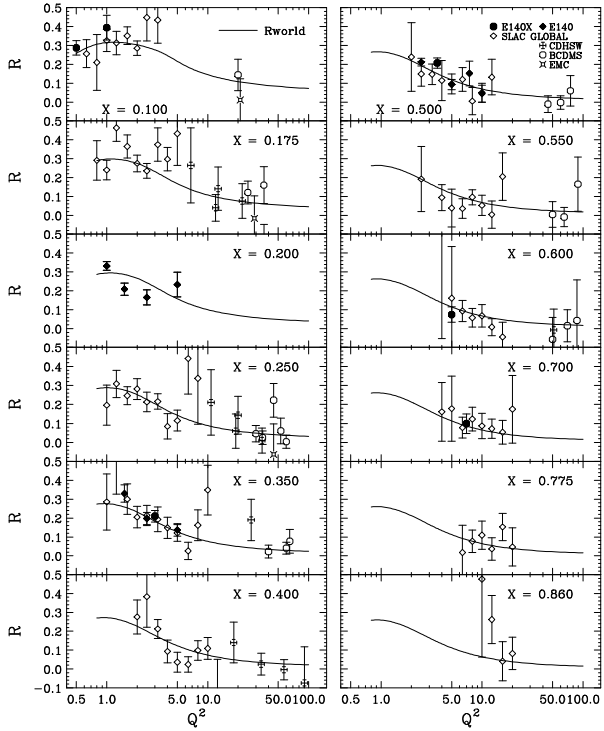,bbllx=50bp,bblly=100bp,bburx=550bp,bbury=700bp,width=3.5in,clip=T}
\caption{Measurements of $R_L$ by neutrino and charged-lepton experiments.
Solid line indicates the ``world'' parameterization of $R_L$.}
\label{RL plot}
\end{figure}

There are very few measurements of $R_L$ from deep inelastic neutrino
scattering. Two experiments with published results are CHARM \cite{CHARMR}
and CDHSW \cite{CDHR}.  
A comparison of the CDHSW measurements of $R_L$ is made to the
charged-lepton scattering results in Fig.~\ref{RL plot}. The best fit to the
world's data on $R_L$ is given by the parameterization ``$R_{World}$''  \cite
{Whitlow}, which is also shown in the plot.

CCFR E744/E770 is in the process of measuring $R_L$ as a function of $x$ and 
$Q^2$. The wide range of CCFR beam energies produces an extended range of
kinematically accessible $y$ values in each $Q^2$ and $x$ bin  \cite{unki}.
Therefore the lever-arm for the fit in most kinematic bins is large.

For the $F_2$ and $xF_3$ results compared in the next section, various
assumptions concerning $R_L$ were made: BEBC and CHARM assumed $R_L=0$, CCFR
E616/E701 used $R_{L,QCD}$, CCFR E744/E770 implemented the $R_{World}$
parameterization, and CDHSW used $R_L=0.1$.

\subsection{Structure Functions from Neutrino Experiments}

This section compares results from precision neutrino structure function
measurements. Small differences in the technique and assumptions used for
extracting the structure functions complicate this comparison. The
assumptions concerning the structure function $R_L$ were discussed in the
previous section. CDHSW and CCFR assume nonzero values for the strange sea,
while CHARM and BEBC assume the strange sea to be zero. All experiments
apply radiative corrections,  but the model and order of the calculation
varies in each case  \cite{DeRujula,rad2}. Finally, CHARM and BEBC have
applied corrections at high $x$ for Fermi smearing, while CDHSW and CCFR
have not. The discussion below is divided between measurements using high-$A$
targets and those from deuterium. The recent data sets presented below are
available in tabular format via the Internet\footnote{%
The CCFR Collaboration E744/770 structure functions are available in tabular
form by contacting seligman@nevis1.columbia.edu. Data from BEBC, CCFR, CDHSW
and many charged-lepton experiments are available in tables from the Durham
Structure Function Archive, http://cpt1.dur.ac.uk/HEPDATA/structure1.html.}.
Older experimental data are summarized in the tables of  \citeasnoun{Diemoz}.

\subsubsection{Heavy Target Experiments}

\label{Heavy Targets}

The experiments with high-$A$ targets have the advantage of high statistics.
In the comparisons between experiments using high-$A$ targets, various
targets were used and no nuclear corrections have been applied.

\begin{figure}[ptbp]
\psfig{figure=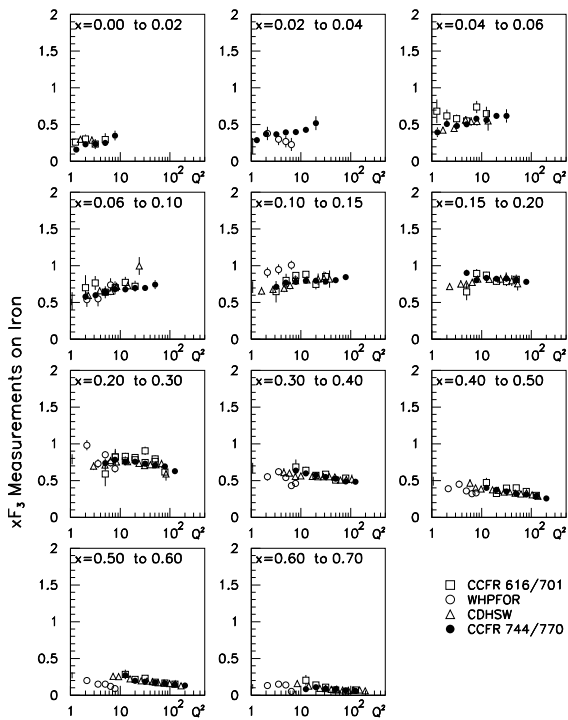,bbllx=0bp,bblly=50bp,bburx=600bp,bbury=800bp,width=3.5in,clip=T}
\caption{Comparison of $xF_3$ data from experiments with iron targets.}
\label{gwxf3fig}
\end{figure}

\begin{figure}[ptbp]
\psfig{figure=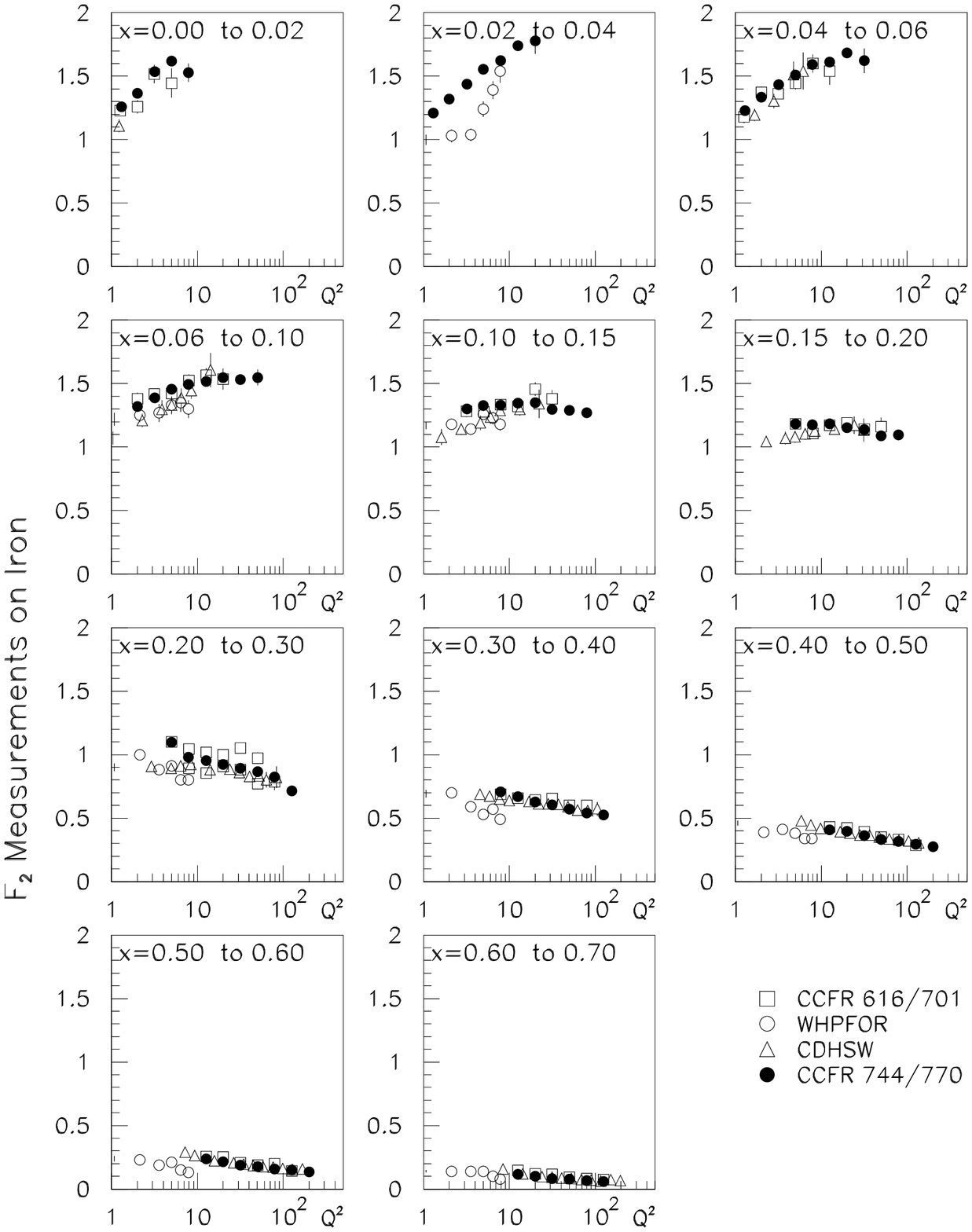,bbllx=0bp,bblly=50bp,bburx=600bp,bbury=800bp,width=3.5in,clip=T}
\caption{Comparison of $F_2$ data from experiments with iron targets.}
\label{gwf2fig}
\end{figure}

Figure ~\ref{gwxf3fig} compares the structure function $xF_3$ from
experiments with iron targets as a function of $Q^2$ for various $x$ bins.
These are from CCFR E616/E701  \cite{Oltman}, CCFR E744/E770  \cite{Seligman},
CDHSW  \cite{CDHSWSF}, and WHPFOR  \cite{Mann}. Of these experiments, CCFR
E744/E770 has the smallest statistical error. The $F_2$ comparison as a
function of $Q^2$, shown in Fig.~\ref{gwf2fig}, indicates good agreement
between the different data sets, except for the case of CDHSW. As shown in
Sec.~\ref{QCD SF Analysis}, the shape of the $Q^2$ dependence as a function
of $x$ for the CDHSW data cannot be described by a QCD-based
parameterization. The source of the disagreement is not understood.

\begin{figure}[ptbp]
\psfig{figure=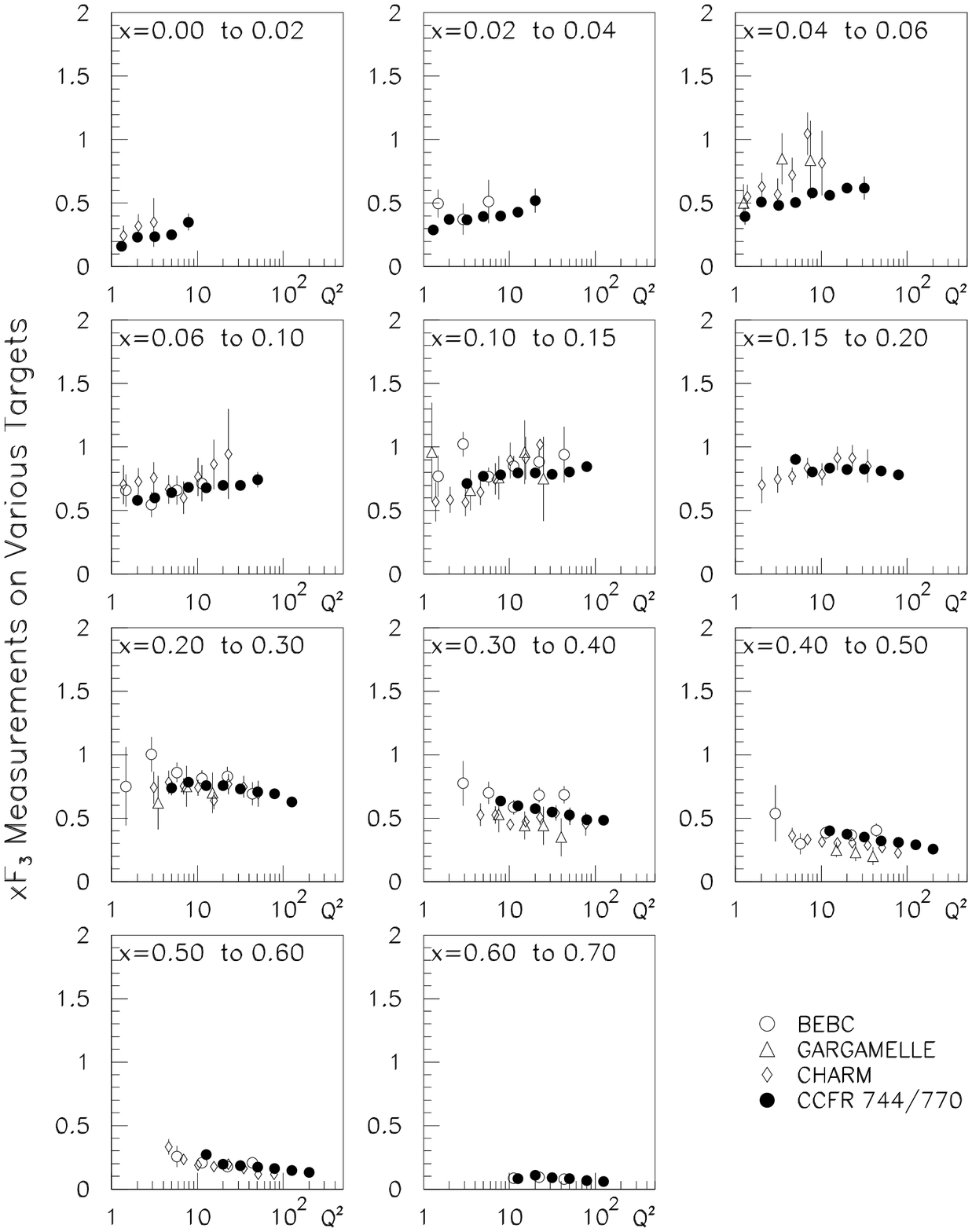,bbllx=0bp,bblly=50bp,bburx=600bp,bbury=800bp,width=3.5in,clip=T}
\caption{Comparison of $xF_3$ data from experiments with various high-$A$
targets.}
\label{gwxf3_other}
\end{figure}

\begin{figure}[ptbp]
\psfig{figure=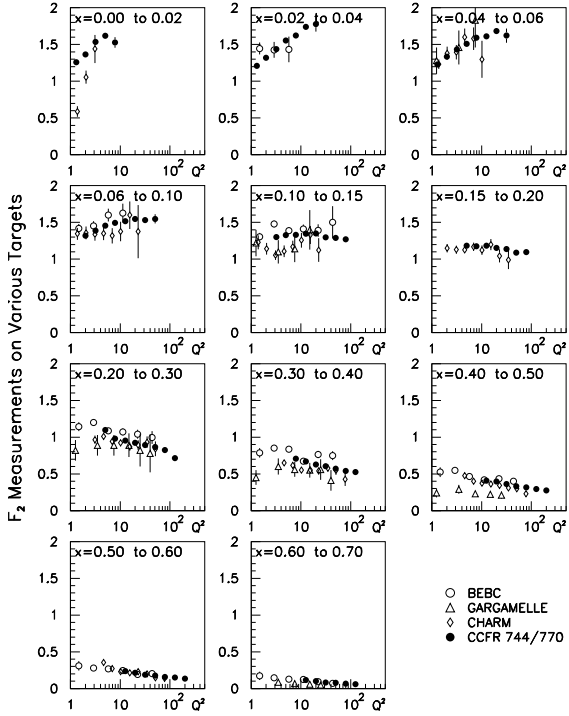,bbllx=0bp,bblly=50bp,bburx=600bp,bbury=800bp,width=3.5in,clip=T}
\caption{Comparison of $F_2$ data from experiments with various high-$A$
targets.}
\label{gwf2_other}
\end{figure}

Figures~\ref{gwxf3_other} and \ref{gwf2_other} compare $xF_3$ and $F_2$
measured on various targets. The CCFR E744/E770 data are used to represent
iron targets. The BEBC  \cite{BEBCNe} data were taken on neon target.
Gargamelle \cite{Morfin}, an earlier and less precise bubble chamber
experiment, used a C$_3$H$_8$-CF$_3$Br mixture. The CHARM data were taken on
marble (CaCO$_2$). The statistical errors are sufficiently large that the
data are consistent, despite the apparent wide spread between experiments.

\begin{figure}[ptbp]
\psfig{figure=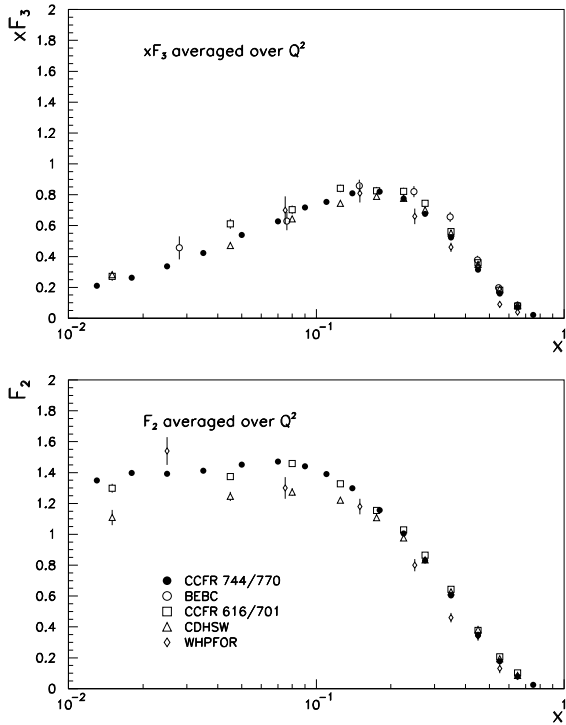,bbllx=0bp,bblly=0bp,bburx=600bp,bbury=800bp,width=3.5in,clip=T}
\caption{Average $xF_3$ and $F_2$ as a function of $x$ from BEBC
(open circles), CCFR E616/E701 (squares), CCFR E744/E770 (closed circles)
and WHPFOR (diamonds).}
\label{sf x depend}
\end{figure}

The qualitative behavior of the structure functions averaged over $Q^2$ as a
function of $x$, shown in Fig.~\ref{sf x depend}, can be interpreted within
QCD.  Quantitative QCD studies, interpretations, and measurements will be
discussed in Sec.~\ref{QCD SF Analysis}.

The structure function $xF_3$ is related to the sum of the valence quarks,
and the measured distribution is peaked at moderate $x$, going to zero as $x$
goes to zero or one. If the proton consisted of three non-interacting
valence quarks, then this distribution would simply reflect the Fermi motion
of the quarks inside the nucleon and would peak at $x=\frac 13$. However,
because these quarks are interacting via gluons, which carry some of the
fractional momentum of the proton, the distribution is smeared further and
peaks at smaller $x$. The position of the peak will move toward smaller $x$
with increasing $Q^2$ due to QCD evolution. Since each neutrino experiment
covers a different range of $Q^2$, the data points on Fig.~\ref{sf x depend}
are not expected to agree.

The $F_2$ structure function is related to the sum of the quark and
antiquark distributions. The $F_2$ distribution goes to zero at high $x$
and, but is approximately constant in the low-$x$ region. Gluon radiation
and gluon splitting into quark-antiquark pairs results in a large number of
low momentum partons that dominate the low-$x$ region. In fact, the
charged-lepton scattering results on $F_2$ from HERA have shown that at very
low $x$, outside of the kinematic range accessible to present neutrino
experiments, $F_2$ rises dramatically \cite{HeraF2a,HeraF2b}.

\subsubsection{Deuterium and Hydrogen Targets}

The BEBC experiment has used neutrino and antineutrino scattering from
deuterium to obtain the best measurement of the structure functions on the
proton and the neutron separately \cite{BEBCD2,BEBC-UvDv}. 
A neutrino or antineutrino
interaction was identified as coming from a neutron if it had either an even
number of prongs, or an odd number of prongs with a proton with momentum
less than 150 MeV. All remaining events, with an odd number of prongs and
hence a net total charge, were classified as interactions with protons.
Misidentifications were corrected on a statistical basis using a Monte Carlo
simulation.

\begin{figure}[ptbp]
\psfig{figure=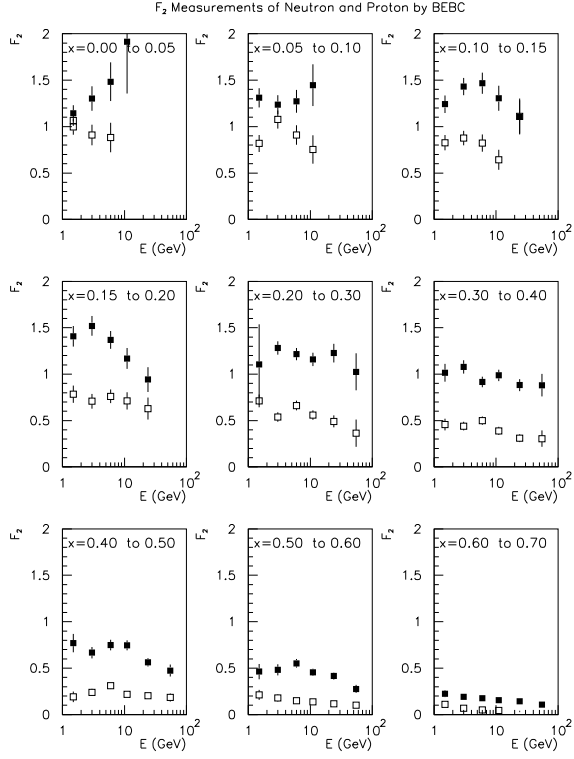,bbllx=0bp,bblly=50bp,bburx=600bp,bbury=800bp,width=3.5in,clip=T}
\caption{$F_2^{\nu p}$ and $F_2^{\nu n}$ as measured by BEBC}
\label{bebcnp}
\end{figure}

The resulting structure functions for neutrino scattering on protons and
neutrons, $F_2^{\nu p}$ and $F_2^{\nu n}$, are shown in Fig.~\ref{bebcnp}.
The most evident feature is that $F_2^{\nu n}\approx 2F_2^{\nu p}$ over most
of the kinematic region. This is because the $W^{+}$ emitted in a neutrino
interaction must interact with a negatively charged quark, which at high $x$
has the highest probability of being the valance $d$ quark. Since the
neutron has twice the number of valance $d$ quarks as the proton, the
neutron structure function is larger. These data clearly indicate the flavor
sensitivity of neutrino scattering.

\subsection{ Comparisons of Neutrino to Charged-Lepton Experiments}

\label{numucomp}

\begin{figure}
\psfig{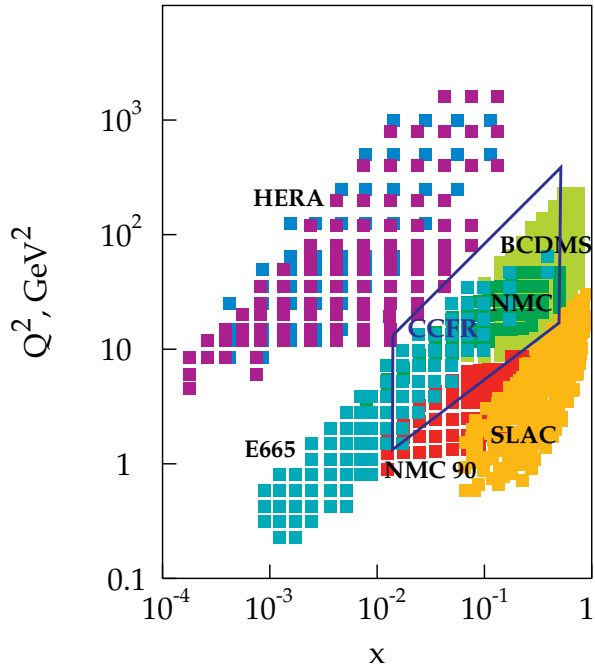}
\caption{The kinematic region accessible to various charged-lepton
experiments compared to the CCFR experiment}
\label{xQ2fig}
\end{figure}

Neutrino deep inelastic scattering experiments now have the statistical
accuracy of charged-lepton scattering experiments. In this section, neutrino
results are compared to charged-lepton results from SLAC ($eN$), NMC ($\mu N$%
), BCDMS ($\mu N$) and E665 ($\mu N$). The kinematic ranges covered by these
experiments is shown in Fig.~\ref{xQ2fig}, with the CCFR region shown for
comparison. The HERA $ep$ data are mainly at high $Q^2 $ and very low $x$,
with only small kinematic overlap with the fixed-target neutrino data.

\subsubsection{Comparison of $F_2$}

The parity-conserving structure function $F_2$ is measured by both charged-
and neutral-lepton scattering experiments. The definitions of $F_2$ in
neutrino and charged-lepton scattering were discussed in Sec.~\ref{QPM
Section}. As a result of the difference in the couplings, a conversion must
be applied in order to compare muon and neutrino experiments. To lowest
order with the charm sea set to zero, the correction is 
\begin{equation}
F_2^{\mu N}={\frac{{5}}{{18}}}F_2^{\nu N}\left[ 1-{\frac{{3(s+\overline{s})}%
}{{5(q+\overline{q})}}}\right] .  \label{five18}
\end{equation}
This relation is true to all orders in the DIS renormalization scheme, which
maintains Eq.~\ref{DIS F_2 definition} as the definition of $F_2$.

Assuming that the strange sea is small at high $x$, one expects $F_2^{\mu
N}/F_2^{\nu N}\rightarrow {\frac{{5}}{{18}}}=0.2778$ as $x\rightarrow 1$.
Comparing the CCFR data and the CDHSW data \cite{Mishra} to the results from
BCDMS \cite{BCDMShighx} at high $x$ yields a ratio of $0.278 \pm 0.010$ which
is in excellent agreement with the expected quark charges.

In the comparisons given below, all corrections for the differences between
the charged-lepton and neutrino data have been applied to the charged-lepton
data. The strange sea ($s$, $\overline{s}$) appears explicitly in Eq.~\ref
{five18} and, therefore, a measurement of this distribution is needed for
the comparison. The strange sea can be measured from the neutrino charm
production cross section which is related to charged-current dimuon
production \cite{Bazarko} described in detail in Sec. \ref{Dimuon Section}.
Additional nuclear corrections must also be made to the charged-lepton
measurements of $F_2$, as discussed in Sec.~\ref{nonpert}\ref{nu nukes}.

\begin{figure}
\psfig{figure=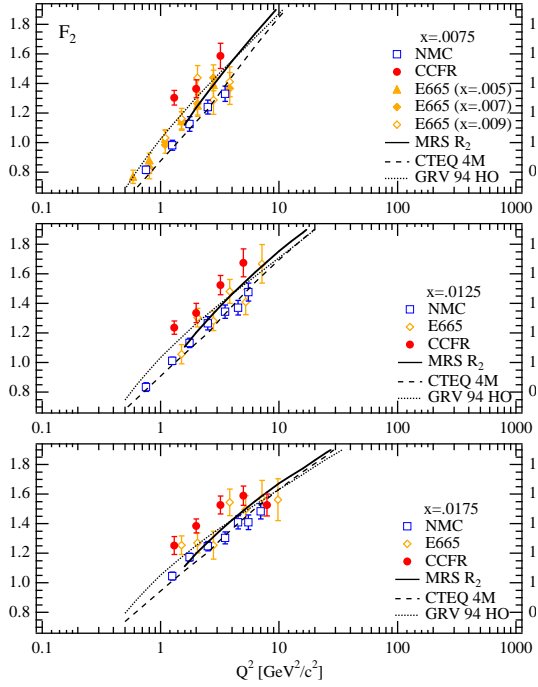,bbllx=100bp,bblly=150bp,bburx=500bp,bbury=750bp,width=2.9in,clip=T}
\caption{Comparison of the CCFR ($\nu$) measurement of $F_2$ to the NMC and
E665 ($\mu$) results for $x=0.0075, 0.0125$ and $0.0175$.}
\label{compare1}
\end{figure}

\begin{figure}
\psfig{figure=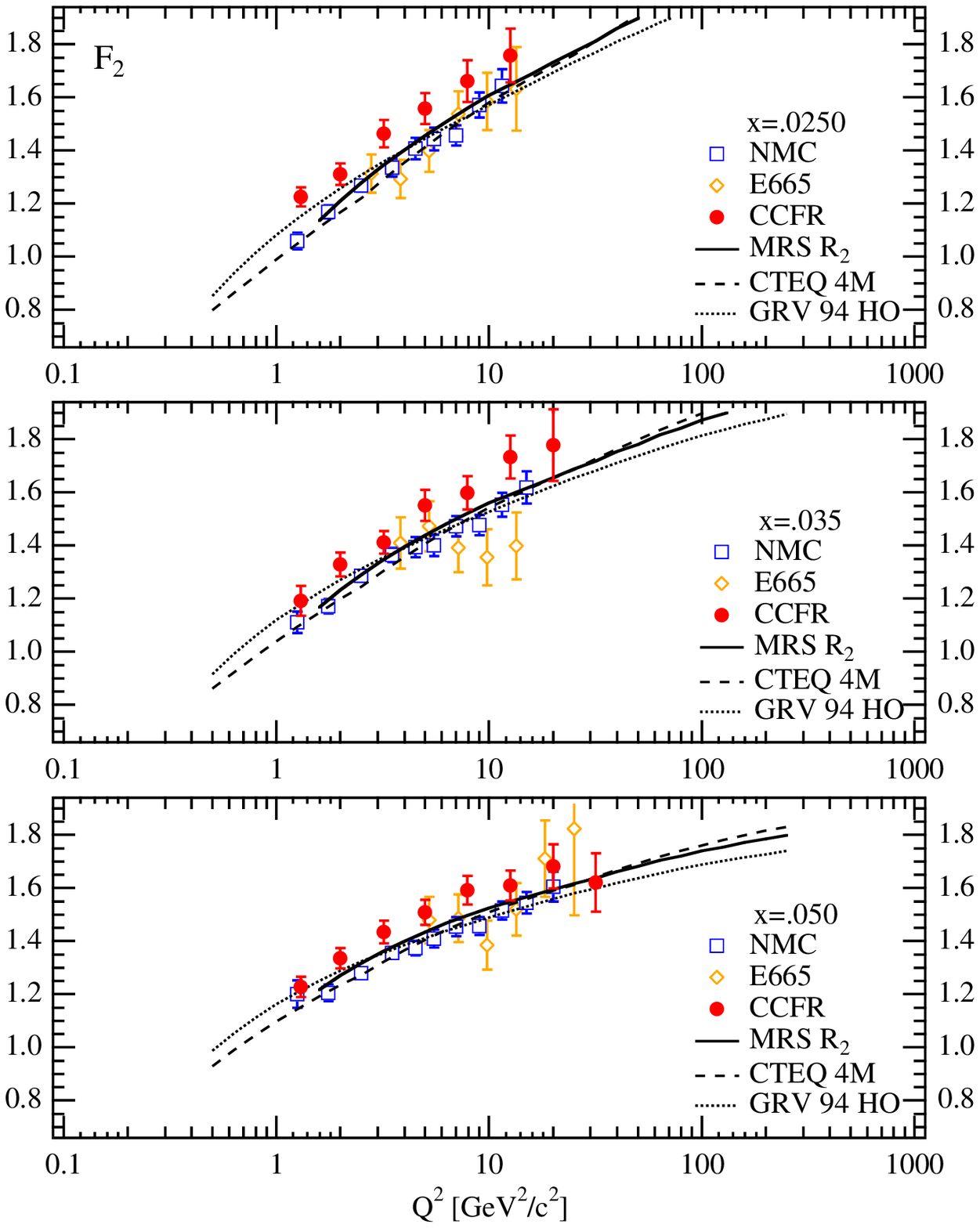,bbllx=100bp,bblly=200bp,bburx=500bp,bbury=750bp,width=2.9in,clip=T}
\caption{Comparison of the CCFR ($\nu$) measurement of $F_2$ to the NMC and
E665 ($\mu$) results for $x=0.0250, 0.035$ and $0.050$}
\label{compare2}
\end{figure}

\begin{figure}[ptbp]
\psfig{figure=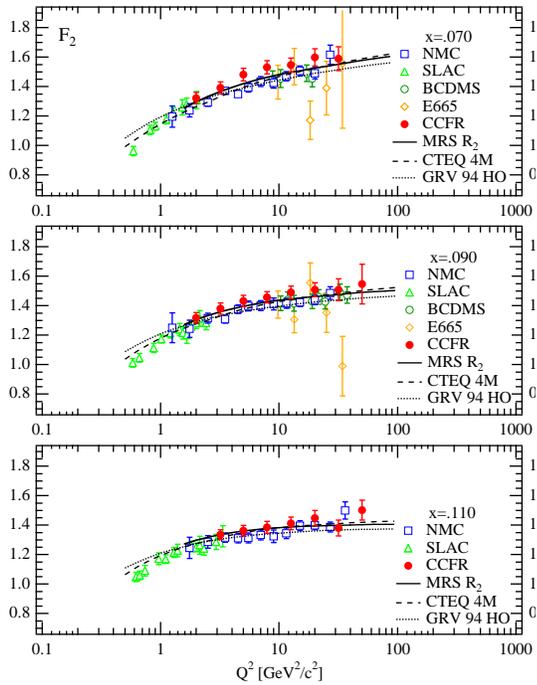,bbllx=100bp,bblly=200bp,bburx=500bp,bbury=750bp,width=2.9in,clip=T}
\caption{Comparison of the CCFR ($\nu$) measurement of $F_2$ to the NMC and
E665 ($\mu$) results and SLAC ($e$) results for $x=0.070, 0.090$ and $0.110$.}
\label{compare3}
\end{figure}

\begin{figure}[ptbp]
\psfig{figure=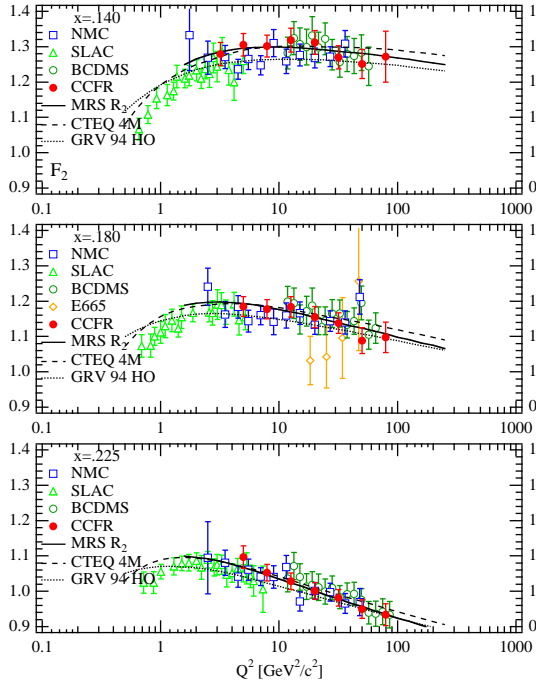,bbllx=100bp,bblly=200bp,bburx=500bp,bbury=750bp,width=2.9in,clip=T}
\caption{Comparison of the CCFR ($\nu$) measurement of $F_2$ to the NMC and
E665 ($\mu$) results for $x=0.140, 0.180$ and $0.225$}
\label{compare4}
\end{figure}

\begin{figure}[ptbp]
\psfig{figure=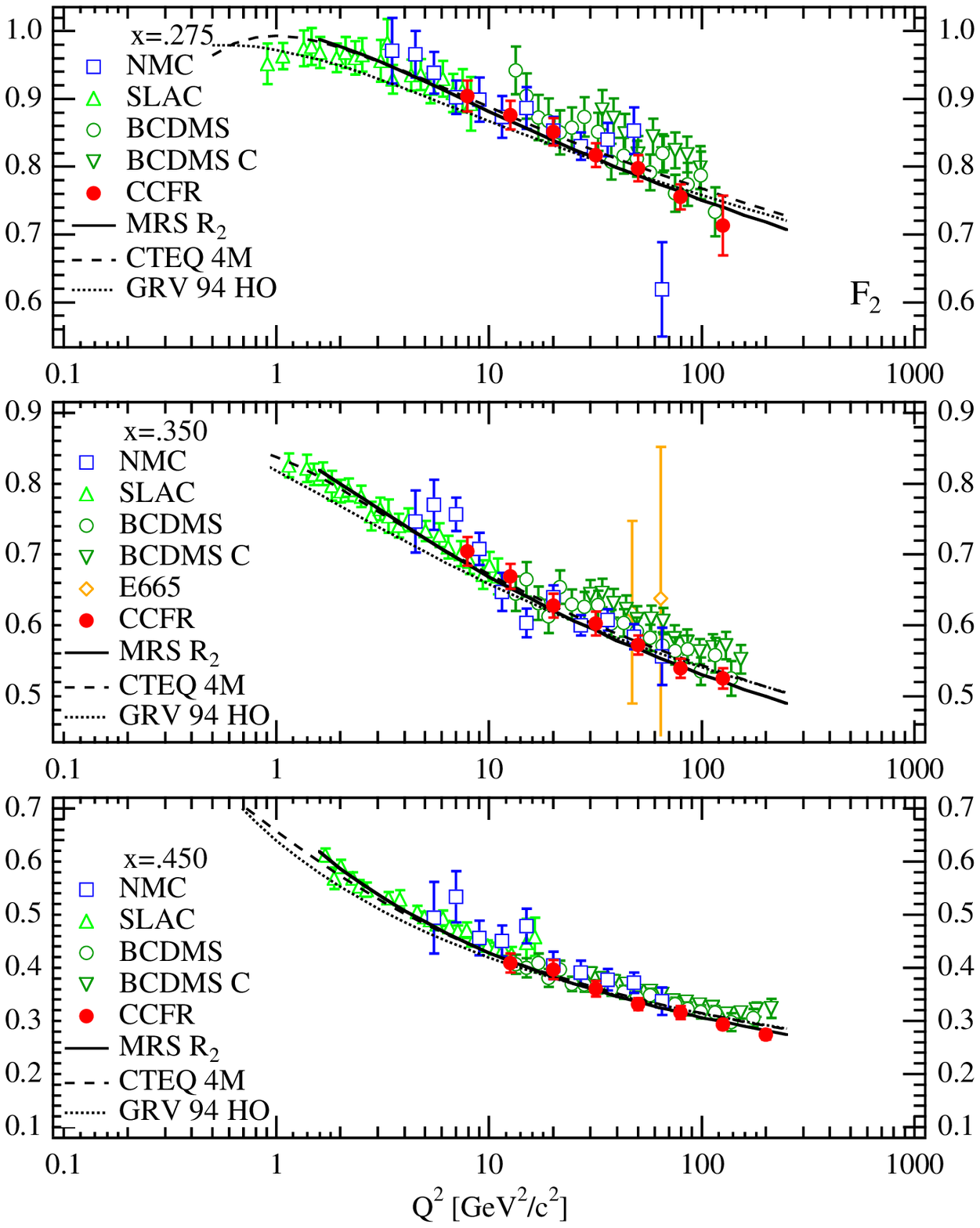,bbllx=100bp,bblly=200bp,bburx=500bp,bbury=750bp,width=2.9in,clip=T}
\caption{Comparison of the CCFR ($\nu$) measurement of $F_2$ to the NMC and
E665 ($\mu$) results for $x=0.275, 0.350$ and $0.450$}.
\label{compare5}
\end{figure}

\begin{figure}[ptbp]
\psfig{figure=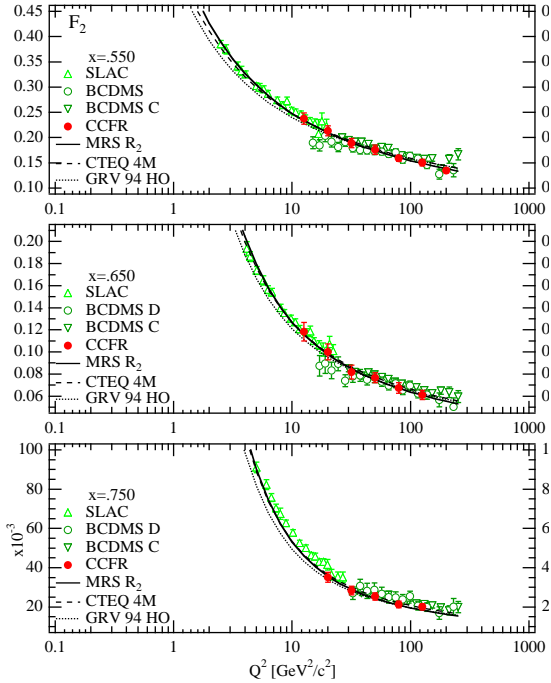,bbllx=100bp,bblly=200bp,bburx=500bp,bbury=750bp,width=2.9in,clip=T}
\caption{Comparison of the CCFR ($\nu$) measurement of $F_2$ to the NMC and
E665 ($\mu$) results for $x=0.550, 0.650$ and $0.750$.} 
\label{compare6}
\end{figure}

Figures \ref{compare1}-\ref{compare6} compare the CCFR $F_2$ measurement to
the measurements from NMC \cite{F2NMC} and E665 \cite{E665rat} over a range of 
$x$ values. The corrections for charge coupling, using the DIS scheme, and
nuclear effects have been applied to the muon data. At low $x$ there appears
to be a disagreement between the charged-lepton scattering data and the
neutrino data. The discrepancy decreases with increasing $x$ until $x\approx
0.070$ where the data are in good agreement to the highest $x$ bins.

The high statistics data in the region of the discrepancy come from CCFR and
NMC. It is possible that one or both experiments have underestimated their
systematic errors. The most important systematic error in the CCFR analysis
is due to the calorimeter calibration. Extensive studies of the test-beam
data were undertaken in order to determine the muon and hadron energy
calibration and the systematic error assigned to this source has been
carefully estimated by the experiment.

Several publications have suggested that the discrepancy is due to an
incorrect strange-sea correction \cite{bigsea,Nick,brodskyma}. The required
distributions to eliminate the discrepancy, as calculated at next-to-leading
order by the CTEQ collaboration \cite{bigsea}, are inconsistent with and
approximately a factor of two larger at low $x$ than the direct CCFR
strange-sea measurement from dimuon data \cite{Bazarko}. The CCFR measurement
of the strange sea, which is to next-to-leading order and includes
corrections for the charm-mass threshold, would have to be incorrect by $%
5\sigma $ to account for the discrepancy. The CCFR strange-sea measurement
and its application to the correction presented here is described in detail
in Sec. \ref{Dimuon Section}.

Another possible cause for the disagreement between the neutrino and charged
lepton measurements of $F_2$ may be the nuclear correction applied to the
charged-lepton deuterium data in order to allow comparison with the neutrino
results. As discussed in Sec.~\ref{nonpert}\ref{nu nukes}, the corrected
charged-lepton data will only agree with the neutrino result if the entire
effect is ascribed to the target material. Propagator effects, such as
fluctuation of the virtual boson to mesons, will not be the same for
neutrino and charged lepton scattering because of the additional axial
component of the $W$. Such ``nuclear shadowing'' would be expected to affect
the low-$x$ (low-$Q^2$) region most strongly, thereby accounting for some of
the disagreement. The data on these nuclear effects in neutrino and
charged-lepton scattering are discussed in the next section.

\subsubsection{Nuclear Effects}

\begin{figure}[ptbp]
\psfig{figure=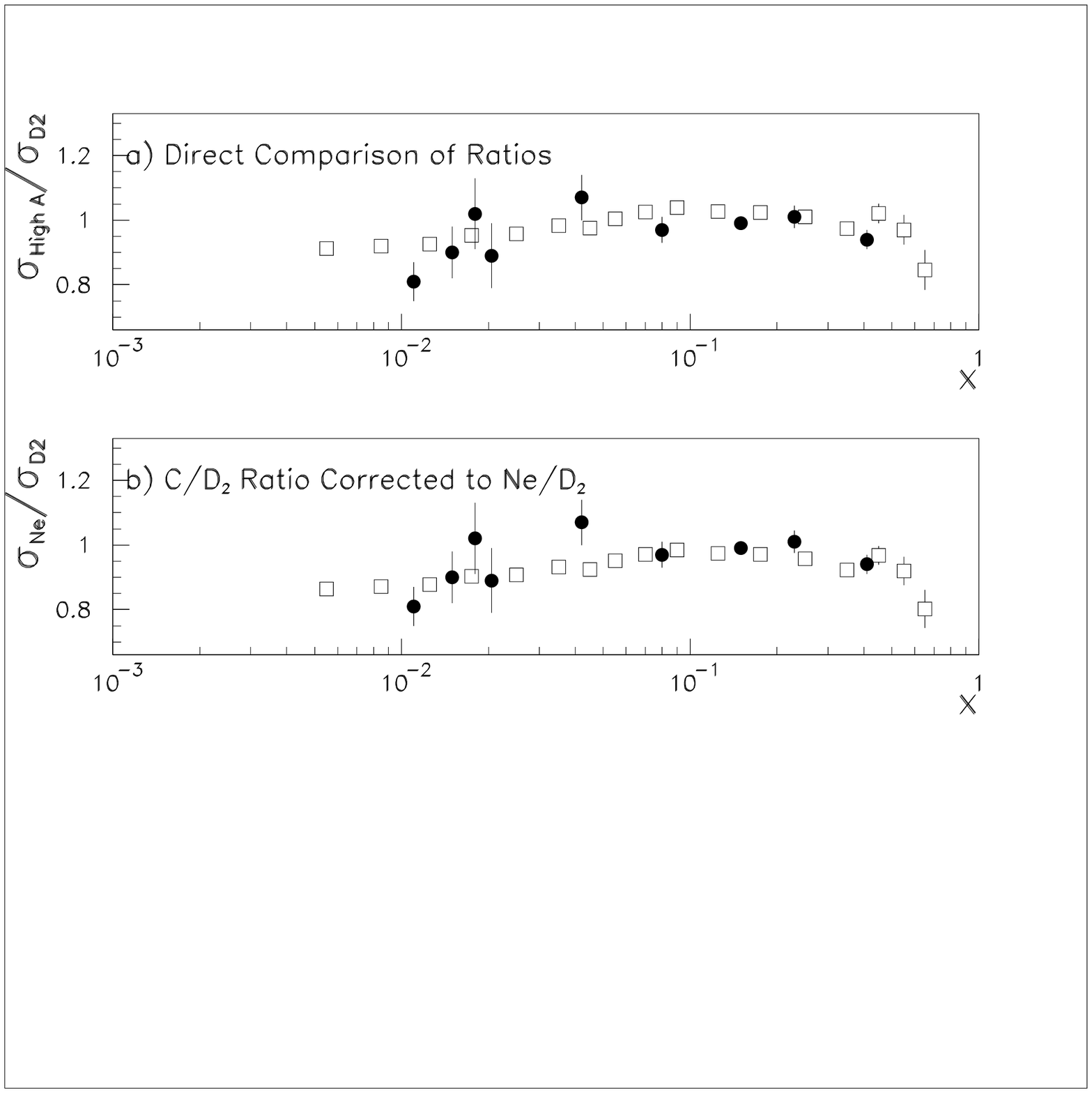,bbllx=22bp,bblly=320bp,bburx=550bp,bbury=680bp,width=3.5in,clip=T}
\caption{Comparison of the ratio of cross sections for high-$A$ targets to
deuterium. Solid circle: BEBC data from neutrino -- neon scattering. Open
boxes: NMC data from muon -- carbon scattering. }
\label{nu A plot}
\end{figure}

Very little data on shadowing in neutrino scattering has been obtained. The
best measurements come from the BEBC collaboration ratios of neon ($A=20$)
to deuterium \cite{BEBCratio}. This can be compared to the muon scattering
results from carbon ($A=12$) as measured by the NMC experiment \cite{NMCC}.
Figure~\ref{nu A plot}a shows Ne/D$_2$ directly compared to C/D$_2$, while
Fig.~\ref{nu A plot}b applies a correction factor for the theoretically
expected $A^{-{\frac 23}}$ dependence of the nuclear cross section to the
carbon ratio \cite{Geesaman}, although it should be noted that some
experimental indicate less $A$ dependence \cite{E665rat}. Data are shown for $%
Q^2>1$ GeV$^2$, except for the three lowest $x$ BEBC data points which are
from $Q^2<0.2$ GeV$^2$, $\left( 0.2<Q^2<0.5\right) $GeV$^2$ and $\left(
0.5<Q^2<1.0\right) $ GeV$^2$, respectively. Within the combined statistical
and systematic errors, the results from charged-lepton and neutrino
scattering appear to agree. These results are also in agreement with an
earlier, lower statistics BEBC measurement of this ratio which used a $3$ m$%
^3$ track-sensitive target filled with deuterium surrounded by neon \cite
{Cooper}. CDHSW has measured the ratio of structure functions in iron to
hydrogen \cite{CDHFeH2} using a 35 m$^3$ hydrogen tank located in front of
the CDHSW detector. This experiment obtained 4,457 $\nu p$ events and 4,178 $%
\overline{\nu }p$ events which were compared to 50,000 neutrino and 150,000
antineutrino events in iron. Although the combined systematic and
statistical errors are too large to establish $A$-dependence, the behavior
of the data is consistent with the behavior observed in charged-lepton
scattering.

Although the NMC and E665 shadowing data which were used to obtain the
nuclear correction are in statistical agreement, there appears to be a
systematic shift between the two data sets as can be seen in Fig.~\ref{nukes}%
. The nuclear correction is dominated by the NMC result, which has much
smaller errors than the E665 data. However, if only the E665 data are used
in a fit which is then applied to the charged-lepton deuterium results, then
the CCFR/NMC discrepancy is reduced by approximately a factor of two for $%
x<0.03$. However, it seems unlikely that the entire $\nu N$ versus $\mu N$
low-$x$ discrepancy can be due to nuclear effects alone.

\subsection{QCD Analyses of the Structure Function Results}

\label{QCD SF Analysis}

An analysis of the structure function data within the QCD framework allows
precise tests of the predicted QCD evolution and the Gross-Llewellyn Smith
and the Adler sum rules. From these studies, measurements of $\alpha _s$ and
the parton distributions can be extracted.

\subsubsection{The Strong Coupling Measured from QCD Evolution}

Perturbative QCD can predict the $Q^2$ evolution of the structure functions
given a starting set of $x$-dependent PDFs at a reference scale $Q_0^2$ \cite
{DGLAP1,DGLAP2,DGLAP3,DGLAP4}, as described in Sec. \ref{QCD SF theory}.
Therefore, an analysis of scaling violations in structure function data
permit a determination of the parton distributions and the QCD parameter $%
\Lambda $. The CCFR E744/E770 analysis, which is the most precise to date,
provides an example of the extraction of these distributions and the QCD
parameter at next-to-leading order for four flavors in the $\overline{MS}$
renormalization scheme ($\Lambda _{\overline{MS}}^{NLO-4f}$).

Two independent QCD evolution analyses were performed by CCFR. The first
used only the parity violating structure function $xF_3$ . The advantage of
this analysis is that the evolution of $xF_3$ is independent of the gluon
distribution, reducing the number of free parameters. The second analysis
used both $xF_3$ and $F_2$ data in the QCD fit. This increases the
statistical power of the fit, but introduces extra parameters to describe
the gluon distribution. On the other hand, the combined $F_2$/$xF_3$ data
can constrain the gluon distribution $G(x,Q^2)$ at moderate values of $x$.
Note that charged-lepton scattering experiments cannot separate gluon
distribution effects from the running of $\alpha _S$ as cleanly because they
do not have access to the parity-violating structure function, $xF_3$.

The QCD fits require initial parameterizations of the parton distributions
at some $Q_0^2$. CCFR E744/E770 used 
\begin{eqnarray}
xq_{NS}(x,Q_0^2) &=&A_{NS}x^{\eta _1}(1-x)^{\eta _2},  \nonumber \\
xq_S(x,Q_0^2) &=&xq_{NS}(x,Q_0^2)+A_S(1-x)^{\eta _S}, \\
xG(x,Q_0^2) &=&A_G(1-x)^{\eta _G},  \nonumber
\end{eqnarray}
where $NS$, $S$, $G$ refer to non-singlet, singlet, and gluon distributions,
respectively; and $A_{NS}$, $A_S$, $A_G$, $\eta _1$, $\eta _2$, and $\eta _G$
are free parameters in the fit. More complicated parameterizations of the
parton distributions were found to yield fit results consistent with the
above parameterizations \cite{Seligman}. The resulting parton distributions
are compared to other measurements in Sec. \ref{QCD PDF Fits}.

The CCFR analysis compared data to a theoretical prediction based on the
next-to-leading order QCD evolution program of Duke and Owens \cite{DO} using
a $\chi ^2$ technique that included the effects of target-mass,
higher-twist, and $R_{L,QCD}$. Only data satisfying $Q^2>5~$GeV$^2$, $W^2>10$
GeV$^2$, and $x<0.7$ were included, and $(x,Q^2)$ bins with statistical
error greater than 50\% were eliminated. The effects of systematic errors
were studied by implementing positive and negative shifts of $F_2$ and $xF_3$
caused by each systematic error and re-extracting $\Lambda $ as well as the
other fit parameters.

The results of NLO QCD fits in the $\overline{MS}$ scheme to the CCFR
structure function data are listed in Table~\ref{tab:QCDfits} and are shown
by the solid line on Fig.~\ref{CCFR-F2-plot} and Fig.~\ref{CCFR-xF3-plot}.
The dashed lines on these figures represents an extrapolation of the
parameterization from the fit to lower $Q^2$ values. As expected, the fit
which uses both the $xF_3$ and $F_2$ data has smaller statistical and
systematic errors because of the increased number of data points used. The
good $\chi ^2$ per degree of freedom indicates consistency of the data with
QCD.

\begin{table}[tbp]
\begin{center}
\begin{tabular}{ccc}
Experiment & Structure Function(s) Fit & $\Lambda^{NLO-4f}_{\overline{MS}}$
\\ \hline
CDHSW & $xF_3$ & $200^{+200}_{-100}$ \\ 
& $F_2$ & $300\pm 80$ \\ 
CHARM & $xF_3$ & $310 \pm 150$ \\ 
CCFR 616/701 & $F_2$ & $340\pm 110$ \\ 
CCFR 744/770 & $xF_3$ & $381 \pm 55$ \\ 
& $F_2$, $xF_3$ & $337 \pm 31$%
\end{tabular}
\end{center}
\caption{ Values of $\Lambda^{NLO-4f}_{\overline{MS}}$ determined from fits
to structure functions measured in various neutrino experiments.}
\label{tab:QCDfits}
\end{table}

\begin{figure}[ptbp]
\psfig{figure=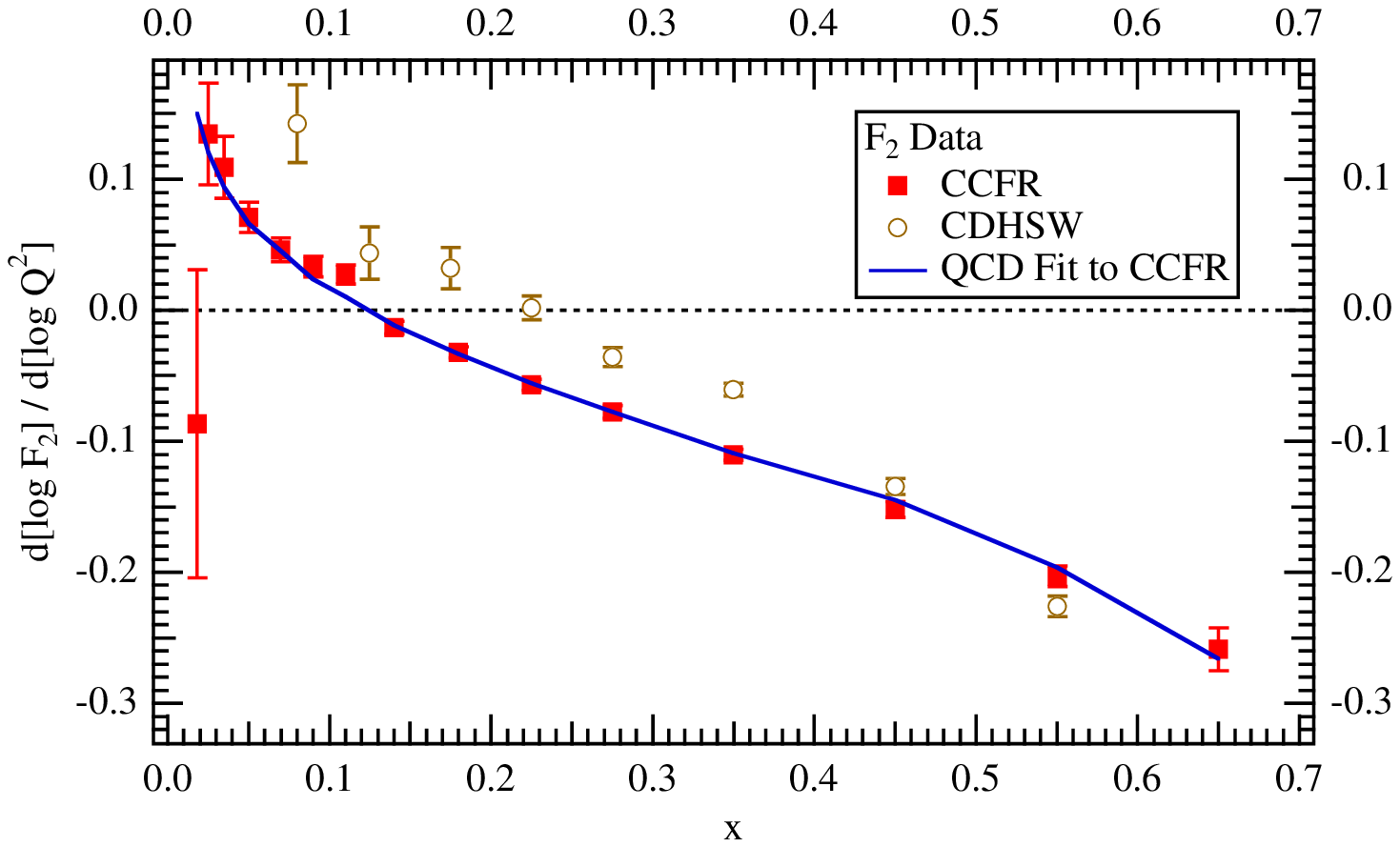,bbllx=0bp,bblly=400bp,bburx=600bp,bbury=750bp,width=3.5in,clip=T}
\psfig{figure=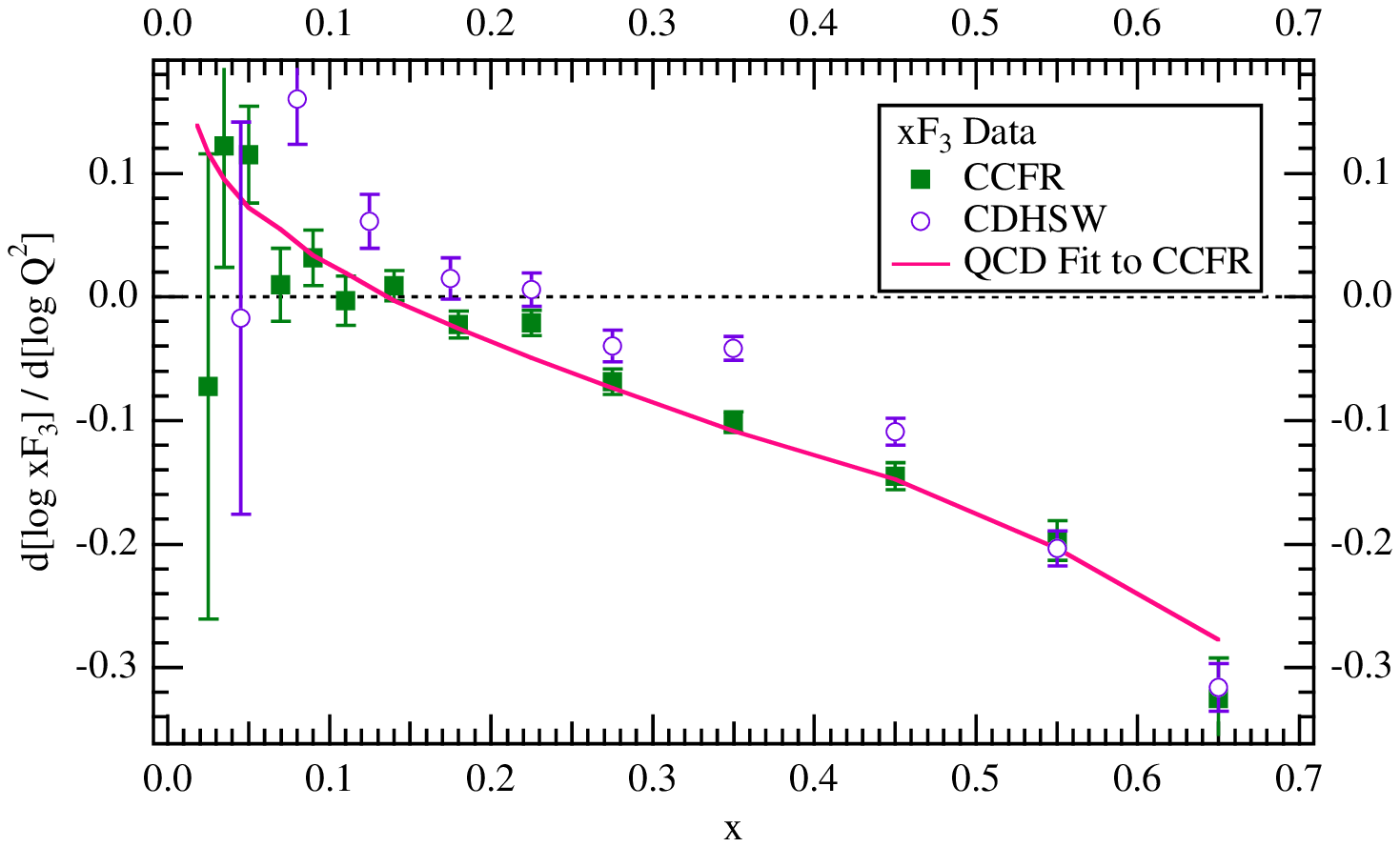,bbllx=0bp,bblly=400bp,bburx=600bp,bbury=750bp,width=3.5in,clip=T}
\caption{CCFR and CDHSW logarithmic slopes of the structure functions. The
line indicates the CCFR QCD fit result.}
\label{CHDSWslopes}
\end{figure}

The CDHSW data are not in agreement with the other neutrino measurements
(see Sec. \ref{Heavy Targets}) and with the expected QCD behavior. Another
way to see this is to consider the logarithmic derivative of the structure
functions with respect to $Q^2$ as a function of $x$. QCD predicts the form
of this distribution up to one free parameter, $\Lambda $, given the parton
distributions as inputs. Figure~\ref{CHDSWslopes} shows the CCFR logarithmic
slopes and the corresponding QCD fit to the data along with the CDHSW data
for comparison.

An alternative and more powerful way to incorporate systematic uncertainties
is to include their effects directly into the QCD fit. For this type of
analysis, a $\chi ^2$ fit to the theoretical prediction for the structure
functions is compared to the data in each $x$ and $Q^2$ bin. The prediction
should be compared to the data using a $\chi ^2$ that includes both
statistical and systematic uncertainties with correlations. The systematic
uncertainties can be handled by introducing a parameter $\delta _k$ for each
uncertainty $k$ into the $\chi ^2$. Defining the structure-function vector $%
{\bf F}=(F_2,xF_3)$ and the structure-function error matrix ${\bf V}=(\sigma
_{ij})$, for $i,j=F_2,xF_3$, then the difference between the theoretical
prediction and data and the $\chi ^2$ are: 
\begin{eqnarray}
{\bf F}^{diff} &=&{\bf F}^{data}-{\bf F}^{theory}+\sum_k\delta _k({\bf F}^k-%
{\bf F}^{data}) \\
\chi ^2 &=&({\bf F}^{diff}){\bf V}^{-1}({\bf F}^{diff})^T+\sum_k\delta _k^2.
\end{eqnarray}
where ${\bf F}^k$ is the value of the extracted structure functions when the 
$k$th systematic parameter is shifted by $1\sigma $. For the CCFR data, the
results of this fitting method are compared to the ``basic'' method
described above in Table~\ref{tab:Billfits}. The ``global fit'' gives a more
precise measure of $\Lambda $ because of the constraint of QCD.

\begin{table}[tbp]
\begin{center}
\begin{tabular}{c|cc|cc}
~ & \multicolumn{2}{c}{from $xF_3$ only} & \multicolumn{2}{c}{from $xF_3$
and $F_2$} \\ \hline
Fit Type: & $\Lambda \pm$ stat $\pm$ sys & $\chi^2/DOF$ & $\Lambda \pm$ stat 
$\pm$ sys & $\chi^2/DOF$ \\ \hline
Basic & $387 \pm 42 \pm 94$ & 81/82 & $381 \pm 23 \pm 58$ & 191/164 \\ 
Global & $381 \pm 41 \pm 37$ & 69/82 & $337 \pm 21 \pm 23$ & 171/164
\end{tabular}
\end{center}
\caption{ Values of $\Lambda^{NLO-4f}_{\overline{MS}}$ determined from fits
to the CCFR data.  The systematic error includes the experimental,
physics model, and higher-twist uncertainties.}
\label{tab:Billfits}
\end{table}

\begin{figure}[ptbp]
\psfig{figure=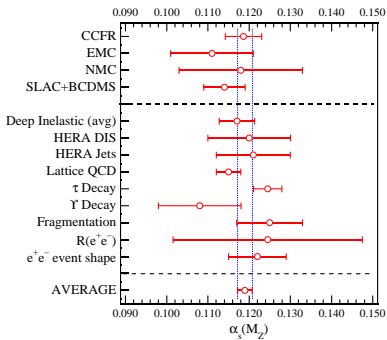,bbllx=10bp,bblly=430bp,bburx=600bp,bbury=800bp,width=3.5in,clip=T}
\caption{Measurement of $\alpha_S$ from various experiments. Results from
other experiments are from Barnett {\it et al.} (1996).} 
\label{alphasfig}
\end{figure}

The value of $\Lambda = 337 \pm 31$ MeV 
from the CCFR global fit using $F_2$ and $xF_3$ is equivalent to an
\[ 
\alpha _s(M_Z^2)=0.119\pm 0.002(exp)\pm 0.004(scale)
\text{\quad (CCFR)}.
\]
The scale error is due to the renormalization and factorization scale
uncertainties \cite{VirchMil}. This result can be compared to other
measurements as reported in  \citeasnoun{PDG}. The CCFR result is higher than
the previous measurement of CCFR, $\alpha _s(M_Z^2)=0.111\pm 0.005$ \cite
{quintas} and is also higher than the previous deep inelastic scattering
average, $\alpha _s(M_Z^2)=0.112\pm 0.005$. Including the new CCFR
measurement, the average for deep inelastic scattering experiments is
\[
\alpha _s(M_Z^2)=0.117\pm 0.004 \text{\quad (DIS Average)}
\]
which is lower than the LEP
measurement from event shapes of $0.122\pm 0.007$. However, all of the above
results are consistent within the errors. Fig.~\ref{alphasfig} compares the
CCFR result to results from other non-neutrino experiments \cite{PDG}.

\subsubsection{Parton Distributions Measured from QCD Evolution}

\label{QCD PDF Fits}

This section considers measurements of the parton distributions from the QCD
evolution fits. The parton distributions are extracted simultaneously with
the fit for $\Lambda $. This method can accurately determine the
distributions for light quarks and gluons. The best method for determining
the strange and charm distributions is through charged- and neutral-current
charm production, respectively, as discussed in Sec. \ref{Dimuon Section}.

\begin{figure}[tbp]
\psfig{figure=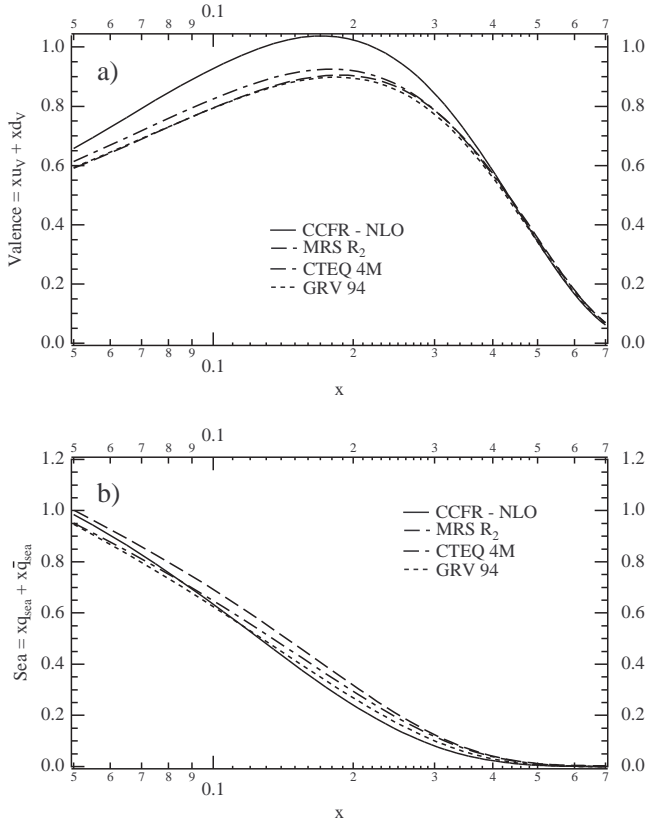,bbllx=20bp,bblly=210bp,bburx=450bp,bbury=770bp,width=3.5in,clip=T}
\caption{The CCFR parton distributions compared to CTEQ 4M, MRS R$_2$,
and GRV 94 at $Q^2 = 5$ GeV$^2$.  a) $xq_{valance}= xu_V +xd_V$, 
b) Total sea $ = \sum\limits_{u,d,s} (xq_{sea} + x\overline{q}_{sea})$. }
\label{PDF plots}
\end{figure}

The parameterizations of the parton distributions found to fit best the CCFR
E744/E770 data are compared with several popular PDF sets available from
PDFLIB  \cite{PDFLIB} in Fig.~\ref{PDF plots}. These PDFs are obtained
through global fits of the world's data, including charged and neutral
lepton scattering, Drell-Yan production, and direct photon production. All
three PDF [CTEQ4M \cite{cteq4m}, MRS-R2 \cite{mrsr2}, and GRV-94 \cite{GRV94}]
have used preliminary CCFR E744/E770 structure function data in the global
fits, so the comparison contains some correlation. The data from other
neutrino experiments have not been included in most global fits because of
poor precision or disagreement with QCD.

\begin{figure}[ptbp]
\psfig{figure=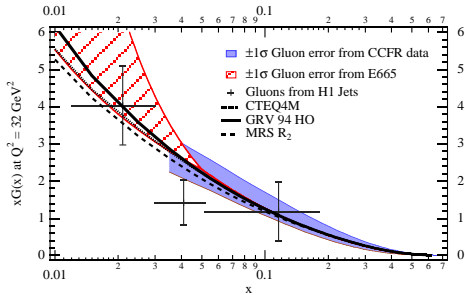,bbllx=10bp,bblly=250bp,bburx=600bp,bbury=600bp,width=3.5in,clip=T}
\caption{The CCFR gluon distribution evolved to $Q^2=32$ GeV$^2$
and compared to gluon distributions measured in other
processes. Shaded
region shows $\pm 1\sigma$ errors. Crosses show results from HERA jet
measurements. Dotted line is CTEQ4M. Solid line is GRV94 HO. Dashed line is
MRS R2. Hatched region is gluon distribution from E665 hadronic energy
distributions.}
\label{glue}
\end{figure}

The gluon distribution is the most poorly constrained of the parton
distribution measurements. This is because in most processes where precision
measurements can be used to extract PDFs, the gluons enter only indirectly
through the QCD predictions for the $Q^2$ evolution. Fits to $xF_3$ and $F_2$%
, described previously, allow one to extract the form of the gluon
distribution. The global fit to the CCFR E744/E770 data yields the
distribution 
\[
xG(x,Q_0^2=5\text{ GeV}^2)=(2.22\pm 0.25)(1-x)^{4.65\pm 0.69}\text{%
\quad(CCFR)}. 
\]
Figure~\ref{glue} shows this distribution as a function of $x$, evolved to $%
Q^2=32~ $GeV$^2$. The shaded region indicates the CCFR $\pm 1\sigma $
errors. For comparison, the crosses show the gluon distribution as measured
from jet production at the H1 experiment \cite{H1}. The hatched region
indicates the E665 gluon distribution measured from the energy distribution
of hadrons produced in deep inelastic muon scattering \cite{Harsh}. Also
shown are the gluon distributions from fits to data from a wide range of
high energy experiments which are available as parton distribution
functions: CTEQ4M \cite{cteq4m} (dotted), GRV 94 HO \cite{GRV94} (solid), and
MRS R2 \cite{mrsr2} (dashed). The gluon distributions from the various
experiments agree in the region of the CCFR data with $x>0.01$.

\subsubsection{The QCD Sum Rules}

QCD predictions exist for a wide variety of ``sum rules,'' which are
integrals over quark densities. Two QCD sum rules can only be measured in
neutrino scattering: the Gross-Llewellyn Smith (GLS) sum rule and the Adler
sum rule. The values for these sum rules are fundamental predictions of QCD.

\paragraph{The GLS Sum Rule.}

The GLS sum rule gives the QCD expectation for the integral of the valance
quark densities. To leading order in perturbative QCD, the integral $\int 
\frac{dx}x(xF_3)$ is the number of valence quarks in the proton and should
equal three \cite{gls}. QCD corrections to this integral result in a
dependence on $\alpha _s$  \cite{gls_higher_order}: 
\begin{eqnarray}
\int_0^1xF_3(x,Q^2)\frac{dx}x &=&  \nonumber \\
3(1-\frac{\alpha _s}\pi - &a(n_f)&(\frac{\alpha _s}\pi )^2-b(n_f)(\frac{%
\alpha _s}\pi )^3).  \label{eq:gls}
\end{eqnarray}
In this equation, $a$ and $b$ are known functions of the number of quark
flavors $n_f$ which contribute to scattering at a given $x$ and $Q^2$. This
is one of the few QCD predictions that are available to order $\alpha _s^3$.

One can determine $\alpha _s(Q^2)$ from $\int F_3dx$ by inverting the GLS
equation. However, for a given $Q^2$ value, there is a limited region in $x$
that is accessible by the data from any one experiment. Two analysis methods
address this problem.

\begin{figure}[pthb]
\psfig{figure=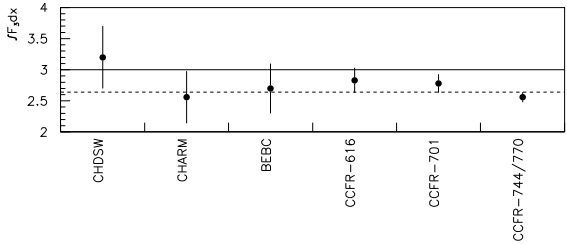,bbllx=10bp,bblly=430bp,bburx=600bp,bbury=800bp,width=3.5in,clip=T}
\caption{Summary of measurements of the GLS sum rule}
\label{GLS plot}
\end{figure}

The first method extrapolates all $xF_3$ data to a single $Q^2$ value,
resulting in data over a sufficiently wide range of $x$ to evaluate
accurately the integral in Eq.~\ref{eq:gls}. However, this introduces
systematic errors involved in the assumptions for the extrapolation. To
evaluate this error, three extrapolation methods can be considered: QCD
evolution, $1/Q^2$ , and a simple power law. The results of this analysis
for several neutrino experiments are shown in Fig.~\ref{GLS plot}. The world
average is 
\begin{equation}
\int F_3dx=2.64\pm 0.06.  \nonumber
\end{equation}
which is consistent with the NNLO evaluation of Eq.~\ref{eq:gls} with $%
\Lambda =250\pm 50~$MeV.

The second method evaluates the integral by combining the $xF_3$
measurements of experiments covering different kinematic regions  \cite
{debbie}. At high $x$, $F_2$ is nearly equal to $xF_3$. Therefore to improve
further the kinematic range at high-$x,$ one can use $F_2$ data from
charged-lepton scattering at for $x>0.5$.

An issue which can be best addressed using the second analysis technique is
that of higher-twist contributions to the GLS integral. Because CCFR
evaluates this integral in the low-$Q^2$ region where higher-twist
contributions can contribute, an additional term $\Delta HT$ must be
included in Eq.~\ref{eq:gls} \cite{gls_higher_order}. This higher-twist
contribution has been computed to be approximately $0.27/Q^2$ based on three
independent calculations of the dynamical power corrections \cite{bktwist}.
The three calculations are consistent with each other to $\sim 50\%$, so an
error of $\pm 0.14/Q^2$ should be assumed for this result. A more recent
result \cite{Ross} improved on the technique of  \citeasnoun{bktwist} by
explicitly retaining extra terms dropped in the earlier analysis. As a
result, the extracted value is more reliable, leading to a smaller
theoretical error, $(0.16\pm 0.01)/Q^2$ \cite{Ross}.

\paragraph{The Adler Sum Rule.}

The Adler Sum Rule predicts the difference between the quark densities of
the neutron and the proton, integrated over $x$. To leading order: 
\begin{eqnarray}
\int_0^1 &&(F_2^{\nu n}(x,Q^2)-F_2^{\nu p}(x,Q^2))\frac{dx}x  \nonumber \\
&=&\int_0^12(u_v(x)-d_v(x))dx=2.  \label{eq:adler}
\end{eqnarray}
The only assumption introduced in this calculation is that the quark and
antiquark seas are equal; there is no assumption concerning the equality of,
for example, the $\overline{u}$ and $\overline{d}$ seas.

\begin{figure}[ptbp]
\psfig{figure=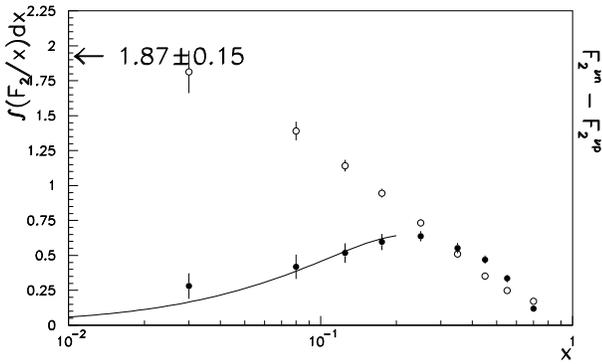,bbllx=10bp,bblly=360bp,bburx=600bp,bbury=738bp,width=3.5in,clip=T}
\caption{Evaluation of the Adler Sum Rule using the data from BEBC evolved
to $Q^2=5~GeV^2$ (closed circles). The solid line shows the fit to the low-$x
$ data. The open circles show the integral from each bin to $x=1$. Using the
fit to evaluate the integral for $x<0.01$, the sum rule is found to be $%
1.87\pm 15$.}
\label{adler.ps}
\end{figure}

Using the $F_2^{\nu p}$ and $F_2^{\nu n}$ data shown in Fig.~\ref{bebcnp},
BEBC has demonstrated that the Adler sum rule is consistent with 2.0  \cite
{BEBCD2}. For this review, we have evolved the $F_2^{\nu n}-F_2^{\nu
p}$ measurements to $Q^2=5~$GeV$^2$ as shown in Fig.~\ref{adler.ps} 
(closed points).  We then evaluate the Adler sum rule under two
assumptions for the low-$x$ behavior.  For the first method, we assume
that the lowest bin is a constant equal to the value of the lowest-$x$
data point.  In this case, the integral is $2.05\pm 0.15$ at $Q^2=5~$.
Alternatively, for the second method, we fit for the low-$x$ behavior
using the 
polynomial shown by the solid line. Using this polynomial to evaluate the
integral for $0<x<0.01$, we obtain $1.87\pm 0.15$, which is also within $%
1\sigma $ of the expectation.

The comparable sum rule for charged-lepton scattering is the Gottfried Sum
Rule \cite{Gottfried theory}: 
\begin{eqnarray}
\int_0^1(F_2^{\nu n}(x,Q^2)&&-F_2^{\nu p}(x,Q^2))\frac{dx}x=  \nonumber \\
&&\int_0^1(u(x)-d(x))(e_u^2-e_d^2)dx={\frac 13},  \label{eq:gott}
\end{eqnarray}
where $e_u^2$ and $e_d^2$ are the charges of the $u$ and $d$ quarks.
Inherent in this form of the Gottfried sum rule is the assumption that the $%
\overline{u}$ distribution is equal to the $\overline{d}$ distribution.
There is no {\it a priori} reason to expect a difference in these
distributions and there was therefore some surprise when the NMC experiment
measured the Gottfried sum to be $0.240\pm 0.016$ \cite{NMCGott}. This result
could be explained if $\overline{u} \ne \overline{d}$. Note that this
assumption $\overline{u}=\overline{d}$ was not required in the Adler sum
rule.

\section{Neutrino Charm Production: Measurements of the Strange Sea, Charm
Sea, and $|V_{cd}|$}

\label{Dimuon Section}

\subsection{Introduction}

As shown in the previous sections, neutrino DIS is particularly well suited
for measuring the parton densities due to the neutrino's ability to resolve
the flavor of the nucleon constituents. In addition, neutrino scattering is
an effective way to study the dynamics of heavy quark production, due to the
light to heavy quark transition at the charged current vertex. In
particular, neutrino charm production can be used to isolate the nucleon
strange quark distributions, $xs(x)$ and $x\overline{s}(x)$, and study the
transition to the heavy charm quark. The strange quark distribution function
is of particular theoretical interest since it may contribute to the low-$%
Q^2 $ properties of the nucleon in the non-perturbative regime. Charm
production is also an important testing ground for NLO perturbative QCD due
to the large contribution from gluon initiated diagrams. In addition,
understanding the threshold behavior associated with the heavy charm mass is
critical to the extraction of the weak mixing angle, $\sin ^2\theta _W$,
from neutrino neutral current data as described in the next section.

One distinctive signature for the production of charmed quarks in neutrino-
and antineutrino-nucleon scattering is the presence of two
oppositely-charged muons. (Studies of same-sign dimuon production have shown
that the observed signal is completely dominated by non-prompt background
sources which preclude any quantitative physics
measurements  \cite{pamela}.)   
In the case of neutrino scattering, the underlying process is a neutrino
interacting with an $s$ or $d$ quark, producing a charm quark that fragments
into a charmed hadron. The charmed hadron's semileptonic decay 
($BR=\approx 10\%$ ) 
produces a second muon of opposite sign from the first. 
\begin{eqnarray}
\nu _\mu \;+\;{\rm N}\;\longrightarrow \;\mu ^{-}\!\! &+&\!c\;+\;{\rm X} \\
&&\!\!\hookrightarrow s\;+\;\mu ^{+}\;+\;\nu _\mu   \nonumber
\end{eqnarray}

The analogous process with an incident antineutrino proceeds through an
interaction with an $\overline{s}$ or $\overline{d}$ antiquark, again
leading to oppositely-signed muons in the final state. 
\begin{eqnarray}
\overline{\nu }_\mu \;+\;{\rm N}\;\longrightarrow \;\mu ^{+}\!\! &+&\!%
\overline{c}\;+\;{\rm X} \\
&&\!\!\hookrightarrow \!\overline{s}+\;\mu ^{-}\;+\;\overline{\nu }_\mu 
\nonumber
\end{eqnarray}

\begin{figure}[pthb]
\psfig{file=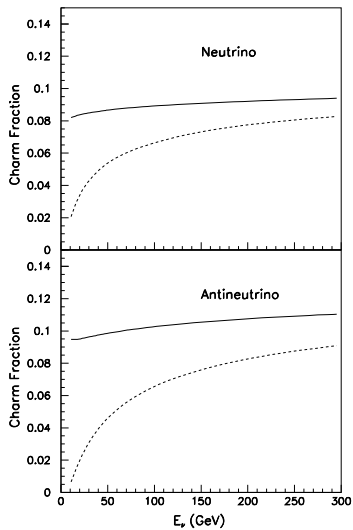,bbllx=50bp,bblly=60bp,bburx=490bp,bbury=697bp,clip=,height=3.0in}
\caption{Fraction of the neutrino (anti-neutrino) 
total cross section that is associated with charm production. 
The dashed curves are from a leading-order calculation 
with $m_c = 1.31$ GeV and the solid are 
from the same calculation with $m_c = 0$.}
\label{fig: sigvsE}
\end{figure}

The heavy charm quark is expected to introduce an energy threshold in the
dimuon production rate. This effect has been described in the past through
the ``slow rescaling'' model \cite{barnetta,Targmass}, in which $\xi ^0$, the
momentum fraction carried by the struck quark, is related to the kinematic
variable $x=Q^2/2M\nu $ by the expression $\xi ^0=x(1+m_c^2/Q^2)$ where $m_c$
is the mass of the charm quark. (A more complete treatment of the kinematics
associated with heavy quark production modifies this model slightly, as
discussed below.) Charm production by neutrinos (or anti-neutrinos) is a
substantial fraction of the total cross section and grows from threshold to
about 10$\%$ at high energy as shown in Figure \ref{fig: sigvsE}.

Neutral current scattering off the charm sea can be isolated and studied by
looking for the production of wrong-sign single muons (WSM) in the
interaction of a relatively pure $\nu _\mu $ beam. This is one of the only
techniques for probing the charm content in the nucleon which is assumed in
the standard calculation to evolve from zero at a threshold $Q^2\approx 4$
GeV$^2$ . The process for study is neutrino neutral current scattering off a
charm quark followed by the semi-leptonic decay of the charm quark into a
wrong-sign muon. The size of this process is therefore dependent on the
number and distribution of charm quarks in the nucleon. The main backgrounds
come from the $\overline{\nu }_\mu $ contamination in the beam and decays of
secondary pions or kaons in the hadronic shower. The signal is smaller than
the background and at present only upper limits on the size of the charm sea
are available.

\subsection{Differential Cross Section for Dimuon Production and Heavy Quark
Effects}

\label{Dimuon Kinematics}

The differential cross section for dimuon production is expressed generally
as 
\begin{equation}
{\frac{d^3\sigma (\nu _\mu N\rightarrow \mu ^{-}\mu ^{+}X)}{d\xi \ dy\ dz}}={%
\frac{d^2\sigma (\nu _\mu N\rightarrow cX)}{d\xi \ dy}\ }D(z)\
B_c(c\rightarrow \mu ^{+}X),  \label{eq:dimuon}
\end{equation}
where the function $D(z)$ describes the hadronization of charmed quarks and $%
B_c$ is the weighted average of the semi-leptonic branching ratios of the
charmed hadrons produced in neutrino interactions.

The heavy charm quark introduces an energy dependence in the charm
production rate and also affects the measured angular (or $y$) distribution
compared to that expected for a massless quark. (Additional thresholds
related to the masses of final-state charmed mesons and baryons are present
but only become important near production threshold.) These are kinematic
effects for which $\xi $, the momentum fraction of the struck quark, is
related to the Bjorken scaling variable $x$, through the expression  \cite
{AOT2} 
\begin{equation}
\xi =\left( {\frac 1{2x}}+\sqrt{{\frac 1{4x^2}}+{\frac{M^2}{Q^2}}}\right)
^{-1}{\frac{Q^2-m_s^2+m_c^2+\Delta (-Q^2,m_s^2,m_c^2)}{2Q^2}},
\end{equation}
where $m_c$ is the charm quark mass and $m_s$ refers to the initial state
quark mass, either the strange quark or the down quark, and $\Delta
(-Q^2,m_s^2,m_c^2)$ is the triangle function, defined by $\Delta
(a,b,c)\equiv \sqrt{a^2+b^2+c^2-2(ab+bc+ca)}$. The full expression for $\xi $
can be simplified by neglecting the small effect of the initial state quark
mass to yield 
\begin{equation}
\xi ^0=x\left( 1+{\frac{m_c^2}{Q^2}}\right) \left( 1-{\frac{x^2M^2}{Q^2}}%
\right) .
\end{equation}
Relating $\xi $ and $x$ through the charm quark mass is referred to as
``slow-rescaling''  \cite{Targmass,barnetta}. A comparison of the full $\xi $
calculation to the simple $\xi ^0$ expression given above yields a
difference of 10\% (2\%) for $x=0.2$ and $Q^2=1.0\,(5.0)$ GeV$^2$; this
difference is much smaller at lower $x$ and higher $Q^2$.

At leading order charm is produced by scattering directly off of  
strange and down quarks in the nucleon. The LO differential cross section
for an isoscalar target, neglecting target mass effects, is given by: 
\begin{eqnarray}
\left\{ {\frac{d^2\sigma (\nu _\mu N\rightarrow cX)}{d\xi \,dy}}\right\}
_{LO} &=&\frac{G_F^2ME_\nu }{\ \pi (1+Q^2/M_W^2)^2}\;\{\;[\xi u(\xi ,\mu
^2)+\xi d(\xi ,\mu ^2)]\;|V_{cd}|^2  \nonumber \\
&&+\;2\xi s(\xi ,\mu ^2)\;|V_{cs}|^2\;\}\left( 1-\frac{m_c^2}{\ 2ME_\nu \xi }%
\right) ,  \label{eq:lo}
\end{eqnarray}
where $\xi u(\xi ,\mu ^2)$, $\xi d(\xi ,\mu ^2)$ and $\xi s(\xi ,\mu ^2)$
represent the momentum distributions of the $u$, $d$ and $s$ quarks within
the proton (the corresponding $\overline{\nu }_\mu $ process has the quarks
replaced by their antiquark partners). $|V_{cd}|$ and $|V_{cs}|$ are the CKM
matrix elements. The dependence of the parton distributions on the scale $%
\mu ^2$ is specified by QCD  \cite{DGLAP1,DGLAP2,DGLAP3,DGLAP4}, but the
precise dependence of $\mu ^2$ on $Q^2$ and $x$ is somewhat arbitrary in
finite-order perturbation theory. The finite charm mass included in the $%
(1-m_c^2/2ME_\nu \xi )$ expression induces an effective Callan-Gross
violation such that $R_L^{LO-charm}(\xi ,Q^2)\simeq 1+m_c^2/Q^2$. In the
modified leading order analysis of  \citeasnoun{sar}, Callan-Gross violation
is included by replacing the term $\left( 1-m_c^2/(2ME_\nu \xi )\right) $ in
Eq.(\ref{eq:lo}) with 
\begin{equation}
1-m_c^2/(2ME_\nu \xi )\rightarrow \frac{1+R_L(\xi ,Q^2)}{1+(2M\xi /Q)^2}%
\left( 1-y-Mxy/(2E_\nu )\,+xy/\xi \right)  \label{eq:RlLO}
\end{equation}
using external measurements of the structure function $R_L(\xi ,Q^2)$  \cite
{rl}. In the next-to-leading order (NLO) formalism, violation of the
Callan-Gross relation emerges as a consequence of the more complete QCD
calculation.

The LO expression illustrates the sensitivity of the process to the strange
quark sea. Charm (anticharm) production from scattering off $d$ $(\overline{d%
})$ quarks is Cabibbo suppressed. In the case of charm produced by
neutrinos, approximately 50\% is due to scattering from $s$ quarks, even
though the $d$ quark content of the proton is approximately ten times
larger. In the case of antineutrino scattering, where $\overline{d}$ quarks
from the sea contribute, roughly 90\% is due to scattering off $\overline{s}$
quarks.

\subsection{Next-to-Leading Order Corrections}

\begin{figure}[pthb]
\psfig{file=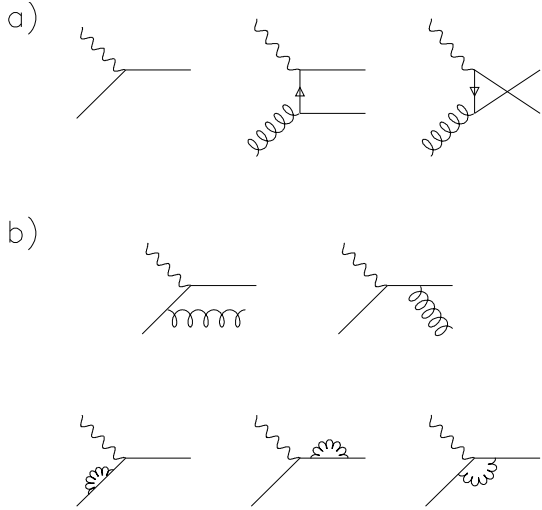,bbllx=82bp,bblly=240bp,bburx=508bp,bbury=576bp,clip=,width=3.5in}
\caption{Mechanisms that contribute to neutrino production of charm up to  
${\cal O}(\alpha_S)$. {\bf a)} The dominant diagrams: the leading-order
quark-initiated diagram, and the t channel and u channel gluon-initiated
diagrams, respectively. {\bf b)} The radiative-gluon and self-energy diagrams.}
\label{fig:NLO} 
\end{figure}

Because neutrino charm production has a large sea quark component at
leading-order, the next-to-leading-order gluon-initiated contributions are
significant  \cite{AOT}. Gluon-initiated production of charm proceeds through
both the $t$ and $u$ channels as shown in Figure \ref{fig:NLO}a. The size of
the gluon distribution, which is an order of magnitude larger than the sea
quark distribution, compensates for the extra power of $\alpha _S$ involved
in the gluon-initiated diagram. The NLO quark-initiated diagrams, shown in
Fig. \ref{fig:NLO}b, in which a gluon is radiated, also enter the
perturbative expansion at ${\cal O}(\alpha _S)$, but the corrections from
these diagrams to the cross section are small. Several calculations
including the next-to-leading-order formalism have been done \cite
{AOT2,kramer,gj}. The treatment of the gluon-initiated diagrams in these has
been cross-checked and found to be consistent.

\begin{figure}[pthb]
\psfig{file=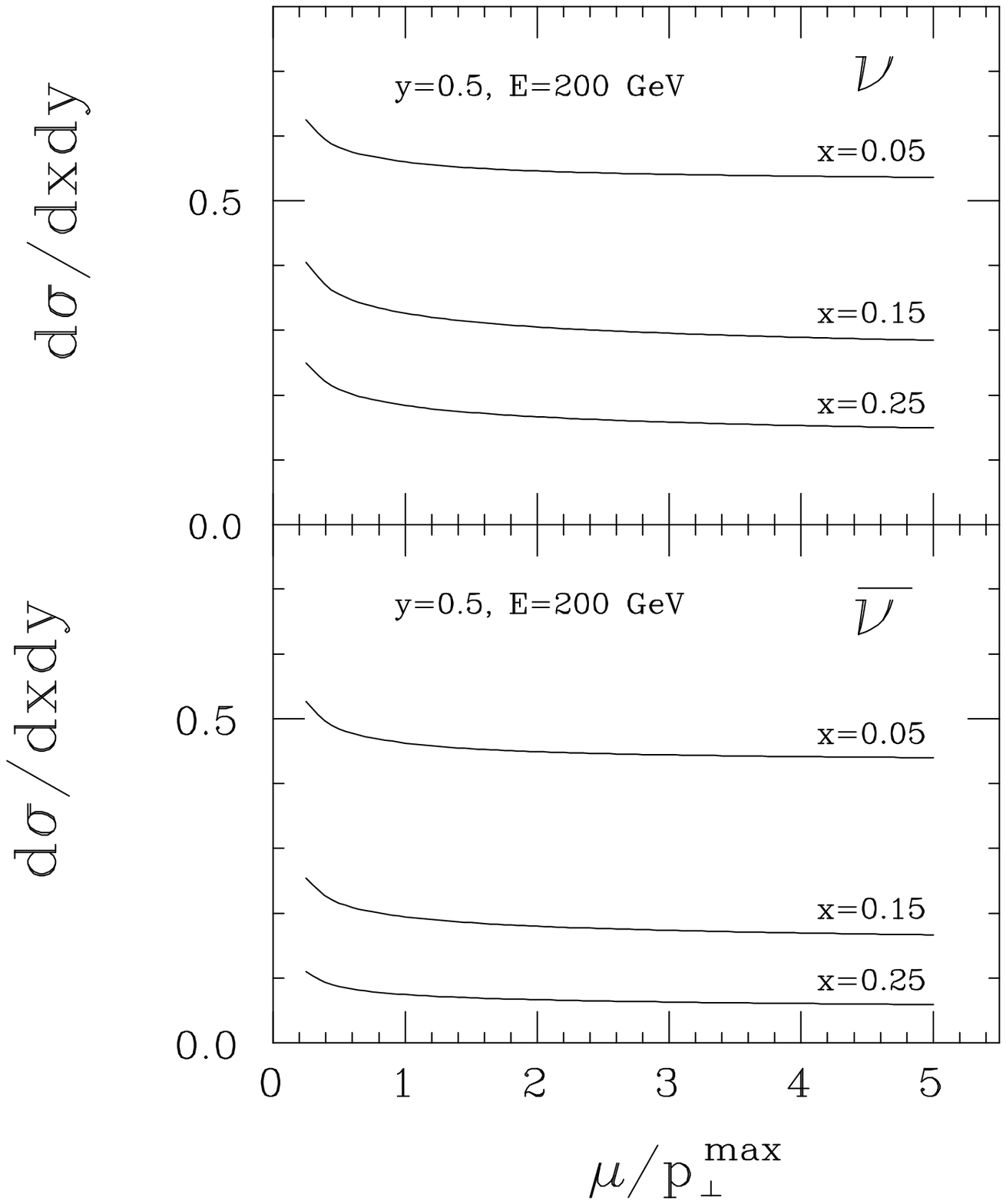,bbllx=72bp,bblly=100bp,bburx=518bp,bbury=627bp,clip=,width=3.5in}
\caption{The $\mu^2$ scale dependence of the differential cross
section for neutrino and antineutrino production of charm, where
$\mu^2$ identifies the factorization and renormalization scales.  
The scale on the abscissa
is in units of $p_\perp^{\rm max}$. For $E=200$ GeV and $y=0.5$, the $x=0.05$, 
0.15, 0.25 lines correspond to $p_\perp^{\rm max}=6.6$, 6.2, 5.8 GeV/c,
respectively.} 
\label{fig:scale}
\end{figure}

As with all applications of perturbative QCD, a theoretical uncertainty is
associated with the choice of factorization and renormalization scales. Some
scale dependence is unavoidable for any calculation done to finite order in $%
\alpha _S$. Most analyses assume that the factorization and renormalization
scales are both equal to a single parameter, $\mu $. The $\mu $ scale is
interpreted as setting the boundary between the collinear and non-collinear
regions of the $p_{\perp }$ integration over the final states where $%
p_{\perp }$ is the transverse momentum of the initial state quark coming
from the gluon splitting. Therefore, a scale proportional to $p_{\perp }^{%
{\rm max}}=\Delta (W^2,m_c^2,M^2)/\sqrt{4W^2}$ is suggested by %
 \citeasnoun{AOT2}. Figure \ref{fig:scale} shows the scale dependence of the
differential cross section for charm production, where the abscissa is in
units of $p_{\perp }^{{\rm max}}$. The scale dependence is weak for values
above one unit of $p_{\perp }^{{\rm max}}$, but there is a strong dependence
when $\mu $ is below this value.

\subsection{Experimental Issues for Dimuon Measurements}

\subsubsection{Data Samples and Event Selection}

Many experiments have investigated neutrino and antineutrino production of
charm since it was first observed in 1974 \cite{HPWF}. The experiments
considered here include the highest statistics experiments using each of the
main experimental techniques. Most of the experiments have been described in
Sec. \ref{Experiments Section} and only details relevant to charm
measurements will be given here. The largest data samples are available from
the CCFR \cite{Bazarko} and CDHS \cite{CDHS:2mu} experiments.  
The CCFR
collaboration used a high-energy, quadrupole-triplet neutrino beam and a
massive, high-density steel detector at the Fermilab Tevatron. The CDHS
collaboration at CERN used a magnetized steel target-calorimeter illuminated
by a horn-focused neutrino beam with a somewhat lower energy spectrum.
Results are also available from the Fermilab 15-ft bubble chamber experiment
(15-ft BC) for $\mu ^{-}e^{+}$ events \cite{ColBNL}. 
In this experiment, the low threshold, 
$p_{e^{+}}>0.3$ GeV, combined with good statistics for E$_\nu <30$ GeV give
good sensitivity to the slow-rescaling threshold behavior.   
The final data
set considered is the E531 neutrino emulsion experiment \cite{E531} at
Fermilab which measures inclusive charm particle production directly by
identifying the charm decay in the emulsion. Thus, this experiment can
provide information on the charm species produced by neutrinos and the
kinematics of the production. Table \ref{2muexps} summarizes the neutrino
charm production data samples from these experiments.

\begin{table}[tbp] \centering%
\begin{tabular}{|l|c|c|c|c|}
\hline
Experiment & E$_\nu $ (GeV) & $\mu ^{-}l^{+}$ events & $\mu ^{+}l^{-}$ events
& Background (\%) \\ \hline
CCFR \cite{Bazarko} & $30-600$ & $5030$ & $1060$ & $15$ \\ 
$\quad p_{\mu _2}>5\text{ GeV}$ & $\left( >100\right) $ & $\left(
3721\right) $ & $\left( 493\right) $ &  \\ 
CDHS \cite{CDHS:2mu} & $30-250$ & $11041$ & $3684$ & $13$ \\ 
$\quad p_{\mu _2}>5\text{ GeV}$ & $\left( >100\right) $ & $\left(
3589\right) $ & $\left( 452\right) $ &  \\ 
15-ft BC \cite{ColBNL} & $1-200$ & $461$ & $-$ & $18$ \\ 
$\quad p_{e^{+}}>0.3\text{ GeV}$ &  &  &  &  \\ 
E531 \cite{E531} & $1-250$ & $122$ & $-$ & $3$ \\ 
$\quad \nu -\text{ Emulsion}$ &  & (Charm Events) &  &  \\ \hline
\end{tabular}
\caption{Summary of data samples from some neutrino charm production
experiments\label{2muexps}.  The CDHS and CCFR experiments detect 
dimuon events, the 15-ft BC measures $\mu\ -  e$ events, and the E531 
experiment detects the inclusive decays of charmed particles in an emulsion target. }%
\end{table}%

\subsubsection{Non-prompt and Other Background Sources of Dimuons}

For the CCFR and CDHS experiments, non-prompt pion and kaon decays
constitute the main background to the charm-initiated signal. The high
density of their target-calorimeters minimizes this contamination due to the
short interaction length of the detector. The background is estimated from
Monte Carlo calculations and test beam measurements of muon production in
hadron showers. The systematic uncertainty in this background has been
estimated \cite{pamela} by the CCFR collaboration to be 15\%(20\%) for $\nu (%
\overline{\nu })$ induced showers and mainly comes from uncertainties in the
secondary particle production spectrum including heavy nuclear target
effects. For the 15-ft BC experiment, the background is from Dalitz decays
of $\pi ^0$ mesons and other $\gamma $ conversions. Finally, the E531
emulsion experiment has a small background from non-charm events that pass
their charm identification cuts. The background level for each experiment is
listed in Table \ref{2muexps}.

\subsubsection{Charm quark production, fragmentation, and experimental
acceptance}

The production and fragmentation of the charm quark into charmed hadrons
must be modeled in order to extract information from the observed dilepton
events. For the LO formalism, the kinematic suppression due to the mass of
the charm quark, $m_c$, is included in Eq. \ref{eq:lo} and the charm quark
direction is defined by the $W^{+}\ q\rightarrow c$ kinematics.

The gluon-initiated diagrams in the NLO calculation proceed by $%
W^{+}g\rightarrow c\overline{s}$ as shown in Fig. \ref{fig:NLO}a. This NLO
production proceeds through both the $t$ and $u$ channels which dominate
different regions of phase space. In the $t$ channel, the gluon splits into
an $s\overline{s}$ pair and the $c$ quark emerges from the $W$-boson vertex.
In the $u$ channel, the legs of the $c$ and $\overline{s}$ quarks are
crossed---the gluon splits into a $c\overline{c}$ pair and the $\overline{s}$
quark emerges from the $W$-boson vertex. In the $W$-boson--gluon center of
mass frame, the $c$ quark is produced at an angle $\theta _c^{*}$ relative
to the $W$-boson direction. The production angle is related to the momentum
of the $c$ quark in the lab. When $\theta _c^{*}$ is small---$t$ channel
dominance---the $c$ quark carries most of the $W$-boson momentum. As $\theta
_c^{*}$ approaches $\pi $---$u$ channel dominance---the $c$ quark emerges
with little momentum in the lab. Acceptance corrections which are dependent
on $\theta _c^{*}$ need to be included in the analysis of any experimental
measurement; these corrections are largest at high $\theta _c^{*}$ where the
decay products of the outgoing $c$ quark are hard to detect. 

The charm quark fragmentation uncertainties are a second important source of
systematic error for the dilepton physics measurements. The fragmentation of
the charm quark into $D$ mesons is typically parameterized using the Collins
and Spiller function \cite{colfrag}, 
\begin{equation}
D(z)=N\left[ (1-z)/z+\epsilon _C(2-z)/(1-z)\right] (1+z)^2[1-(1/z)-\epsilon
_C/(1-z)]^{-2};
\end{equation}
older measurements also employed the 
Peterson function \cite{peterson},   
\begin{equation}
D(z)=N{z[1-(1/z)-\epsilon _P/(1-z)]^2}^{-1},
\end{equation}
where $z=p_D/p_D^{{\rm max}}$ is the fraction of its maximum possible
momentum that the $D$ meson carries and the $\epsilon $ parameters determine
the shape of the distribution. 
In principle, $e^{+}e^{-}$ collider data could provide 
precise data on the $\epsilon$ parameters.
However, because of uncertainties due to the different quark environment
in  $e^{+}e^{-}$ relative to neutrino scattering, 
the $\epsilon $ parameters in the
fragmentation models are best determined from fits to neutrino charm
production data.    
Fits to the charm particle spectrum in the
E531 data \cite{E531} give 
$\epsilon_P=0.18\pm 0.06$ and fits to the CCFR, $z_{\text{vis%
}}=E_{l_2}/(E_{l_2}+E_{had})$ distribution yield $\epsilon _C=0.81\pm 0.14$ 
\cite{Bazarko}
(or $\epsilon _P=0.22\pm 0.05$ \cite{sar}). These can be compared to the CLEO
measurements \cite{cleo}: The Collins and Spiller function fits the CLEO data
well with $\epsilon _C=0.64\pm 0.14$, but the Peterson function does not
reproduce the data for any value of $\epsilon _P$ . The CDHS collaboration
in their dimuon analysis \cite{CDHS:2mu} parameterized the charm
fragmentation phenomenologically by functions ranging from a delta function
at $z=0.68$ to a flat distribution.

The semileptonic decay of the charmed particles also needs to be
modeled and the effect of missing energy from the outgoing neutrino
needs to be included.  For this modeling, the three-body decay
kinematics is taken from $e^{+}e^{-}$ collider studies including the
differences for the various charmed particle species. 
The composition of charmed particles produced in neutrino interactions has
been measured by the E531 $\nu -$emulsion experiment \cite{E531}. With an E$%
_\nu >30$ GeV cut, the production is dominated by neutral and charged $D$
mesons. From a recent re-analysis \cite{Bolton:CKMa,Bolton:CKMb} of the
neutrino produced charm particle data, the production fractions are $0.58\pm
0.06$, $0.26\pm 0.06$, $0.07\pm 0.05$, and $0.07\pm 0.04$ for D$^0$, D$^{+}$%
, D$_s^{+}$, and $\Lambda _c^{+}$ respectively for this energy range.

\subsection{Dimuon Measurements and Results}

\subsubsection{Charm Production Rate\label{sec:dimurates}}

\begin{figure}[pthb]%
\psfig{file=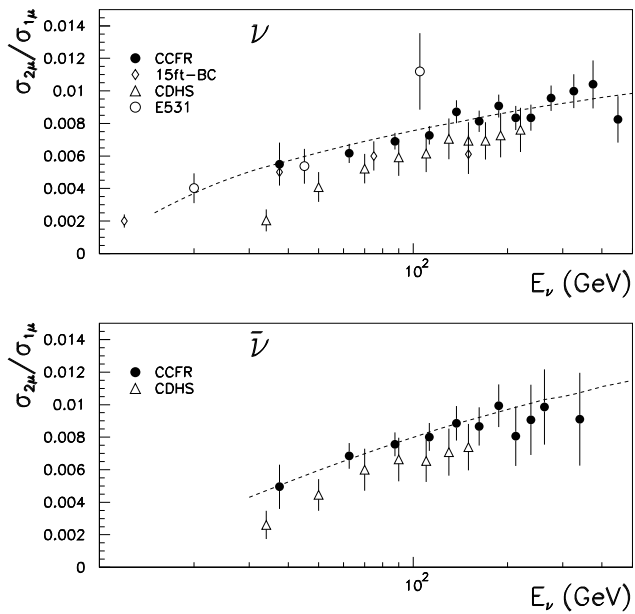,bbllx=20bp,bblly=150bp,bburx=550bp,bbury=656bp,clip=,width=3.5in}
\caption{The dilepton to single muon rate, corrected for acceptance, 
smearing, and 
minimum second lepton cut versus neutrino energy for the 
CCFR, CDHS, and 15-ft BC
experiments. Also, shown is the E531 emulsion experiment 
where the charm rate has been 
multiplied by a semileptonic branching ratio of 0.11.  The dashed 
curve is a LO ``slow-rescaling''
model prediction with m$_c = 1.31$ GeV. (The errors on the 
CDHS points include uncertainties
in fragmentation and normalization; the other points do not.) }
\label{fig:dimurate}   
\end{figure}

The energy dependence of the neutrino charm production rates provides a
direct test of the modeling of the heavy charm quark production and the
thresholds appropriate for the charm particles in the final states. At low
energy, E$_\nu <20$ GeV , quasi-elastic scattering, such as $\nu _\mu
\,n\rightarrow \mu ^{-}\,\Lambda _c^{+}$ , is a large fraction of the charm
production and accounts for more than 25\% of the charm cross section. Above
50 GeV, the exclusive final state effects become small (below $5\%$) and the
simple parton model approach in which the production and hadronization
factorizes becomes valid. Comparing the predicted energy dependence expected
from the ``slow-rescaling'' model with measured rates corrected for
acceptance, smearing, and other kinematic cuts shows good agreement between
the data and model for the CCFR, 15-ft BC, and E531 experiments as shown in
Fig. \ref{fig:dimurate}. The CDHS data, however, is somewhat below the
prediction and the other data, especially at low energy.

\begin{figure}[pthb]
\psfig{file=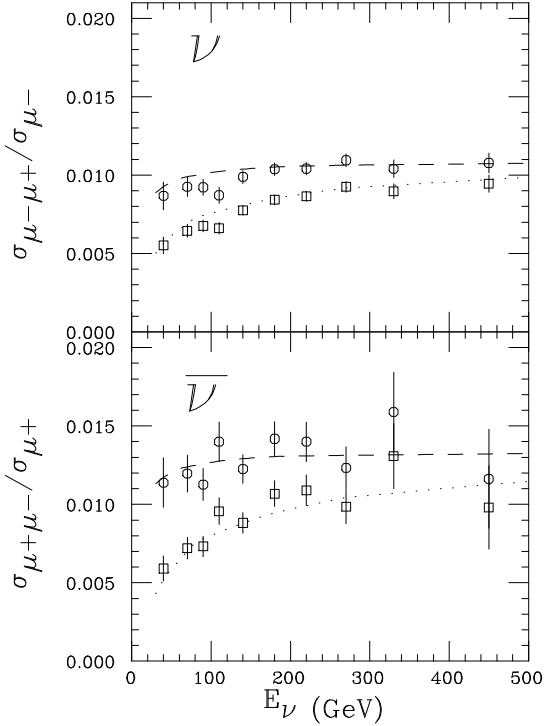,bbllx=80bp,bblly=130bp,bburx=500bp,bbury=640bp,clip=,width=3.5in}
\caption{Opposite-sign dimuon rates versus E$_\nu$ 
for the CCFR $\nu_\mu$ (top) and 
$\overline{\nu}_\mu$ (bottom) data.  Rates corrected for acceptance, 
smearing, and kinematic cuts are indicated by the squares.  
Those corrected for slow rescaling
with m$_c = 1.31$ GeV are given by the circles. The curves indicate the 
LO model prediction with m$_c = 1.31$ GeV before (dotted) and after (dashed) 
correcting for the finite charm mass but including 
charm particle mass effects.}
\label{fig:dimucorr} 
\end{figure}

After correcting for the ``slow-rescaling'' threshold corresponding to an
appropriate $m_c$, the rates become less dependent on E$_\nu $, exhibiting
only the sharp, low energy threshold behavior associated with the production
of heavy charmed mesons $[W^2>(M_D+M_{proton})^2]$. The CCFR data shown in
Fig. \ref{fig:dimucorr} is in good agreement with this model for a charm
quark mass of $m_c=1.31$ GeV/$c^2$. In the next section, the energy
dependence of charm production will be used to determine $m_c$.

\subsubsection{Extraction of the Strange Sea, $xs(x,Q^2)$, and Charm mass, $%
m_c$.}

The CCFR collaboration have done both a leading order \cite{sar} and
next-to-leading \cite{Bazarko} order analysis of their dimuon data samples.
For these analyses, the single-muon and dimuon events were simulated using
Monte Carlo techniques and compared to the observed dimuon distribution to
extract the various physics parameters. Quark and antiquark momentum
densities were obtained from the measured CCFR structure functions \cite
{quintas,leung} using the following procedure.

The $F_2$ and $xF_3$ structure functions are used to determine the singlet, $%
xq_{SI}(x,\mu ^2)=xq(x,\mu ^2)+x\overline{q}(x,\mu ^2)$, non-singlet, $%
xq_{NS}(x,\mu ^2)=xq(x,\mu ^2)-x\overline{q}(x,\mu ^2)$, and gluon, $%
xg(x,\mu ^2)$, distributions. These distributions are obtained from
leading-order or next-to-leading-order QCD fits to the structure function
data, described in  \citeasnoun{Seligman}, using the QCD evolution programs
of Buras-Gaemers \cite{Buras:Gaemers} or Duke and Owens  \cite{DO}
respectively.

To resolve the strange component of the quark sea, the singlet and
non-singlet quark distributions are separated by flavor. Insofar as isospin
is a good symmetry, the experiment is insensitive to the exact form of the
up and down valence and sea quark distributions, because the neutrino target
is composed of iron which is nearly isoscalar. An isoscalar correction
accounts for the 5.67\% neutron excess in the iron target. The proton
valence quark content, $xq_V(x,\mu ^2)=xq_{NS}(x,\mu ^2)$, is parameterized
by 
\begin{eqnarray}
xq_V(x,\mu ^2) &=&xu_V(x,\mu ^2)+xd_V(x,\mu ^2),  \nonumber \\
xd_V(x,\mu ^2) &=&A_d(1-x)xu_V(x,\mu ^2),
\end{eqnarray}
where the shape difference for $xd_V(x)$ better fits charged-lepton
scattering measurements of $F_2^{{\rm n}}/F_2^{{\rm p}}$  \cite{NMC Gottfried}%
. $A_d$ is fixed by demanding that the ratio of the number of $d$ to $u$
valence quarks in the proton is 1/2. The non-strange quark and antiquark
components of the sea are assumed to be symmetric, so that $x\overline{u}%
(x,\mu ^2)=xu_S(x,\mu ^2)$, $x\overline{d}(x,\mu ^2)=xd_S(x,\mu ^2)$. The
isoscalar correction is applied assuming $x\overline{u}(x,\mu ^2)=x\overline{%
d}(x,\mu ^2)$.

The strange component of the quark sea is allowed to have a different
magnitude and shape from the non-strange component. The strange quark
content is set by the parameter 
\begin{equation}
\kappa ={\frac{\int_0^1[xs(x,\mu ^2)+x\overline{s}(x,\mu ^2)]\,dx}{\int_0^1[x%
\overline{u}(x,\mu ^2)+x\overline{d}(x,\mu ^2)]\,dx}},
\end{equation}
where $\kappa =1$ would indicate a flavor SU(3) symmetric sea. The shape of
the strange quark distribution is related to that of the non-strange sea by
a shape parameter $\alpha $, where $\alpha =0$ would indicate that the
strange sea has the same $x$ dependence as the non-strange component of the
quark sea. Assuming that $xs(x,\mu ^2)$ and $x\overline{s}(x,\mu ^2)$ are
the same, the sea quark distributions are then parameterized by: 
\begin{eqnarray}
x\overline{q}(x,\mu ^2) &=&2\left[ {\frac{x\overline{u}(x,\mu ^2)+x\overline{%
d}(x,\mu ^2)}2}\right] +xs(x,\mu ^2),  \nonumber \\
xs(x,\mu ^2) &=&A_s(1-x)^\alpha \left[ {\frac{x\overline{u}(x,\mu ^2)+x%
\overline{d}(x,\mu ^2)}2}\right] ,
\end{eqnarray}
where $A_s$ is defined in terms of $\kappa $ and $\alpha $.

\begin{figure}[pthb]
\psfig{file=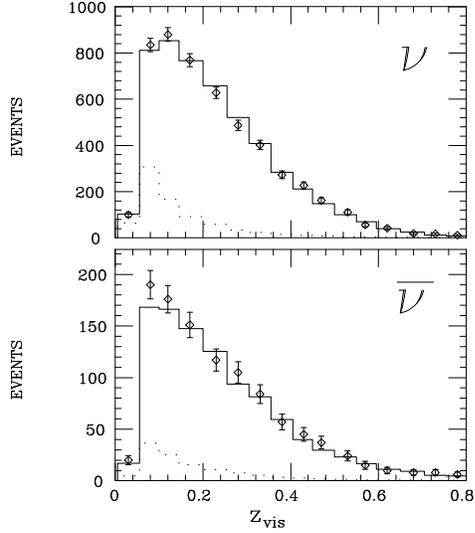,bbllx=75bp,bblly=144bp,bburx=550bp,bbury=576bp,clip=,width=3.5in}
\caption{The $z_{\rm vis}$ distribution
for the CCFR $\nu$- and $\overline\nu$-induced dimuon events. 
Data are given
by the points and the solid histogram is the result of fitting the dimuon
event simulation. The dotted histogram is the background contribution 
to the former from pion and kaon decay. }
\label{fig:zvis}   
\end{figure}

\begin{figure}[pthb]
\psfig{file=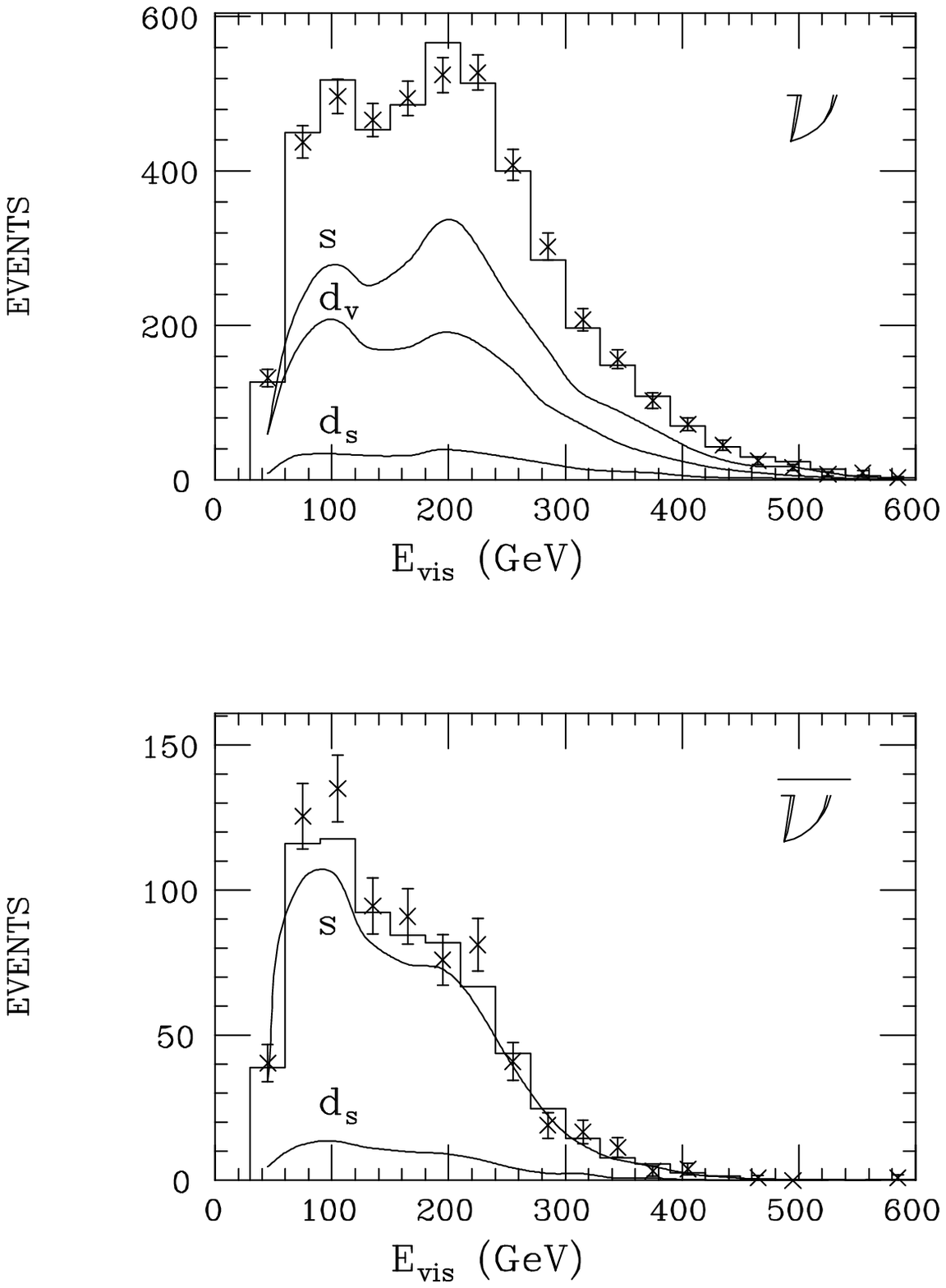,bbllx=75bp,bblly=144bp,bburx=550bp,bbury=650bp,clip=,width=3.5in}
\caption{The $E_{\text{vis}}$ distribution 
for the CCFR $\nu$- and $\overline{\nu}$-induced dimuon events 
after correction for $\pi/K$ decay
backgrounds and $\nu - \overline{\nu}$ misidentification. Data are given
by the points and the solid histogram is the result of fitting the dimuon
event simulation. The quarks contributing to the 
histogram are indicated by the curves 
where s, d$_v$, and d$_s$ are the strange sea, d valence, d sea
components, respectively.  }
\label{fig:evis}   
\end{figure}

\begin{figure}[pthb]
\psfig{file=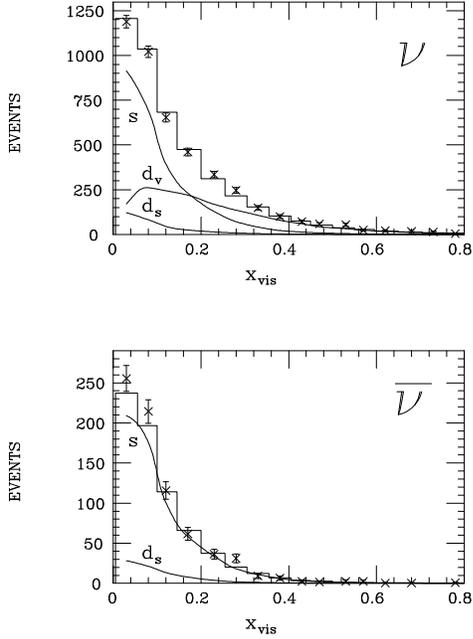,bbllx=75bp,bblly=144bp,bburx=550bp,bbury=650bp,clip=,width=3.5in}
\caption{ The $x_{\text{vis}}$ distribution 
for the CCFR $\nu$- and $\overline{\nu}$-induced dimuon events
after correction for $\pi/K$ decay
backgrounds and $\nu - \overline{\nu}$ misidentification. Data are given
by the points and the solid histogram is the result of fitting the dimuon
event simulation. The quarks contributing to the 
histogram are indicated by the curves 
where s, d$_v$, and d$_s$ are the strange sea, d valence, d sea
components, respectively.}
\label{fig:xvis}   
\end{figure}

A $\chi ^2$ minimization is performed to find the strange sea parameters $%
\kappa $ and $\alpha $, the values of $B_c$ and $m_c$ that appear in Eq.'s 
\ref{eq:dimuon} and \ref{eq:lo}, and the fragmentation parameter $\epsilon $%
, by comparing the data and Monte Carlo $x_{{\rm vis}}$, $E_{{\rm vis}} $
and $z_{{\rm vis}}$ event distributions. Projections of data for each of
these variables are compared with the NLO fits \cite{Bazarko} in Figs. \ref
{fig:zvis}, \ref{fig:evis} and \ref{fig:xvis}. Taking $|V_{cd}|=0.221\pm
0.003$ and $|V_{cs}|=0.9743\pm 0.0008$  \cite{PDG} as input values, the
extracted CCFR LO and NLO parameters with their statistical and systematic
errors are presented in Table \ref{tab:ccfr}.

{\small
\begin{table}[tbp] \centering%
\begin{center}
\begin{tabular}{|c|c|c|c|c|c|c|}
\hline
& Fragmentation & $\chi ^2/$dof & $\kappa $ & $\alpha $ & $B_c$ & $m_c$
(GeV) \\ \hline
CCFR \cite{Bazarko} & Collins-Spiller & $52.2/$ & $0.477$ & $-0.02$ & $%
0.1091$ & $1.70$ \\ 
NLO fit & $\epsilon =0.81\pm 0.14$ & $65$ & $\pm 0.045\pm 0.024\;$ & $\pm
0.57\pm 0.27\;$ & $\pm 0.0078\pm 0.0057\;$ & $\pm 0.17\;\pm 0.09$ \\ \hline
CCFR \cite{Bazarko} & Peterson & $41.2/$ & $0.468$ & $-0.05$ & $0.1047$
& $1.69$ \\ 
NLO fit & $\epsilon _P=0.20\pm 0.04$ & $46$ & $\pm 0.053\pm 0.025\;$ & $\pm
0.47\pm 0.27\;$ & $\pm 0.0076\pm 0.0057\;$ & $\pm 0.16\pm 0.11$ \\ \hline
CCFR \cite{sar} & Peterson & $42.5/$ & 0.373 & 2.50 & 0.1050 & 1.31 \\ 
LO fit & $\epsilon _P=0.20\pm 0.04$ & $46$ & $\pm 0.045\pm 0.018$ & $\pm
0.58\pm 0.31$ & $\pm 0.007\pm 0.005$ & $\pm 0.21\pm 0.12$ \\ \hline
CDHS \cite{CDHS:2mu} & $\delta $-function &  & $0.478$ & Not fit & $%
0.0839 $ & Not fit \\ 
LO fit & $(z=0.68)$ &  & $\pm 0.094$ & Assumed $0.0$ & $\pm 0.0147$ & Varied 
$1.2-1.8$ \\ \hline
\end{tabular}
\end{center}
\caption[]{
Next-to-leading-order and leading-order fit results, assuming
$xs(x)=x\overline s(x)$. Errors are statistical and systematic, except that the
errors on the fragmentation parameters are statistical only.\label{tab:ccfr}}
\end{table}%
}

The CDHS collaboration \cite{CDHS:2mu} extracts the combination $%
|V_{cd}|^2B_c $ from the slow-rescaling and acceptance corrected dimuon
rates at high energy, E$_\nu >80$ GeV. The $\overline{\nu }_\mu $ rate is
used to remove the strange quark contribution to the $\nu _\mu $ rate,
leaving the down to charm quark contribution as follows: 
\begin{equation}
|V_{cd}|^2B_c=\frac 23\frac{R_\nu -r\,R_{\overline{\nu }}}{(1-r)}
\end{equation}
where $R_{\nu (\overline{\nu })}=\sigma _{\nu (\overline{\nu })}^{2\mu
}/\sigma _{\nu (\overline{\nu })}^{1\mu }$ and $r=\sigma _{\overline{\nu }%
}^{1\mu }/\sigma _\nu ^{1\mu }=0.48\pm 0.02$.

In the CDHS analysis, the combination $\kappa |V_{cs}|^2/|V_{cd}|^2$is found
by a shape analysis of the observed $x_{vis}$ distribution for neutrino
induced dimuons where the distribution is separated into scattering off $d$
or $s$ quarks. The antineutrino dimuon data gives the strange quark
distribution and their measured charged-current structure functions gives
the distribution for down quarks. After including the fragmentation
uncertainties, a 10\% scale error, and variations in the charm mass between
1.2 and 1.8 GeV/$c^2$, the results are: 
\begin{eqnarray}
|V_{cd}|^2B_c &=&\left( 4.10\pm 0.70\pm 0.16\right) \times 10^{-3}\quad 
\text{(CDHS),}  \nonumber \\
\kappa \frac{|V_{cs}|^2}{|V_{cd}|^2} &=&9.3\pm 1.6\pm 0.9\quad \text{(CDHS).}
\nonumber
\end{eqnarray}
where the first error is the total error except the $m_c$ variation which is
given as the second error. Using the same CKM matrix elements as above, $%
|V_{cd}|=0.221\pm 0.003$ and $|V_{cs}|=0.9743\pm 0.0008$ \cite{PDG}, these
values can be converted to the measurements of $B_c$ and $\kappa $ shown in
Table \ref{tab:ccfr}.

\subsubsection{Discussion of Results}

Parton distributions are defined to a given order and scheme in QCD.
Therefore, the magnitude of a given parton distribution differs between
leading-order and next-to-leading-order. The parameters $\alpha $ and $m_c$
show shifts between LO and NLO, but $\kappa $ and $B_c$ are similar. For the
CCFR NLO analysis, the nucleon strange quark content is found to be $\kappa
=0.477\pm 0.051$, indicating that the sea is not SU(3) symmetric. This is
qualitatively the same result as from the LO analysis of CCFR and CDHS.

\begin{figure}[pthb]
\psfig{file=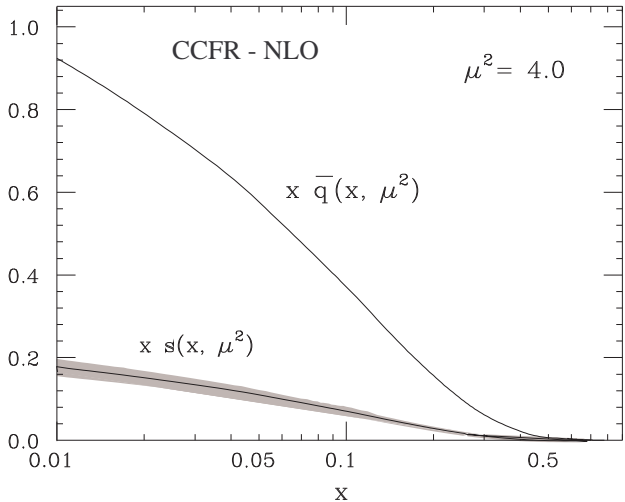,width=3.5in}
\caption{The quark and strange sea distributions, $x\overline{q}(x,\mu^2)$ and 
$xs(x,\mu^2),$ at $\mu^2=4\,$GeV$^2$ determined from the 
CCFR next-to-leading order 
analysis.  The band around the NLO strange sea distribution 
indicates the $\pm 1\sigma $ 
uncertainty in the distribution.}
\label{fig:qsbar}  
\end{figure}

Since a nonzero value of $\alpha $ would indicate a shape difference between 
$x\overline{q}(x)$ and $xs(x)$, the CCFR NLO value $\alpha
=-0.02\;_{-\;0.60}^{+\;0.66}$ indicates no shape difference at NLO. At
leading order, CCFR finds the strange quarks softer than the overall quark
sea. This comes about in part because the CCFR LO analysis includes a
longitudinal component, $R_L(\xi ,Q^2)$ , in both the inclusive
charged-current and charm production differential cross section as shown in
Eq. \ref{eq:RlLO}. This procedure is an approximation and does not correctly
treat the differences in the longitudinal component for heavy and light
quark production. On the other hand, the NLO formalism does include these
mass effects correctly and should give reliable measurements of the $x%
\overline{q}(x)$ and $xs(x)$ seas. Fig. \ref{fig:qsbar} compares the results
for the quark and strange seas determined by the NLO CCFR analysis which
shows that the magnitude is different but the shape is the same.

The charm quark mass parameter from the CCFR NLO fit is $1.70\pm 0.19$ GeV
which differs from the leading-order result. As shown in Section \ref
{sec:dimurates}, the LO analysis with the ``slow-rescaling'' correction for
the heavy charm quark is able to parameterize the threshold behavior of the
measured process. On the other hand, the NLO value of $m_c$, while also
describing the threshold behavior well, correctly includes the kinematic
effects associated with heavy quark production and provides a better
comparison with measurements derived from other processes involving similar
higher-order perturbative QCD calculations. For example, the
photon-gluon-fusion analysis of photoproduction data of  \citeasnoun{photo}
finds $m_c=1.74\,_{-\,0.18}^{+\,0.13}$ $GeV$ which is in agreement with the
NLO neutrino value.

\begin{figure}[]
\psfig{file=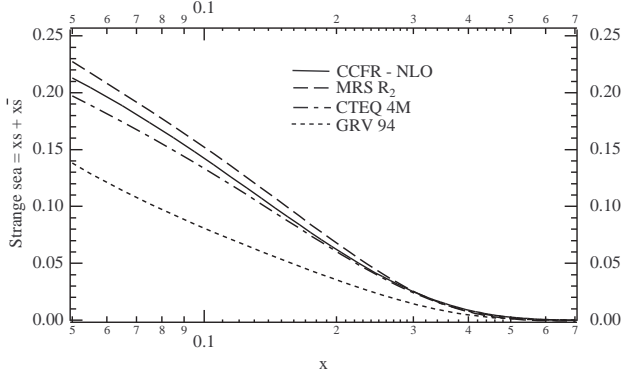,width=3.5in}
\caption{The CCFR NLO strange sea distributions compared to CTEQ 4M, MRS R$_2$,
and GRV 94 at $\mu^2 = 5$ GeV$^2$.} 
\label{strangenew}   
\end{figure}

Gl\"{u}ck, Kretzer, and Reya \cite{GKReya} have erroneously claimed that the
acceptance correction was applied incorrectly in the CCFR NLO analysis \cite
{Bazarko}. In the analysis procedure of  \citeasnoun{AOT2}, the cross section
is divided into three terms: LO, NLO, and subtraction. The CCFR analysis
consistently corrects for all experimental acceptance effects using the
proper kinematics for each term. For the LO and subtraction term, the
underlying distribution corresponds to the appropriate $W^{+}s\rightarrow c$
kinematics. The acceptance for the NLO term uses the $W^{+}g\rightarrow c%
\overline{s}$ kinematics with the calculated matrix element from %
 \citeasnoun{AOT2}. 
Gl\"{u}ck, Kretzer, and Reya also claim that the NLO to LO difference is not
supported by the currently available parton distribution functions \cite
{cteq4m,GRV94,mrsr2}. As stated above, the CCFR LO analysis includes higher
order longitudinal cross section components that make their parton
distributions different from strict LO analyses. The most recent and
complete CCFR NLO analysis \cite{Seligman} gives parton distributions that
are consistent with those from the various global fits as shown in Fig. \ref
{PDF plots}. It is interesting to note that the most recent GRV 94 \cite
{GRV94} strange sea distributions are much smaller than the CCFR NLO
measurements and the other global fit results as shown in Fig. \ref
{strangenew}. Publication in the future of acceptance and smearing corrected
neutrino charm production cross sections or structure functions by the CCFR
group will allow independent analyses and global fits using the different
NLO calculations.

The CCFR collaboration has also investigated the dependence of their NLO
results on factorization and renormalization scale uncertainties. Their
standard results quoted in Table \ref{tab:ccfr} use $\mu =2p_{\perp }^{{\rm %
max}}$ for both scales and find the scale uncertainty by varying $\mu $
between $p_{\perp }^{{\rm max}}$ and $3p_{\perp }^{{\rm max}}$. Fit results
with various choices of the common factorization and renormalization scale
are presented in Table \ref{tab:scale}. These results indicate that the data
favor $\mu ^2$ scales with smaller magnitudes but that the values of the fit
parameters are fairly insensitive to the choice of scale.

{\renewcommand{\arraystretch}{1.5} 
\begin{table}[tbp] \centering%
\begin{tabular}{|c|c|c|c|c|c|}
\hline
choice of scale, $\mu ^2$ & $\chi ^2$ & $\kappa $ & $\alpha $ & $B_c$ & $m_c$
$({\rm GeV})$ \\ \hline\hline
$(\,p_{\perp }^{{\rm max}})^2$ & 50.4 & 0.513 & $0.18$ & 0.0987 & 1.71 \\ 
\hline
$(2\,p_{\perp }^{{\rm max}})^2$ & 52.2 & 0.477 & $-0.02$ & 0.1091 & 1.70 \\ 
\hline
$(3\,p_{\perp }^{{\rm max}})^2$ & 54.4 & 0.460 & $-0.10$ & 0.1142 & 1.68 \\ 
\hline\hline
$Q^2$ & 51.7 & 0.423 & $-0.37$ & 0.1074 & 1.80 \\ \hline
$(2Q)^2$ & 56.1 & 0.410 & $-0.46$ & 0.1159 & 1.71 \\ \hline
$(3Q)^2$ & 59.4 & 0.408 & $-0.54$ & 0.1206 & 1.73 \\ \hline\hline
$Q^2+m_c^2$ & 52.5 & 0.421 & $-0.03$ & 0.1066 & 1.65 \\ \hline
$4\,(Q^2+m_c^2)$ & 57.1 & 0.409 & $-0.16$ & 0.1154 & 1.64 \\ \hline\hline
$Q^2+(2m_c)^2$ & 52.8 & 0.428 & $0.00$ & 0.1068 & 1.62 \\ \hline
$4\,[Q^2+(2m_c)^2]$ & 57.3 & 0.415 & $-0.15$ & 0.1161 & 1.63 \\ \hline
\end{tabular}
\caption{Central values of the CCFR fit parameters for various choices of the QCD 
scale $\mu^2$. Each fit contains 65 degrees of freedom. \label{tab:scale}} 
\end{table}%
}

\subsubsection{Comparisons to other strange sea measurements}

\begin{figure}[pthb]
\psfig{file=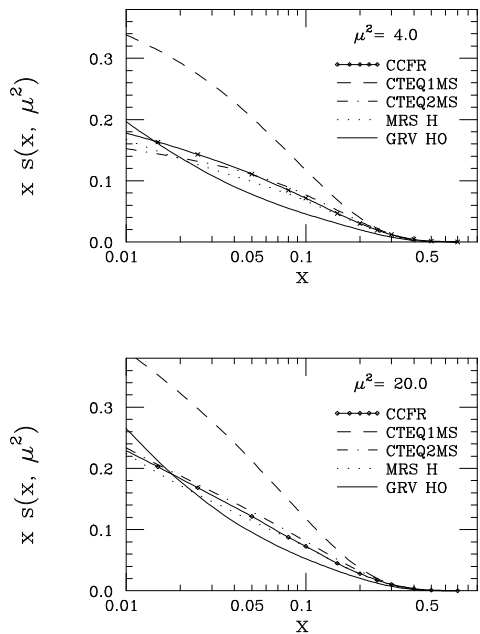,bbllx=75bp,bblly=144bp,bburx=550bp,bbury=650bp,clip=,width=3.5in}
\caption{Strange quark distributions $x\,s(x,\mu^2)$ from the 
CCFR experiment and previous  
global fits by CTEQ, MRS and GRV for $\mu^2 = 4$ and $20$ GeV$^2$. }
\label{fig:strangesea}   
\end{figure}

The comparison of neutrino and charged-lepton measurements of the $F_2$
structure function (See Section \ref{numucomp}) can also be used to study
the magnitude of the strange sea. In the DIS renormalization scheme, the
strange sea is related to the two $F_2$ measurements by 
\begin{equation}
\frac 12\left( x\,s(x,Q^2)+x\,\overline{s}(x,Q^2)\right) \simeq \frac 56%
F_2^{\nu N}\left( x,Q^2\right) -3\,F_2^{\mu N}\left( x,Q^2\right)
\label{Eq:f2nu-mu}
\end{equation}
The CTEQ collaboration in their CTEQ-1MS parton distributions \cite{bigsea}
tried increasing the strange sea at low $x$ to be consistent with Eq. \ref
{Eq:f2nu-mu}. The strange sea from this procedure is much larger than the
direct CCFR NLO measurements \cite{Bazarko} and other parton distributions%
 \cite{bigsea,mrs,grvho} as shown in Fig. \ref{fig:strangesea}. It therefore
seems unlikely that a larger strange sea is the explanation for the
neutrino/charged-lepton difference at low $x$. The same conclusion has been
put forward by  \citeasnoun{GKReya} who state that if the CCFR  \cite{Seligman}
and NMC \cite{F2NMC} data are correct, this discrepancy ``will constitute a
major problem which cannot be solved within our present understanding of the
so far successful perturbative QCD.''

\subsubsection{Tests of {\it $xs(x)\neq x\overline{s}(x)$}}

In principle, the momentum distributions of {\it $s$} and {\it $\overline{s}$%
} quarks need not be the same as long as the strangeness content of the
nucleon is constrained to have equal numbers of {\it $s$} and {\it $%
\overline{s}$} quarks. Theoretical work has explored the possibility that
the nucleon contains a sizable heavy quark component at moderate $x$---the
possibility of so-called intrinsic heavy quark states within the nucleon 
 \cite{brodsky,brodskyma}. Postulating intrinsic strange particle states,
such as a $K^{+}\Lambda $ component of the proton wave function, leads to
the prediction that the $s$ quark momentum distribution will be harder than
the $\overline{s}$ quark distribution  \cite{burkardt}.

The CCFR collaboration has explored this possibility by performing an
analysis \cite{Bazarko} in which the momentum distributions of the $s$ and $%
\overline{s}$ quarks are allowed to be different. For this study the sea
quark distributions are parameterized by: 
\begin{eqnarray}
x\overline{q}(x,\mu ^2) &=&2\left( {\frac{x\overline{u}(x,\mu ^2)+x\overline{%
d}(x,\mu ^2)}2}\right) +{\frac{xs(x,\mu ^2)+x\overline{s}(x,\mu ^2)}2}, 
\nonumber \\
xs(x,\mu ^2) &=&A_s(1-x)^\alpha \left[ {\frac{x\overline{u}(x,\mu ^2)+x%
\overline{d}(x,\mu ^2)}2}\right] ,  \nonumber \\
x\overline{s}(x,\mu ^2) &=&A_s^{\prime }(1-x)^{\alpha ^{\prime }}\left[ {%
\frac{x\overline{u}(x,\mu ^2)+x\overline{d}(x,\mu ^2)}2}\right] .
\end{eqnarray}
With the $s$ and $\overline{s}$ distributions constrained to have the same
number, $\int_0^1s(x,\mu ^2)\,dx=$ $\int_0^1\overline{s}(x,\mu ^2)\,dx$, $%
A_s $ and $A_s^{\prime }$ are defined in terms of $\kappa $, $\alpha $ and $%
\alpha ^{\prime }$. In the CCFR analysis, four parameters, $\kappa $, $%
\alpha $, $\Delta \alpha =\alpha -\alpha ^{\prime }$ and the charm quark
mass $m_c$, are determined from the data with the charm hadron branching
ratio fixed to the value obtained from other measurements, $B_c^{{\rm I}%
}=0.099\pm 0.012$ (see Section \ref{sec:Vcd}). The results are: 
\begin{eqnarray}
\kappa &=&0.536\pm 0.030\pm 0.037\;_{+\;0.098}^{-\;0.064},  \nonumber \\
\alpha &=&-0.78\pm 0.40\pm 0.75\pm 0.98,  \nonumber \\
\Delta \alpha &=&-0.46\pm 0.42\pm 0.40\pm 0.65,  \nonumber \\
m_c &=&1.66\pm 0.16\pm 0.07\;_{-\;0.01}^{+\;0.04}\;\text{GeV/c}^2\text{,}
\end{eqnarray}
where the first error is statistical, the second is systematic, and the
third is due to the uncertainty in $B_c^{{\rm I}}$. The value of $\Delta
\alpha =-0.46\pm 0.87$ indicates that the momentum distributions of $s$ and $%
\overline{s}$ are consistent, and the difference in the two distributions is
limited to $-1.9<\Delta \alpha <1.0$ at the 90\% confidence level. %
 \citeasnoun{brodskyma} have compared their intrinsic strangeness model to
these CCFR fits. The simple application of the model in their paper uses a
scaling of the $d$ valence quark distributions to do the QCD evolution of
the intrinsic strange quark sea. The results of the calculation including a
30\% extrinsic strange quark component are inconsistent with the CCFR
analysis as shown in Fig. 4 of the  \citeasnoun{brodskyma} paper. From the
figure, the predicted ratio is $x\overline{s}(x)/xs(x)=1.7$ at $x=0.07$ and
falls to $1$ at $x=0.22$. This difference corresponds to a $\Delta \alpha
\approx -3$ which is inconsistent with the CCFR limit. 

The CCFR fits assume a simple power law relation of the strange to
non-strange sea and may not have sufficient flexibility to encompass an
intrinsic strange quark contribution at high $x$. An additional high $x$
component would show up in the CCFR analysis as a modification of $B_c$ (and 
$|V_{cd}|^2$) for the neutrino charm production and as an unexplained excess
at high $x$ for the antineutrino production. The agreement of the
antineutrino data (Fig. \ref{fig:xvis}) and the agreement of the fit $%
B_c=0.1091\pm 0.0097$ with $B_c^{\text{I}}=0.099\pm 0.012$ (see Section \ref
{sec:Vcd}) extracted from other measurements limits the size of any
intrinsic contribution to below 20\%. Further analyses will be needed to
directly test the intrinsic strangeness models more precisely. Since the
intrinsic component mainly contributes at large $x$, it is unlikely that an 
{\it $s$} / {\it $\overline{s}$} asymmetry could explain the discrepancy
between the charged-lepton and neutrino measurements of $F_2$ in the small-$x
$ region.

\subsubsection{Measurements of {\bf $|V_{cd}|$}}

\label{sec:Vcd} If the CKM matrix elements are not assumed, then the CCFR
NLO results \cite{Bazarko} can be interpreted in terms of $\alpha $, $m_c$
and the following products : 
\begin{eqnarray}
|V_{cd}|^2B_c &=&(5.34\;\pm 0.39\;\pm 0.24\;_{-\;0.51}^{+\;0.25})\times
10^{-3}\quad \text{(CCFR - NLO),}  \nonumber \\
{\frac \kappa {\kappa +2}}|V_{cs}|^2B_c &=&(2.00\pm 0.10\;\pm
0.06\;_{-\;0.14}^{+\;0.06})\times 10^{-2},  \label{ccfr2mu}
\end{eqnarray}
where the first error is statistical, the second systematic, and third from
the QCD scale uncertainty.

As shown in Table \ref{tab:ccfr}, there is little difference in the
parameters if the analysis is performed to leading order or next-to-leading
order. Accordingly, we may combine the CCFR result with the LO result from
the CDHS collaboration \cite{CDHS:2mu}, 
\[
|V_{cd}|^2B_c=\left( 4.1\pm 0.7\,_{-0.39}^{+0.19}\right) \times 10^{-3}\quad 
\text{(CDHS - LO)}, 
\]
where the first error is the total experimental error and the second is the
QCD scale error which is not given by the original analysis but assumed to
be the same as in the CCFR measurement. Combining the two results, assuming
that all of the experimental errors are uncorrelated, yields 
\begin{equation}
|V_{cd}|^2B_c=\left( 4.92\pm 0.52\right) \times 10^{-3}\quad \text{%
(CCFR/CDHS).}  \label{vcdcomb}
\end{equation}

These combinations can be used to extract $|V_{cd}|^2$ and $\kappa
|V_{cs}|^2 $ when $B_c$ is determined from other data. $B_c$ is determined
by combining the charmed particle semileptonic branching ratios measured at $%
e^{+}e^{-}$ colliders  \cite{PDG} with the neutrino-production fractions
measured by the Fermilab E531 neutrino-emulsion experiment  \cite{E531}.
Using an $E_{\text{vis}}>30$ GeV cut, E531 determined the following
production fractions: $52\pm 6$\% $D^0$, $42\pm 6$\% $D^{+}$, $1\pm 2$\% $%
D_s^{+}$, and $5\pm 3$\% $\Lambda _c^{+}$. In the E531 analysis, events that
could not be unambiguously identified as $D^{+}$ or $D_s^{+}$ were all
categorized as $D^{+}$ events. To remove this small bias, a re-analysis was
performed that included updated values of the charmed hadron lifetimes  \cite
{Bolton:CKMa,Bolton:CKMb}. This re-analysis finds the following production
fractions with an $E_{\text{vis}}>30$ GeV cut: $60\pm 6$\% $D^0$, $26\pm 6$%
\% $D^{+}$, $7\pm 5$\% $D_s^{+}$, and $7\pm 4$\% $\Lambda _c^{+}$. These
production fractions are consistent with those measured by $e^{+}e^{-}$
experiments  \cite{cleo}. With these particle fractions, $B_c^{\text{I}%
}=0.099\pm 0.012$ \cite{Bolton:CKMa,Bolton:CKMb} and when encorporated with
the combined measurement of Eq. \ref{vcdcomb} gives the value of the CKM
matrix element 
\[
|V_{cd}|=0.220\;_{-\;0.021}^{+\;0.018} \quad \text{ (CCFR/CDHS),} 
\]
where the error indicates all sources of uncertainty including the $\mu ^2$
scale uncertainty. This value compares well with the PDG value, $%
|V_{cd}|=0.221\pm 0.003$, which is determined from measurements of the other
matrix elements and the unitarity constraint on the CKM matrix assuming
three generations. The errors on the direct $|V_{cd}|$ measurement from $\nu 
$ charm production are currently at the $\pm 9\%$ level which precludes a
precise test of the CKM unitarity.

The CKM parameter $|V_{cs}|$ requires an independent determination of the
strange sea fraction, $\kappa $, which is currently unavailable. Using the
value of $B_c^{\text{I}}$ given above with the result of Eq. \ref{ccfr2mu}
and making the conservative assumption that $\kappa \leq 1$ implies that 
\[
|V_{cs}|>0.69\;\text{at 90\% C.L. \quad (CCFR). }
\]
In the future, $\kappa $ could be determined from a measurement of the $xF_3$
difference between neutrinos and antineutrinos, $xF_3^\nu (x,Q^2)-xF_3^{%
\overline{\nu }}(x,Q^2)=4x\left[ s(x,Q^2)-c(x,Q^2)\right] $.  
It should be possible with this technique to measure $\kappa $ to $20\%$
which would allow the quantity $|V_{cd}|/|V_{cs}|$ to be determined to $10\%$
from the dimuon measurements.

\subsection{Charm Sea Measurements from WSM Production}

WSM production can be used to probe the charm content in the nucleon through
the neutral current scattering of a $\nu _\mu $ off a charm quark which
subsequently decays into a $\mu ^{+}$ . A charm component is expected in QCD
through gluon splitting into $c\overline{c}$ pairs. In a typical QCD
calculation, the charm sea is set to zero for $Q^2<Q_0^2\approx 4$ GeV$^2$
and is then allowed to evolve via the DGLAP equations. In the $\nu _\mu $
interaction, NC scattering off a charm quark is suppressed due to the
kinematic effects related to producing two heavy charm quarks $(c\overline{c}%
)$ in the final state. 
\begin{eqnarray}
\nu _\mu +c\;(+\;\overline{c}\,)\longrightarrow \;\nu _\mu \!\!+\,\!
&c&\,(+\;\overline{c}\,) \\
&\hookrightarrow &s\;+\;\mu ^{+}\;+\;\nu _\mu  \nonumber
\end{eqnarray}
For a mean energy of 100 GeV, this suppression introduces a factor of $\sim
0.3$ relative to massless quark production. In addition, to go from the
charm production rate to the WSM rate, one must incorporate the
semi-leptonic branching ration for the final state charmed particle ($\sim
10\%$) and the acceptance for the muon to be measured in the detector
(typically $\sim 20\%$). The size of the cross section for neutral current
scattering off charm quarks in the nucleon can be estimated using the charm
sea calculations. Using the NLO parton distributions of the CTEQ
collaboration  \cite{bigsea}, the cross section for neutral current charm
scattering relative to the total neutrino CC scattering cross section is a
factor of $0.005$ lower. Combining all these factors, one would expect a WSM
rate which was approximately $3\times 10^{-5}$ of the normal CC rate.

The CCFR collaboration has made a study \cite{CCFR-WSM} of WSM production
using the data from their narrow-band E616/E701 experiment. The dominant
background comes from the $\overline{\nu }_\mu $ contamination in the beam.
CCFR minimizes this background by demanding the observed energy in the WSM
events to have $E_{vis}>100$ GeV which reduces the background rate to $%
2.3\times 10^{-4}$. Other background sources ($\nu _e$ induced dilepton
production, mis-identified dimuon events, and NC interactions with a $\pi /K$
decay in the hadron shower) contribute an additional background at the $%
1.5\times 10^{-4}$ level. For $E_{vis}>100$ GeV, the CCFR experiment
observes $43.0 \pm 6.6$ WSM events with a calculated background of $26.9\pm
5.1$ events leading to a $2\sigma $ excess of $16.1\pm 8.3$ events or a rate
of $\left( 2.3\pm 1.2\right) \times 10^{-4}$ . The CCFR collaboration
presents the excess as a $90\%$ confidence level (CL) upper limit on the WSM
rate of $4.3\times 10^{-4}$ which is much larger than the estimate of $%
1.4\times 10^{-5}$ for NC charm scattering with $E_{vis}>100 $ GeV.

\section{Electroweak Physics with Neutrino Beams}

\label{EW Section}

Lepton scattering experiments permit the study of electroweak interactions
at space-like momentum transfers of $10^{-2}\leq Q^2\leq 10^{+2}$ GeV$^2$ in
a fixed target environment with $Q^2\rightarrow 10^4$ GeV$^2$ at the HERA\ $%
ep$ collider possible in the future. As such they complement the large $%
\sqrt{s}$ time-like measurements now performed with extraordinary precision
at LEP, SLD, and the Tevatron. It is sometimes not appreciated that
electroweak measurements performed in fixed target lepton scattering
experiments can also still compete with the collider results in probing the
Standard Model.

Neutrino scattering, in particular, has contributed to our understanding of
electroweak physics over roughly four historical phases since the early
1970's: (I) confirmation of the existence of neutral currents through first
observation of $\nu _\mu e^{-}\rightarrow \nu _\mu e^{-}$ at CERN \cite{NC
Discovery}; (II) early measurements of $\sin ^2\theta _W$ that provided a
critical ingredient to the successful prediction of the masses of the $W$
and $Z$ bosons \cite{WZ Discovery}; (III) second generation weak mixing angle
determinations that probed one loop corrections to the Standard Model and
provided among the first useful upper and lower limits on the top mass; and,
finally, (IV) third generation experiments that seek to test the internal
consistency of electroweak theory through precise coupling constant
extrapolations that can be compared to other precision electroweak
measurements at colliders. In this Review we will summarize the current
experimental state of electroweak measurements with neutrino probes (period
III), and give short and intermediate term prospects for current and future
experiments (period IV).  
Detailed discussions of results obtain prior to direct observation of the $W$
and $Z$ may be obtained elsewhere  \cite{Old NC A,Old NC B}{}. The remainder
of this section divides into four parts: Theoretical Motivation,
Experimental Neutrino-Electron Scattering, Experimental Neutrino-Nucleon
scattering, and Summary and Outlook.

\subsection{Theoretical Motivation}

Neutrino-nucleon and neutrino-electron scattering depends on the left and
right handed $\nu -Z^0$ and $Z^0$ -target fermion couplings. Because
all of the parameters of the Standard Model 
except the Higgs mass are now well
known, the couplings measured in neutrino scattering are predicted to high
accuracy. Any significant deviation between measurement and prediction would
thus indicate new physics. Other experimental programs (LEP/SLD, LEP II,
CDF/D0) also probe for new physics through precision electroweak tests, and
neutrino experiments complement these high energy efforts along
several lines: 

\begin{itemize}
\item  By measuring different combinations of couplings than collider
experiments, in particular, couplings to light quarks.

\item  By measuring cross sections at moderate space-like momentum transfer,
as opposed to the large time-like scattering explored at colliders.

\item  By extracting cross sections with ``orthogonal'' analysis tools: for
example, different radiative correction packages and very limited
quark-gluon fragmentation model dependence.
\end{itemize}

Electroweak measurements are frequently summarized by quoting a value of the
weak mixing angle $\sin ^2\theta _W$; however, this parameter can be defined
in many different ways, e.g., 
\begin{equation}
1-M_W^2/M_Z^2\equiv \sin ^2\theta _W({\rm {on-shell})}  \label{on-shell}
\end{equation}
and 
\begin{equation}
A_{LR}(Z^0)\equiv \frac{\left( \frac 12-\sin ^2\theta _W({\rm {eff}))}%
\right) ^2-\sin ^4\theta _W({\rm {eff})}}{\left( \frac 12-\sin ^2\theta _W(%
{\rm {eff})}\right) ^2+\sin ^4\theta _W({\rm {eff})}},
\end{equation}
with the differing definitions appropriate for different experiments. A less
confusing way to compare the sensitivity of different experiments to
electroweak physics is to translate all measurements to an equivalent
determination of a common physical parameter, for which a convenient choice
is the $W$ boson mass\footnote{%
These translations use SM electroweak theory with the fine structure
constant $\alpha $, the Fermi constant $G_F$, the $Z^0$ mass $M_Z$, and the
top quark mass $M_{top}$, as input. Low energy electroweak observables can
be calculated to sufficient accuracy without knowing the Higgs boson mass $%
M_H$.}. It follows simply that a $\pm 0.002$ on-shell mixing angle
measurement in neutrino scattering (see Section \ref{sm corr} \ref{EW RadCor}%
) is equivalent to a $\pm 100$ MeV $W-$mass measurement, which compares well
with expected results from LEP II and CDF/D0.

\subsubsection{Basic Cross Sections}

Muon neutrino scattering cross sections off light fermion targets can be
expressed in tree-level as 
\begin{eqnarray}
\frac{d\sigma (\nu _\mu f\rightarrow \nu _\mu f)}{dy} &=&\frac{G_F^2s}\pi
\left( \ell _f^2+r_f^2(1-y)^2\right) \left( 1+\frac{sy}{M_Z^2}\right) ^{-2},
\\
\frac{d\sigma (\bar{\nu}_\mu f\rightarrow \bar{\nu}_\mu f)}{dy} &=&\frac{%
G_F^2s}\pi \left( \ell _f^2(1-y)^2+r_f^2\right) \left( 1+\frac{sy}{M_Z^2}%
\right) ^{-2},
\end{eqnarray}
where $f=e^{-},u,d,s,c$; $G_F=\left( 1.16639\pm 0.00002\right) \times
10^{-5} $GeV$^{-2}$ is the Fermi constant, $M_Z=91.187\pm 0.007$ GeV is the $%
Z^0$ mass; $s$ is the effective center of mass energy, which depends on $f$; 
$y$ is the inelasticity; and $\ell _f,r_f$ are left and right handed
coupling constants summarized in Table \ref{NC couplings}. At fixed target
energies, the propagator term $\left( 1+sy/M_Z^2\right) ^{-2}$ seldom
differs from unity by more than one percent for quark targets and is
entirely negligible for electron targets. Although all experiments apply
propagator corrections, this factor will be omitted for brevity in
subsequent formulas.

\begin{table}[tbp] \centering%
\begin{tabular}{|c|c|c|}
\hline
$f$ & $\ell_f$ & $r_f$ \\ \hline
$e^{-}$ & $-\frac 12+\sin ^2\theta _W$ & $\sin ^2\theta _W$ \\ 
$u,c$ & $\frac 12-\frac 23\sin ^2\theta _W$ & $-\frac 23\sin ^2\theta _W$ \\ 
$d,s$ & $-\frac 12+\frac 13\sin ^2\theta _W$ & $\frac 13\sin ^2\theta _W$ \\ 
\hline
\end{tabular}
\caption{Neutral current standard model couplings
for electron and quark targets.\label{NC couplings}}%
\end{table}%

Electron-neutrino-electron ($\nu _ee^{-}$ and $\bar{\nu}_ee^{-}$) scattering
cross sections contain additional charged-current and interference terms.
These processes have been observed  \cite{nue-e A,nue-e B,nue-e C,nue-e
D,nue-e E,nue-e F,nue-e G}; and experimental results have provided important
qualitative confirmation of the SM, for example, the elimination of the
four-fold sign ambiguity in axial and vector coupling constants of the
electron. However,  limited statistics preclude precision electroweak tests.
Similar comments currently apply to very high $Q^2$ NC scattering
measurements at HERA  \cite{ZEUS NC-I,ZEUS NC-II,H1 NC}; although there is
considerable hope in this case for future improvement \citeaffixed{Future
HERA}{See for example,}. Accordingly, we will restrict our discussion to
scattering experiments using $\nu _\mu $ and $\bar{\nu}_\mu $ beams.

\subsubsection{Neutrino Electron Scattering}

For electrons, $s_e=2m_eE_\nu $ with $m_e$ the electron mass and $E_\nu $
the neutrino energy. The total NC scattering cross sections then follow as 
\begin{eqnarray}
\frac{d\sigma (\nu _\mu e^{-}\rightarrow \nu _\mu e^{-})}{dy} &=&\sigma
_0^{\nu e}E_\nu \left( \ell _e^2+r_e^2(1-y)^2\right) , \\
\frac{d\sigma (\bar{\nu}_\mu e^{-}\rightarrow \bar{\nu}_\mu e^{-})}{dy}
&=&\sigma _0^{\nu e}E_\nu \left( \ell _e^2(1-y)^2+r_e^2\right) ,
\end{eqnarray}
with $\sigma _0^{\nu e}=\frac{2G_F^2m_e}\pi =17.2\times 10^{-42}$ cm$^2/$%
GeV. The electron mass sets the scale for the very small cross section and
also drives the kinematics. With a maximum momentum transfer $Q_{\max
}^2=2m_eE_\nu \simeq 0.001E_\nu $ GeV$^2$, with $E_\nu $ in GeV, the
electron always scatters nearly parallel to the neutrino beam.

The charged current process $\nu _\mu e^{-}\rightarrow \nu _e\mu ^{-}$ ,
sometimes called inverse muon decay, represents an electroweak test in its
own right. The tree level cross section is 
\begin{equation}
\frac{d\sigma (\nu _\mu e^{-}\rightarrow \nu _e\mu ^{-})}{dy}=\sigma _0^{\nu
e}E_\nu \left( 1-\frac{m_\mu ^2}{2m_eE_\nu }\right) ^2,  \label{IMD XC}
\end{equation}
with the kinematic threshold occurring at $E_\nu \cong 11$ GeV. The small
center-of-mass energy forces the muon to be emitted in the forward direction.

Muon decay measurements do not measure final state neutrino helicities and
thus cannot distinguish between vector left-handed couplings, $g_{LL}^V$,
and scalar couplings $g_{LL}^S$; the inverse muon decay cross section is
proportional to $\left| g_{LL}^V\right| ^2$, and thus resolves this
ambiguity. Muon decay {\it and} inverse muon decay are required to establish
rigorously the $V-A$ character of the charged current.

Electroweak radiative corrections to $\nu _\mu e^{-}$ and $\bar{\nu}_\mu
e^{-}$ NC \cite{Okun} and CC \cite{Bardin nu-e} scattering have been
calculated. An apparently coincidental cancellation between the two dominant
radiative corrections in NC\ scattering causes the effective weak mixing
angle $\sin ^2\theta _W^{\nu e}$ to be very close to $\sin ^2\theta _W($eff$%
) $ extracted from asymmetry measurements at LEP/SLD.

\subsubsection{Neutrino-Nucleon Scattering}

For quarks, the effective center-of-mass energy is $s_q=2ME_\nu \xi $, with $%
M$ the nucleon mass and $\xi $ interpretable as the fraction of the
nucleon's 4-momentum carried by struck quark. An experiment can only
measure, in practice, the average over $\xi $ of the sum over all quark
targets inside the nucleon (and in fact, these distributions are further
averaged over incident neutrino energy distribution): 
\begin{eqnarray}
\frac{d\sigma (\nu _\mu p\rightarrow \nu _\mu X)}{dy} &=&\frac{2G_F^2ME_\nu }%
\pi \int_0^1d\xi \sum_{q=u,d,s,c}\left\{ \left[ \left( \ell
_f^2+r_f^2(1-y)^2\right) \xi q(\xi )\right. \right.  \nonumber \\
&&\left. +\left( \ell _f^2(1-y)^2+r_f^2\right) \xi \bar{q}(\xi )\right] , \\
\frac{d\sigma (\bar{\nu}_\mu p\rightarrow \bar{\nu}_\mu X)}{dy} &=&\frac{%
2G_F^2ME_\nu }\pi \int_0^1d\xi \sum_{q=u,d,s,c}\left\{ \left[ \left( \ell
_f^2(1-y)^2+r_f^2\right) \xi q(\xi )\right. \right.  \nonumber \\
&&\left. +\left( \ell _f^2+r_f^2(1-y)^2\right) \xi \bar{q}(\xi )\right] .
\end{eqnarray}
Neutron cross sections follow from isospin arguments with the substitutions $%
u(\bar{u})\rightarrow d(\bar{d})$, $d(\bar{d})\rightarrow u(\bar{u})$, $s(%
\bar{s})\rightarrow s(\bar{s})$, and $c(\bar{c})\rightarrow c(\bar{c})$ (but
see Section \ref{sm corr} \ref{isovector}). Contributions from $b$ and $t$
quarks are assumed to be negligible.

For isoscalar targets, omitting second generation quark contributions for
the moment,

\begin{eqnarray}
\frac{d\sigma (\nu _\mu N\rightarrow \nu _\mu X)}{dy} &=&\frac{G_F^2ME_\nu }%
\pi \int_0^1d\xi \left[ \left( \ell _u^2+\ell _d^2+\left( r_u^2+r_d^2\right)
(1-y)^2\right) \left( \xi u(\xi )+\xi d(\xi )\right) \right.   \nonumber \\
&&+\left. \left( \left( \ell _u^2+\ell _d^2\right)
(1-y)^2+r_u^2+r_d^2\right) \left( \xi \bar{u}(\xi )+\xi \bar{d}(\xi )\right)
\right] , \\
\frac{d\sigma (\bar{\nu}_\mu N\rightarrow \bar{\nu}_\mu X)}{dy} &=&\frac{%
G_F^2ME_\nu }\pi \int_0^1d\xi \left[ \left( \left( \ell _u^2+\ell
_d^2\right) (1-y)^2+r_u^2+r_d^2\right) \left( \xi u(\xi )+\xi d(\xi )\right)
\right.   \nonumber \\
&&+\left. \left( \ell _u^2+\ell _d^2+\left( r_u^2+r_d^2\right)
(1-y)^2\right) \left( \xi \bar{u}(\xi )+\xi \bar{d}(\xi )\right) \right] .
\end{eqnarray}
At this level, one notes that NC cross sections can be re-expressed as 
\begin{eqnarray}
\frac{d\sigma (\nu _\mu N\rightarrow \nu _\mu X)}{dy} &=&\left( \ell
_u^2+\ell _d^2\right) \frac{d\sigma (\nu _\mu N\rightarrow \mu ^{-}X)}{dy} \\
&&+\left( r_u^2+r_d^2\right) \frac{d\sigma (\bar{\nu}_\mu N\rightarrow \mu
^{+}X)}{dy},  \nonumber \\
\frac{d\sigma (\bar{\nu}_\mu N\rightarrow \nu _\mu X)}{dy} &=&\left(
r_u^2+r_d^2\right) \frac{d\sigma (\nu _\mu N\rightarrow \mu ^{-}X)}{dy} \\
&&+\left( \ell _u^2+\ell _d^2\right) \frac{d\sigma (\bar{\nu}_\mu
N\rightarrow \mu ^{+}X)}{dy},  \nonumber
\end{eqnarray}
a result that follows more generally from isospin conservation \cite
{Llewellyn Smith}. When integrated over $y$, the above imply the Llewellyn
Smith (LS) relations 
\begin{eqnarray}
R^\nu  &\equiv &\frac{\int dy\frac{d\sigma (\nu _\mu N\rightarrow \nu _\mu X)%
}{dy}}{\int dy\frac{d\sigma (\nu _\mu N\rightarrow \mu ^{-}X)}{dy}} \\
&=&\left( \ell _u^2+\ell _d^2\right) +r\left( r_u^2+r_d^2\right) , \\
&=&\frac 12-\sin ^2\theta _W+(1+r)\sin ^4\theta _W; \\
R^{\bar{\nu}} &\equiv &\frac{\int dy\frac{d\sigma (\bar{\nu}_\mu
N\rightarrow \bar{\nu}_\mu X)}{dy}}{\int dy\frac{d\sigma (\bar{\nu}_\mu
N\rightarrow \mu ^{+}X)}{dy}} \\
&=&\left( \ell _u^2+\ell _d^2\right) +r^{-1}\left( r_u^2+r_d^2\right) , \\
&=&\frac 12-\sin ^2\theta _W+(1+r^{-1})\sin ^4\theta _W;
\end{eqnarray}
with 
\begin{equation}
r\equiv \frac{\int dy\frac{d\sigma (\bar{\nu}_\mu N\rightarrow \mu ^{+}X)}{dy%
}}{\int dy\frac{d\sigma (\nu _\mu N\rightarrow \mu ^{-}X)}{dy}}
\end{equation}
the $\nu _\mu $ to $\bar{\nu}_\mu $ CC charged current ratio. Because the LS
relationships also hold at the differential cross section level, $r$ can be
defined to incorporate experimental hadron energy acceptance. For an ideal
full acceptance experiment, $r\simeq 0.5$, whereas typical experimental cuts
reduce this value to $r\simeq 0.35-0.45$. The LS\ formula is valid for an
idealized isospin zero target composed only of first generation quarks.
Corrections must be applied in real experiments to take into account the
effects of non-isoscalar targets and, especially, heavy quarks involved in
the scattering.

The left handed coupling can be isolated through a linear combination of $%
R^\nu $ and $R^{\bar{\nu}}$: 
\begin{eqnarray}
R^{-} &=&\frac{R^\nu -rR^{\bar{\nu}}}{1-r}, \\
&=&\frac{\frac{d\sigma (\nu _\mu N\rightarrow \nu _\mu X)}{dy}-\frac{d\sigma
(\bar{\nu}_\mu N\rightarrow \bar{\nu}_\mu X)}{dy}}{\frac{d\sigma (\nu _\mu
N\rightarrow \mu ^{-}X)}{dy}-\frac{d\sigma (\bar{\nu}_\mu N\rightarrow \mu
^{+}X)}{dy}} \\
&=&\left( \ell _u^2+\ell _d^2\right) \\
&=&\frac 12-\sin ^2\theta _W.
\end{eqnarray}
This final line expresses the Paschos-Wolfenstein(PW) relationship \cite
{Paschos-Wolfenstein}. It retains its accuracy even if heavy quark
contributions to the NC and CC\ cross section are included, provided that
these contributions are the same for $\nu _\mu $ and $\bar{\nu}_\mu $. This
latter feature makes this quantity the most superior of all electroweak
observables in the neutrino sector.

\subsubsection{SM Corrections to $\nu _\mu (\bar{\nu}_\mu )N$ Cross Sections}

\label{sm corr}

Several corrections must be applied before SM couplings can be extracted
from the LS or PW relations or their variants.

\paragraph{\label{EW RadCor}SM Electroweak Radiative Corrections}

Radiative corrections approximately factor into two parts: QED  final state
radiation \cite{DeRujula}, for which final state muon bremsstrahlung diagrams
are most important, and purely weak corrections which effectively shift
coupling constants \cite{SirlinAndMarciano}. A complete treatment that
combines the two effects exists \cite{Bardin nu N}.

The effect of weak corrections can be represented schematically by the
following: 
\begin{eqnarray}
\ell _f &\rightarrow &(1+\Delta \rho )\ell _f, \\
r_f &\rightarrow &(1+\Delta \rho )r_f, \\
\sin ^2\theta _W &\rightarrow &(1+\Delta \kappa )\sin ^2\theta _W.
\end{eqnarray}
Calculation of $\Delta \rho $ and $\Delta \kappa $ require a precise
definition of $\sin ^2\theta _W$. While any definition is possible, it has
become conventional to use the on-shell or Sirlin definition (Eq. \ref
{on-shell}) to describe $\nu N$ scattering. With this convention, the
leading contributions to $\Delta \rho $ and $\Delta \kappa $ which arise
from the large top mass, 
\begin{eqnarray}
\Delta \kappa  &=&\frac{3G_F}{8\sqrt{2}\pi ^2}\cot {}^2\theta _WM_{top}^2+%
{\cal O}\ln \left( \frac{M_H^2}{M_W^2}\right) +...  \label{KappaRadCor} \\
\Delta \rho  &=&\frac{3G_F}{8\sqrt{2}\pi ^2}M_{top}^2+{\cal O}\ln \left( 
\frac{M_H^2}{M_W^2}\right) ...,  \label{RhoRadCor}
\end{eqnarray}
largely cancel \cite{SirlinAndMarciano} \cite{Stuart}; and a $\sin ^2\theta _W$
measurement from $\nu N$ scattering can be thought of as effective $W-$mass
measurement\footnote{%
Strictly speaking, this is true only if $\sin ^2\theta _W$ is extracted
using the LS formalism, and even then, only if experimental acceptance
effects do not move the effective value of $r$ too far from $r\simeq 0.4-0.5$%
. These restrictions have no consequence now that $M_{top}$ corrections can
be made.} 
\begin{equation}
\sin ^2\theta _W(\nu N)\simeq \sin ^2\theta _W({\rm {on-shell}).}
\end{equation}
Since the direct measurement of the top mass  \cite{Mtop-CDF,Mtop-D0}, the
on-shell definition is less motivated, although it remains convenient so
that the precision of neutrino measurements can be compared to that of
colliders.

QED radiative corrections are numerically large and must be handled with
care. Their dominant effect is to harden the observed $y$ distributions for $%
\nu _\mu N$ and $\bar{\nu}_\mu N$ CC\ cross sections relative to their
expected theoretical forms; good control of acceptance corrections minimizes
their effect.

\paragraph{Structure Functions and Higher Order QCD Corrections}

The use of NC to CC cross section ratios coupled with the approximate
validity of the LS and PW relations obviates the need to correct for light
quark structure function effects. Similarly, uncertainties in $\alpha
_S(Q^2) $ contribute negligible error in the extraction of electroweak
parameters.

\paragraph{Heavy Quark Production}

The most uncertain parts of $\nu N$ CC and NC cross section are due to
scattering from and production of heavy flavor, 
\begin{eqnarray}
\nu _\mu s &\rightarrow &\nu _\mu s, \\
\nu _\mu c &\rightarrow &\nu _\mu c, \\
\nu _\mu d &\rightarrow &\mu ^{-}c, \\
\nu _\mu s &\rightarrow &\mu ^{-}c, \\
\nu _\mu s &\rightarrow &\mu ^{-}u;
\end{eqnarray}
here $s$ is counted as ``heavy''. The phenomenology and experimental status
of neutrino induced heavy quark production is described in Sec. \ref{Dimuon
Section}. For the purpose of NC analyses, heavy flavor production introduces
uncertainties in electroweak analyses through two main effects:

\begin{itemize}
\item  The neutral current processes are not simply related to analogous
charged current cross sections through an LS type formula due to the role of
the charm mass, CKM matrix elements, and large flavor asymmetry ($s(\xi
)\neq c(\xi )$), 
\begin{eqnarray}
&&2\left[ \left( \ell _d^2+r_d^2\right) \xi s(\xi )+\left( \ell
_u^2+r_u^2\right) \xi c(\xi )\right] \left[ 1+\left( 1-y\right) ^2\right] \\
&\neq &R_\nu \left[ \left| V_{cd}\right| ^2\left( \xi u(\xi )+\xi d(\xi
)\right) \right.  \nonumber \\
&&\left. +2\left| V_{cs}\right| ^2\xi s(\xi )+2\left( 1-y\right) ^2\xi \bar{c%
}(\xi )\right] .  \nonumber
\end{eqnarray}

\item  The threshold behavior of charm production is not known well enough,
independent of the model used to describe it. In the naive leading order
``slow rescaling'' model where the threshold dependence in CC charm
production is described by a single effective charm mass $m_c^{eff}$, one
only knows $m_c^{eff}=1.31\pm 0.24$ GeV \cite{sar}, an 18\% error. Better
models of charm production exist, but these do not improve the experimental
uncertainty. Threshold effects in charm production persist to very high
neutrino energies because of the important role of sea quark mechanisms
which predominate at low $\xi $, where the effective center-of mass-energy $%
2ME_\nu \xi $ is not necessarily large compared to $\left( m_c^{eff}\right)
^2$.
\end{itemize}

\paragraph{\label{isovector}Isovector Effects}

Two kinds of isospin-violating effects are present; these can be thought of
as nuclear and nucleon isospin violating effects.

Heavy targets, such as iron, used in high statistics scattering experiments
have a small ($\sim 6\%$) neutron excess. This excess creates cross section
terms proportional to $u-\bar{u}+d-\bar{d}\equiv u_V-d_V$ and $\bar{u}-\bar{d%
}$, referred to as valence and sea contributions, respectively. The valence
contribution has only directly been measured in relatively low statistics $%
\nu p$ and $\nu D$ bubble chamber experiments \cite{BEBC-UvDv}; however,
global fits \citeaffixed{cteq4m,mrsr2}{see, for example,} to parton
distributions that compare high statistics $\nu N$ CC\ scattering to $ep$, $%
\mu p$, $eD$, and $\mu D$ electroproduction provide tight indirect
constraints. The sea quark contribution, now known not to vanish \cite{NMC
Gottfried}, is not well-constrained, but is numerically small.

Isospin violation exists even in an isoscalar target because electromagnetic
and quark mass effects break the equality between $u$(proton) and $d$%
(neutron) quark distributions. These effects produce corrections to $\sin
^2\theta _W$ of $\sim 0.002$ \cite{Sather}.

\paragraph{Longitudinal Cross Section}

The simple decomposition of $\nu N$ cross sections into terms proportional
to $1$ and $(1-y)^2$ is broken by the existence of a non-zero longitudinal
structure function $R_L(\xi ,Q^2)$. As in the case of other higher order QCD
effects, the longitudinal structure function modifies the $y$ distribution
but not the LS or PW\ relations. Influence on neutrino experiments arises
only because the $y$ distribution slightly affects CC/NC separation
experimentally. With sufficient care, these acceptance affects due to $%
R_L(\xi ,Q^2)$ can be applied with negligible uncertainty added to coupling
extractions.

\paragraph{Higher Twist Effects}

Higher twist contributions to neutrino cross sections fall as $1/Q^{2n}$,
with $n\geq 1$. This kinematic dependence alone severely suppresses their
effects on NC analyses at high energies. Further suppression arises from the
similarity of NC to CC\ cross section ratios from deep inelastic and higher
twist sources, provided scattering occurs from an isoscalar target (a
consequence of the LS relation). For example, purely elastic scattering, the
``highest twist'' of all, has a NC to CC cross section ratio of $\sim 0.35$
as compared to the DIS total of $\sim 0.31$.

In fact, higher twist effects are arguably absent in any experiment that is
in a regime where logarithmic QCD evolution of structure functions dominates
($Q^2> 1-10$ GeV$^2$, depending on $\xi $). In order to exhibit this
approximate scaling behavior, the cross section must consist of a very large
number of high twist terms; but, by duality arguments, many high twist terms
added together are equivalent to the simpler quark scattering picture.

Pumplin \citeyear{Pumplin} has argued to the contrary based on an application
of vector/axial-vector dominance. In his model, the $a_1-\rho $ mass
difference generates substantial and uncertain violations of the LS
relationship (but not the PW relationship, since vector dominance produces
equal $\nu _\mu $ and $\bar{\nu}_\mu $ cross sections). However, his model
fails to describe CC differential cross sections unless its parameters are
constrained to the level where effects on the NC to CC\ cross section ratios
are small.\footnote{%
Specifically, we have compared Pumplin's model to high precision charged
lepton scattering and have concluded, conservatively, that his parameter $S_0
$ is constrained to $S_0<2$ GeV$^2$ at 90\% confidence level. Pumplin's
original analysis allowed $S_0\rightarrow \infty $.}  Furthermore, from the
arguments above, it may simply be incorrect to decompose an inclusive cross
section displaying approximate scaling into a DIS contribution and a single
higher twist term. More theoretical study of this issue would be useful.

\subsubsection{New Physics Effects in $\nu _\mu (\bar{\nu}_\mu )e^{-}$ and $%
\nu _\mu (\bar{\nu}_\mu )N$ Cross Sections}

Direct effects of physics beyond the SM can appear in $\nu _\mu (\bar{\nu}%
_\mu )N$ scattering through higher order loop corrections (SM and non-SM
Higgs, fourth generation quarks), new propagator effects (heavy $Z^0$,
leptoquarks, compositeness), and lepton mixing (mirror fermions) \cite
{Langacker}. Somewhat surprisingly, the advantage enjoyed by neutrino
measurements in probing certain classes of new physics derives often from
the low energy scale of the process. For example, an unmixed new $Z^{\prime }
$ would produce linear shifts in couplings in $\nu _\mu N$ NC scattering
since both the $Z^0$ and $Z^{^{\prime }}$ are far off-shell. In contrast,
the effect of a heavy, unmixed $Z^{\prime }$ in $e^{+}e^{-}$ scattering on
the $Z^0$ pole would be to produce quadratic shifts in observed coupling
since the $Z^{\prime }$ would not interfere directly with the $Z^0$ at the
pole. An advantage is also enjoyed over direct searches at the Tevatron,
which have excellent sensitivity to constructive interference between $Z^0$
and $Z^{\prime }$, which increases lepton pair cross sections over
expectations, but relatively poor sensitivity to destructive interference,
which has the opposite effect.

Other new physics not directly related to NC processes can manifest itself
through shifts in the apparent values of $R^\nu ,R^{\bar{\nu}}$ measured by
experiment. For example, $\nu _\mu \rightarrow \nu _e,\nu _\tau $
oscillations would increase the value of $R^\nu ,R^{\bar{\nu}}$ relative to
SM expectations because both $\nu _e$ and $\nu _\tau $ charged current
interaction produce an experimental $\nu _\mu $ NC signature \cite{Donna and
Kevin}.

\subsection{$\nu _\mu e^{-}$ and $\bar{\nu}_\mu e^{-}$ Scattering: Experiment
}

The experimental challenge in measuring NC processes is to separate genuine
scattering from electrons from much larger cross section elastic and
quasi-elastic $\nu _eN$ and $\bar{\nu}_eN$ reactions 
\begin{equation}
\nu _eN\rightarrow e^{-}(n\pi ^0)X,
\end{equation}
where $X$ consists of hadrons with too low energy to be measured; and from
the rare, but kinematically similar, coherent pion production reactions: 
\begin{equation}
\nu _\mu N\rightarrow \nu _\mu \pi ^0N.
\end{equation}
Background rejection arises primarily from application of cuts that exploit
the small momentum transfer inherent in scattering from electron targets.
This demands good electron energy and angle measurement, which in turn
mandates fine-grained low $Z$ detectors. Since measurement of electron
charge is not possible in a practical device, separate, high purity,
sign-selected beams are required. Corrections must be applied for the
wrong-sign neutrino content, and for the few percent component of $\nu _e$
or $\bar{\nu}_e$ in the beam.

Charged current scattering is also optimally performed in a low $Z$, fine
grained detector. This allows superior rejection of the primary background, $%
y\rightarrow 0$ inclusive $\nu _\mu N$ scattering. At high energies, inverse
muon decay can also be observed in dense detectors as the quasi-elastic
processes become proportionally less important and the final state muon is
easily observed. The low $y$ $\nu _\mu N$ background processes can be
measured from inclusive $\bar{\nu}_\mu N$ scattering under the assumption
that $\frac{d\sigma (\nu _\mu N\rightarrow \mu ^{-}X)}{dy}=\frac{d\sigma (%
\bar{\nu}_\mu N\rightarrow \mu ^{+}X)}{dy}$ as $y\rightarrow 0$.

\subsubsection{Neutral Current Scattering}

CHARM II \cite{CHARM II Final EW}\ recorded $2677\pm 82$ $\nu _\mu e^{-}$ and 
$2752\pm 88$ $\bar{\nu}_\mu e^{-}$ events, after corrections, using a 690
ton instrumented glass detector \cite{CHARM II Detector} (Fig. \ref{CHARM II
Det-pict} ) in a $2.1\times 10^{19}$ proton-on-target exposure to the 450
GeV CERN horn beam. The detector has excellent electron identification and
measurement capabilities (Fig. \ref{CHARM II event} ). Average energies for $%
\nu _\mu $ and $\bar{\nu}_\mu $ were 23.7 and 19.2 GeV, respectively.
Backgrounds, mainly from coherent $\pi ^0$ production in elastic nuclear and
nucleon scattering and quasi-elastic $\nu _eN$ scattering, were at the $50\%$
level. The CHARM II\ statistics ensure that this experiment dominates
previous efforts at Brookhaven \cite{BNL nu-e-A,BNL nu-e-B,BNL nu-e-C};
hence, this section is essentially a resume of their results.

\begin{figure}[pthb]
\psfig{file=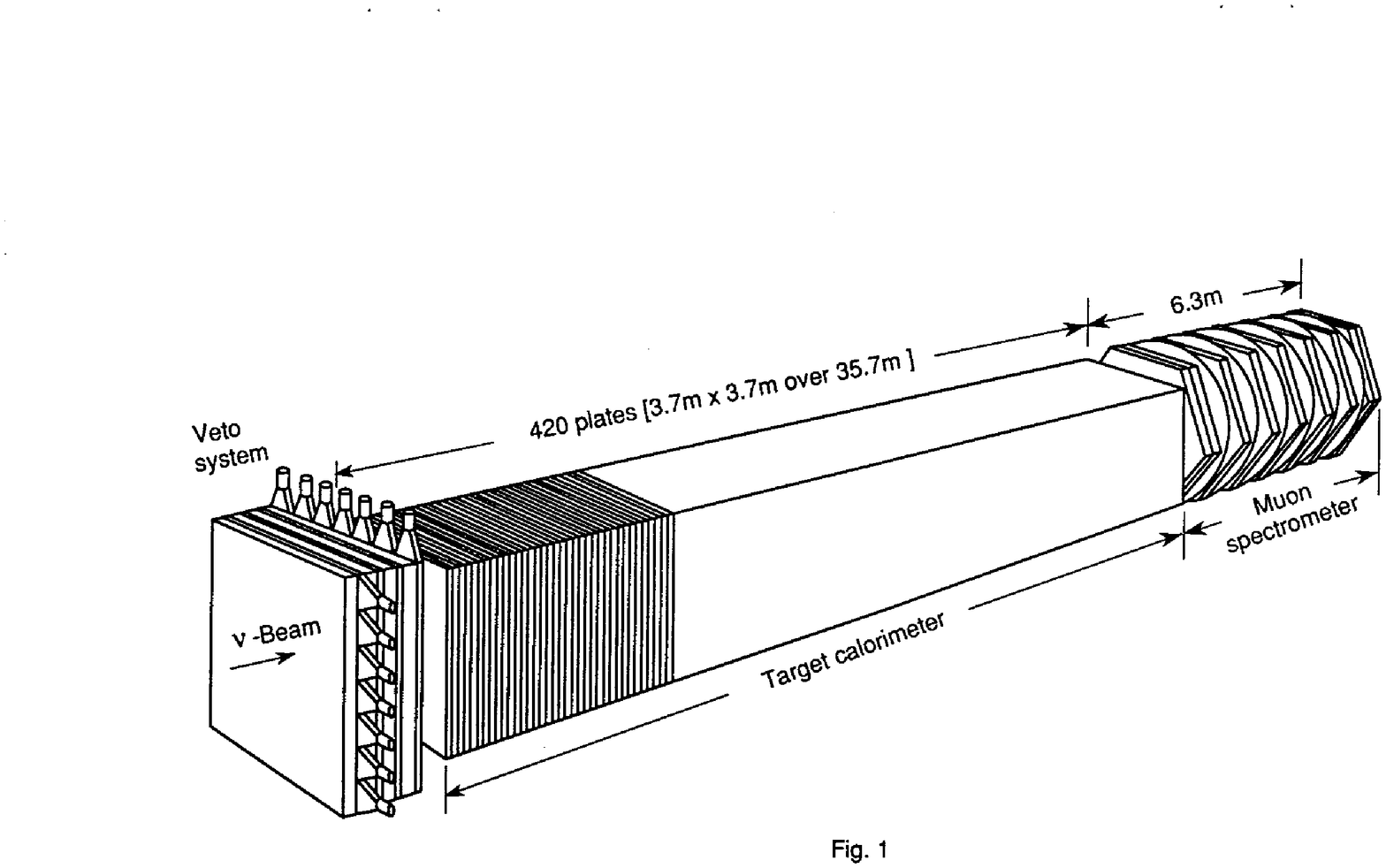,clip=,angle=270,height=3.5in}
\caption{The CHARM II detector.}
\label{CHARM II Det-pict}
\end{figure}

\begin{figure}[pthb]
\psfig{file=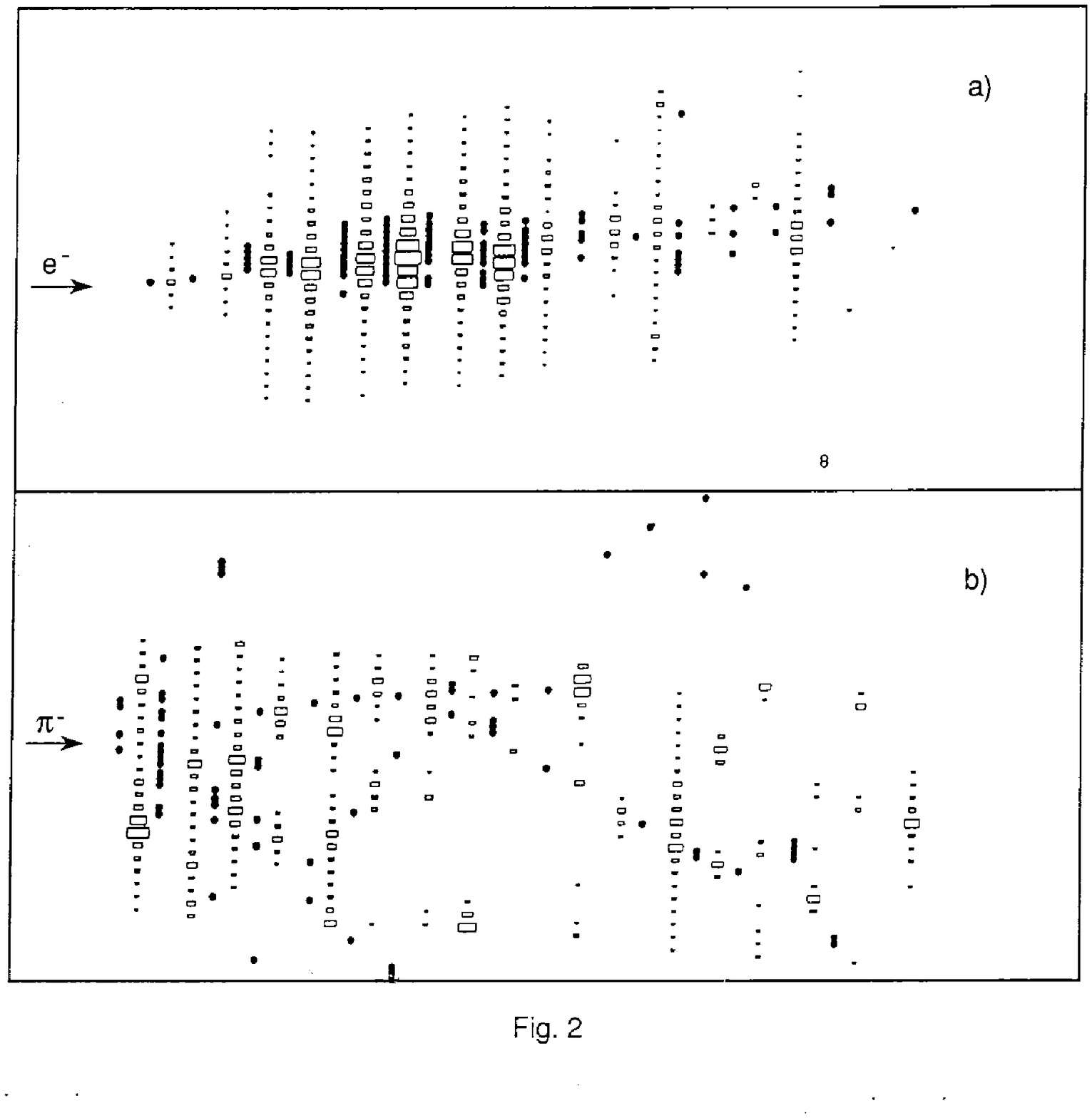,clip=,angle=270,height=3.5in}
\caption{Display of two test beam events in CHARM II detector showing
response to 10 GeV electron (a) and 10 GeV pion (b). The small dots
represent tube hits and the area of the squares is a measure of energy
deposition.}
\label{CHARM II event}
\end{figure}

\begin{figure}[pthb]
\psfig{file=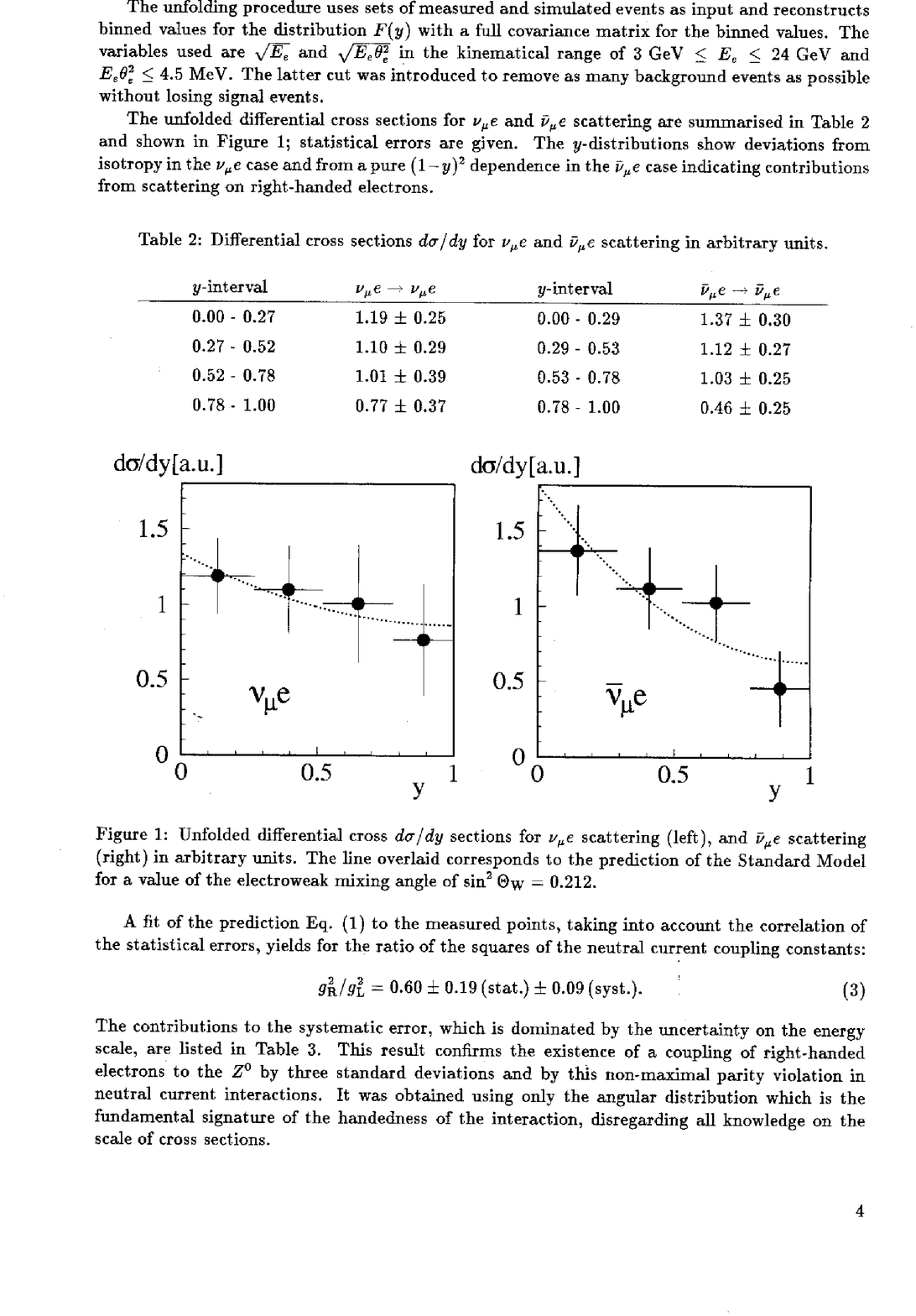,clip=,width=5.5in}
\caption{ Neutral current $\nu_\mu$ and $\bar\nu_\mu$ differential cross
section shapes measured by Charm II.}
\label{Charm II XC}
\end{figure}

CHARM II published their final results in terms of vector and axial vector
couplings, $g_V^{\nu e}$ and $g_A^{\nu e}$, which are simply related to left
and right handed electron couplings, assuming the neutrino is purely left
handed, by $\ell _e=\frac 12\left( g_V^{\nu e}+g_A^{\nu e}\right) $, $r_e=%
\frac 12\left( g_V^{\nu e}-g_A^{\nu e}\right) $. The results are 
\begin{eqnarray*}
g_V^{\nu e} &=&-0.035\pm 0.012\pm 0.012 \quad {\rm (CHARM~II),} \\
g_A^{\nu e} &=&-0.503\pm 0.006\pm 0.016 \quad {\rm (CHARM~II),}
\end{eqnarray*}
where the first uncertainty listed is the statistical error and the second
is the systematic error. These measurements were then combined to yield an
effective weak mixing angle for $\nu e$ scattering, $\sin ^2\theta _W^{\nu
e} $ , which, as noted earlier, is fortuitously close, theoretically, to the
effective mixing angle measured at LEP and SLD: 
\[
\sin ^2\theta _W^{\nu e}=0.2324\pm 0.0058\pm 0.0059 \quad {\rm (CHARM~II),} 
\]
with the first error statistical and the second systematic. These results,
based on an analysis of the absolute differential cross sections, were
confirmed with an earlier study based only on the shape of the $y$%
-distributions \cite{CHARM II Ydis} (Fig. \ref{Charm II XC}). Excellent
agreement with LEP/SLD results exists.

The CHARM II data also provided, after combination with LEP data and lower
energy $\nu _ee^{-}$ scattering, an elegant demonstration of the flavor
independence of the $Z^0$ coupling to neutrinos \cite{CHARM II Flavor}.

\begin{figure}[pthb]
\psfig{file=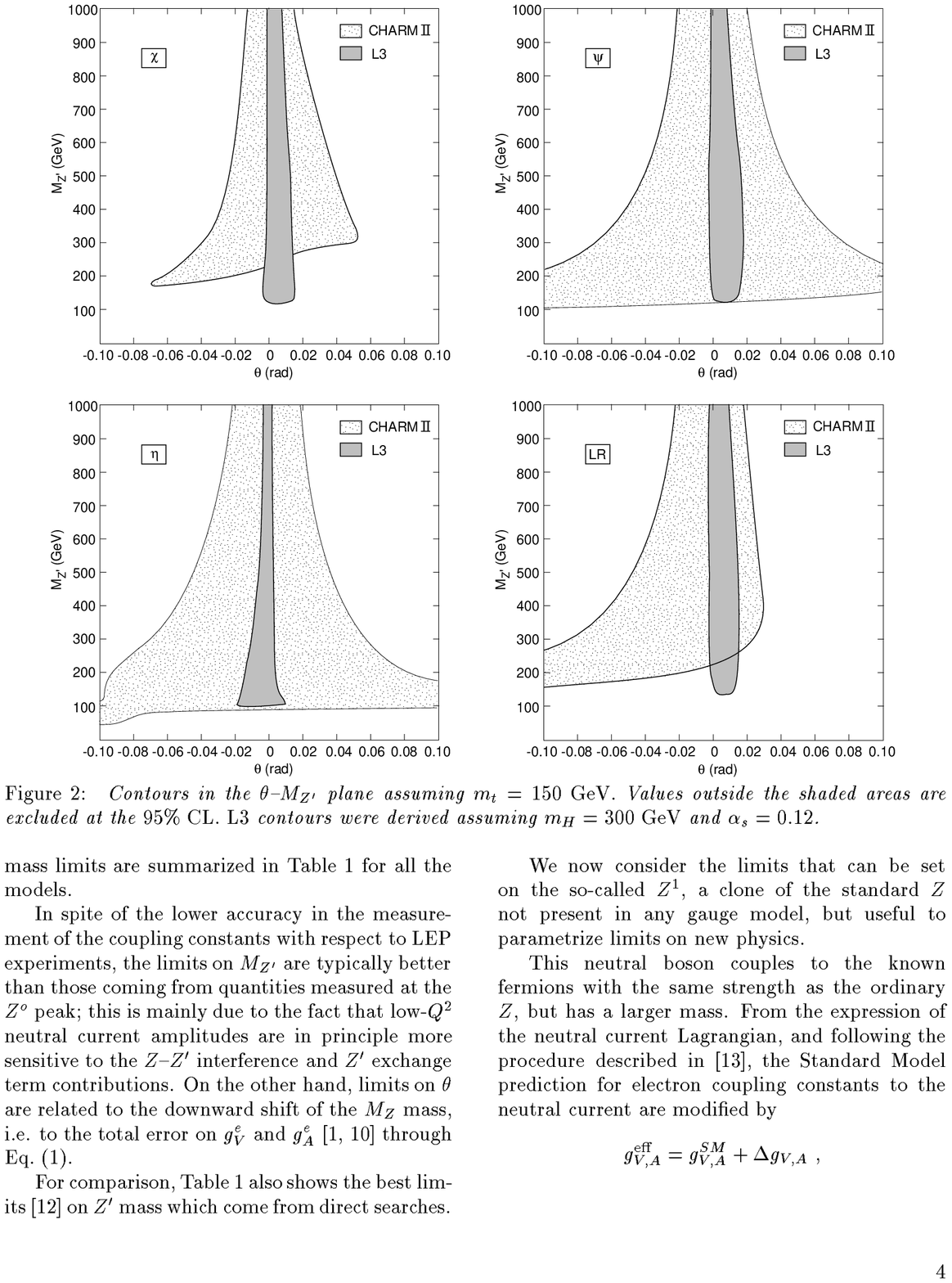,clip=,height=3.5in}
\caption{ Limits of $Z^\prime$ set by CHARM II from a comparison of electron
couplings with the SM predictions}
\label{Charm II Z' plot}
\end{figure}

Finally, the comparison of CHARM II\ coupling measurements to SM electroweak
theory allowed limits to be set on masses of new heavy $Z^{\prime }$ bosons.
For the optimal case of no mixing with the ordinary $Z^0$ boson, the
following limits hold \cite{CHARM II Z'}: 
\begin{eqnarray*}
M_{Z(\chi )} &>&262{\rm ~GeV~at~95\%~C.L. \quad (CHARM~II),} \\
M_{Z(\psi )} &>&135{\rm ~GeV~at~95\%~C.L. \quad (CHARM~II),} \\
M_{Z(\eta )} &>&100{\rm ~GeV~at~95\%~C.L. \quad (CHARM~II),} \\
M_{Z(LR)} &>&253{\rm ~GeV~at~95\%~C.L. \quad(CHARM~II).}
\end{eqnarray*}
More general limits are summarized in Fig. \ref{Charm II Z' plot}. While
these results have largely been superceded by direct searches and other
precision measurements, it is nonetheless remarkable that such constraints
follow from the modest statistics and low energy scale of CHARM II.

\subsubsection{Inverse Muon Decay}

\begin{figure}[pthb]
\psfig{file=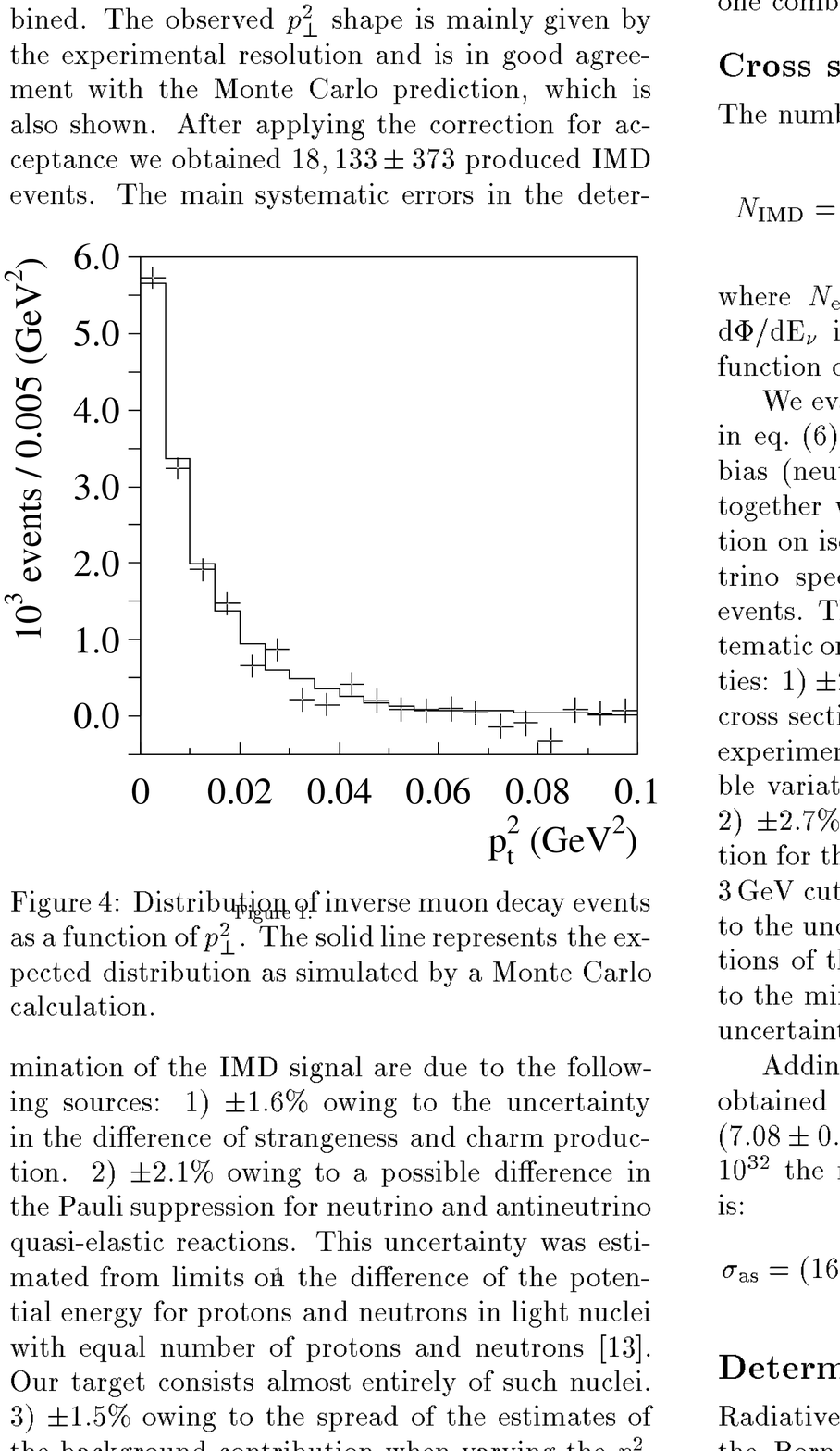,clip=,height=3.5in}
\caption{ Distribution of inverse muon decay events as a function of 
squared transverse momentum $p_t^2$ from
CHARM II. The solid line represents the expected distribution as simulated
by a Monte Carlo calculation.}
\label{CHARM II IMD plot}
\end{figure}

CCFR (FNAL\ E744/770) \cite{CCFR IMD} has a sample of $3248\pm 148$ corrected
events from an exposure corresponding to $3\times 10^6$ ordinary CC events
at an average energy $E_\nu =160$ GeV. Charm II  \cite{Charm II IMD-A,Charm
II IMD-B} recorded a signal of $15758\pm 324$ events from a much larger
exposure ($13.3\times 10^6$ CC events) at an average energy $E_\nu =23$ GeV.
Signal to background for this reaction was approximately $1:2.5$ for both
experiments. Figure \ref{CHARM II IMD plot} shows the background-subtracted
transverse momentum spectrum from CHARM II.

The $E_\nu \rightarrow \infty $ asymptotic cross section slope (Eq. \ref{IMD
XC}) $\sigma _0^{\nu e}$ measured by the two experiments agree, and the
total measurement errors are comparable. CCFR's error is dominated by
statistics, whereas CHARM II, with their lower energy beam, is more affected
by model uncertainties in making the background subtraction. The quoted
cross section results, 
\begin{eqnarray*}
\sigma _0^{\nu e} &=&\left( 16.93\pm 0.85\pm 0.52\right) \times 10^{-42}{\rm %
cm}^2\quad {\rm (CCFR)}{,} \\
\sigma _0^{\nu e} &=&\left( 16.01\pm 0.33\pm 0.83\right) \times 10^{-42}{\rm %
cm}^2\quad {\rm (CHARM~II)}{,}
\end{eqnarray*}
also agree with the SM prediction $\sigma _0^{\nu e}=17.2\times 10^{-42}$ cm$%
^2$.

Consistency with the SM allows the two groups to set limits in the scalar
couplings coupling of 
\begin{eqnarray*}
\left| g_{LL}^S\right| ^2 &<&0.30{\rm ~at~90\%~C.L. \quad (CCFR),} \\
\left| g_{LL}^S\right| ^2 &<&0.475{\rm ~at~90\%~C.L. \quad (CHARM~II);}
\end{eqnarray*}
and CHARM II\ also quotes 
\[
\left| g_{LL}^V\right| ^2>0.881{\rm {~at~90\%~C.L. \quad (CHARM~II).} } 
\]
These results confirm the $V-A$ character of the charged current interaction.

It should be noted that CHARM II\ applied SM radiative corrections before
setting their limits whereas CCFR apparently did not. The effect of the
radiative correction was to increase $\sigma _0^{\nu e}$ by $\sim 3\%$.
Because the radiative corrections would (fortuitously) move the CCFR\
measurement closer to the SM, actual limits on scalar couplings are probably
tighter.

\subsection{$\nu _\mu N$ and $\bar{\nu}_\mu N$ Scattering: Experiment}

\subsubsection{Method}

\begin{figure}[pthb]
\psfig{file=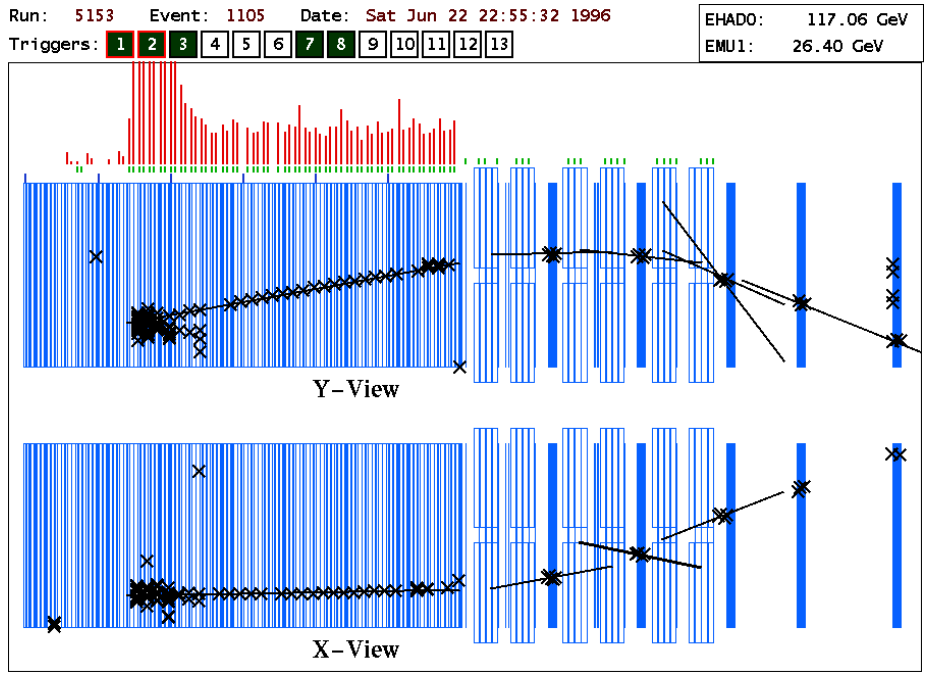,clip=,height=3.5in}
\caption{ A charged current event candidate in the CCFR detector. }
\label{CCFR CC}
\end{figure}

\begin{figure}[pthb]
\psfig{file=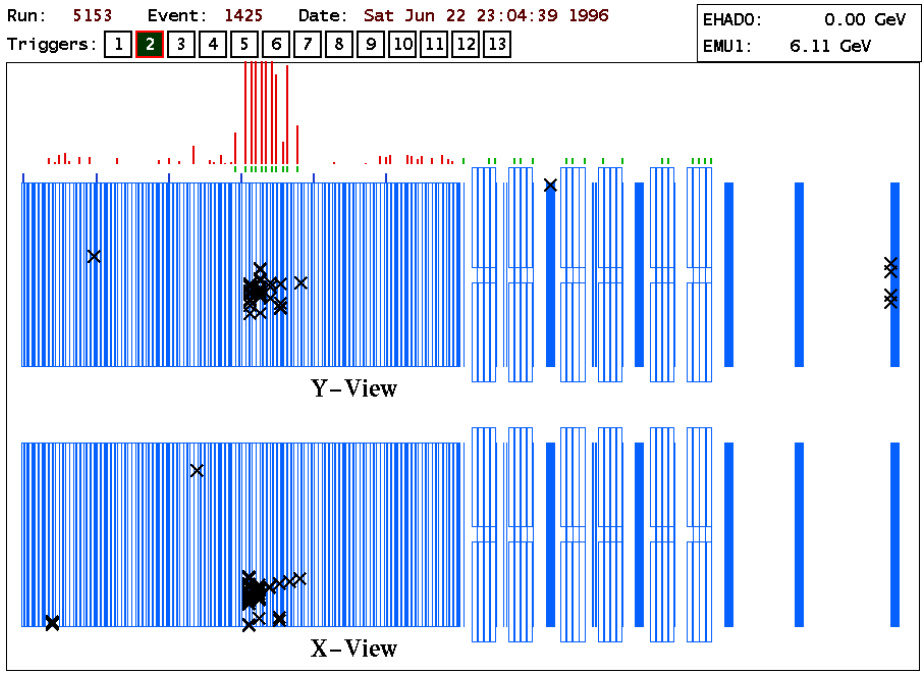,clip=,height=3.5in}
\caption{ A neutral current event candidate in the CCFR detector. }
\label{CCFR NC}
\end{figure}

The critical issue in neutrino-nucleon NC to CC ratio measurements is the
experimental separation between NC and CC\ events. This separation
essentially depends on the presence or absence of a muon in a particular
event, as can be seen in Figs. \ref{CCFR CC} and \ref{CCFR NC}, which show
CC and NC candidate events in the CCFR detector.

\begin{figure}[pthb]
\psfig{file=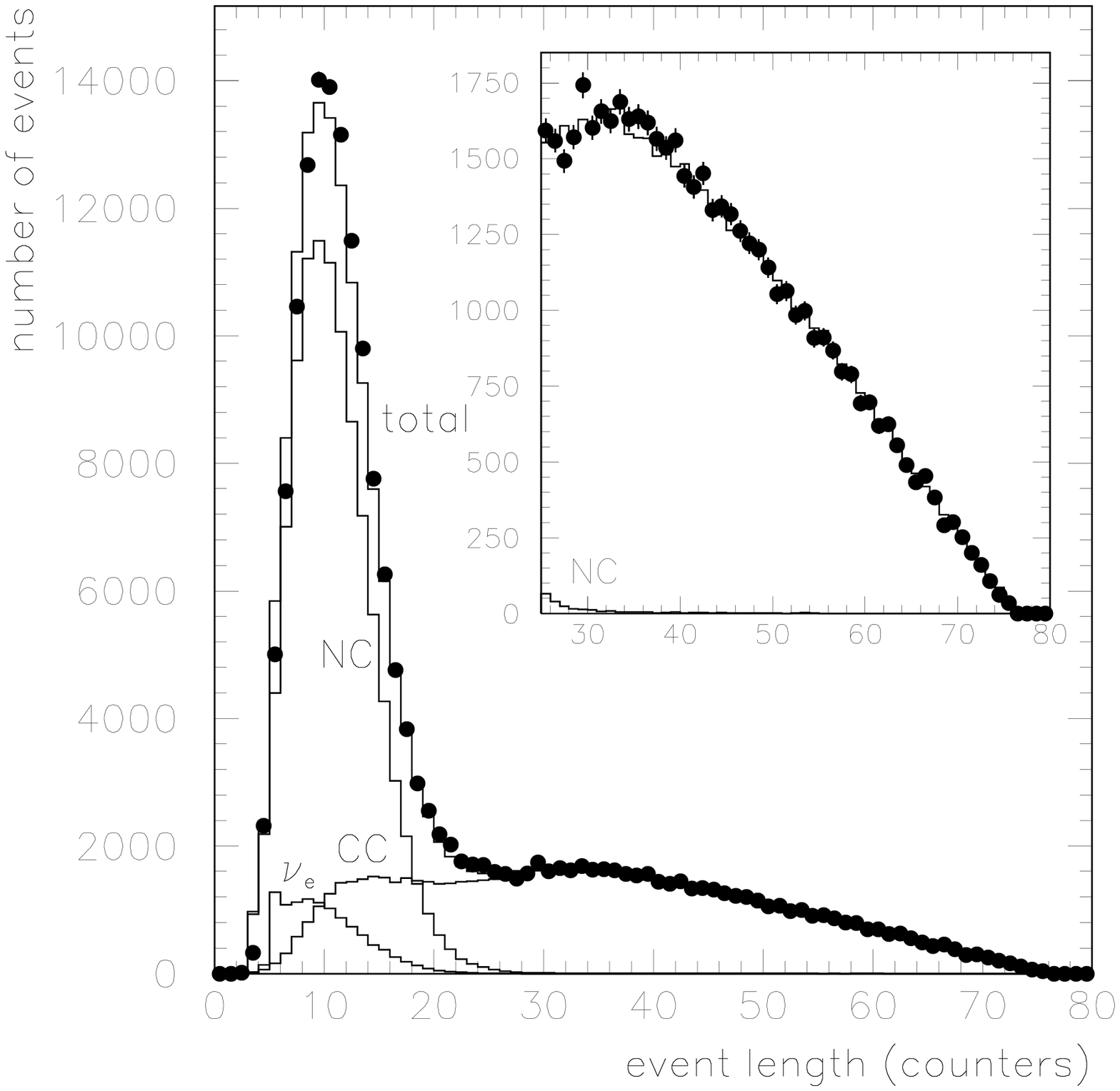,clip=,angle=90,height=3.5in}
\caption{ Event length distribution for CCFR neutral current analysis.
Length is measured in units of scintillation counters exhibiting energy
deposition in the event, with one counter corresponding to approximately
10cm of steel. The large peak at short event lengths consists mainly of $%
\nu_\mu$ neutral current interactions, with a background of $\nu_e$ NC and
CC interactions and high $y$ $\nu_\mu$ CC interactions. }
\label{CCFR Ldis}
\end{figure}

\begin{figure}[pthb]
\psfig{file=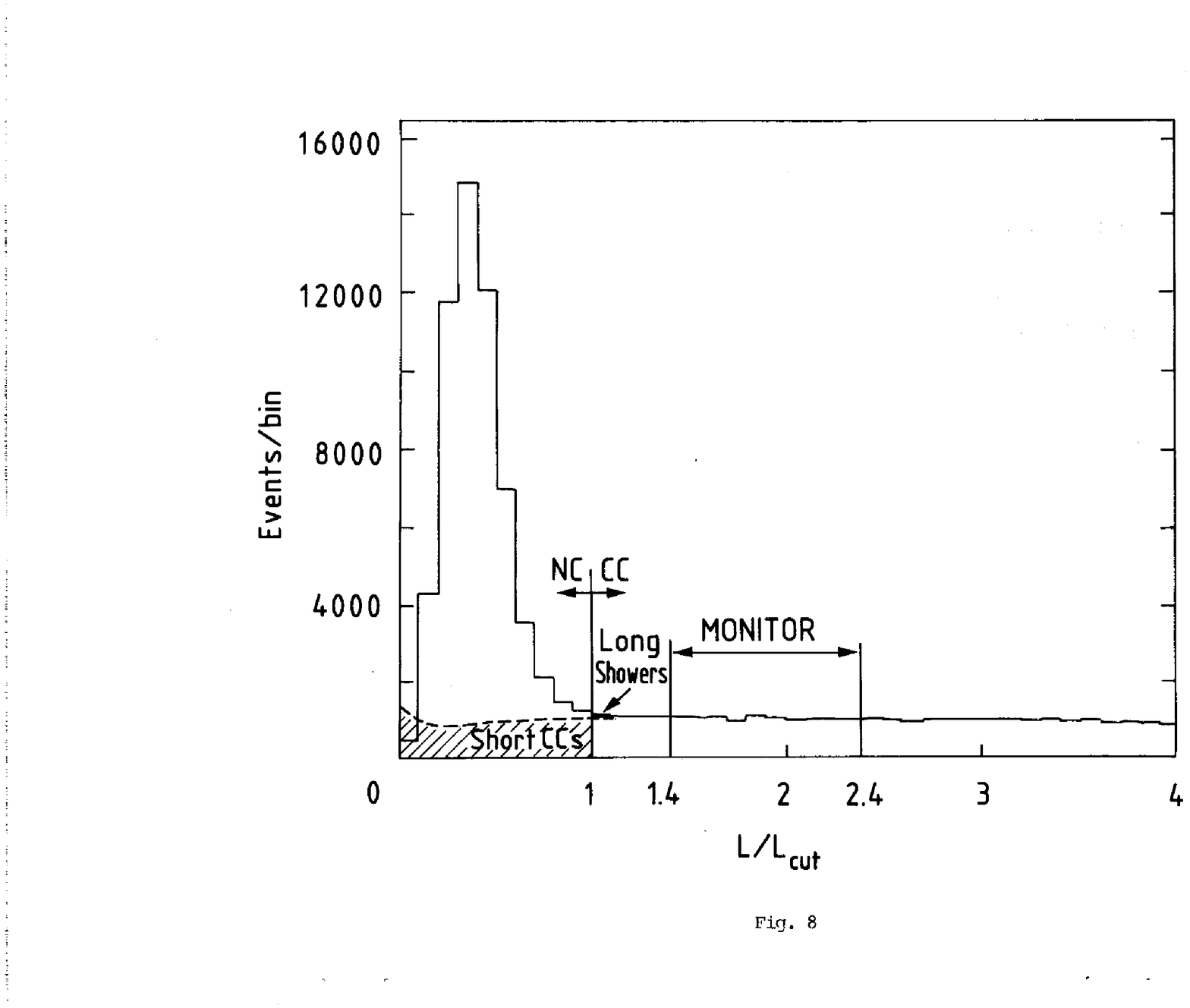,clip=,angle=270,height=3.5in}
\caption{ Scaled event length distribution for CDHS neutral current
analysis. The shape of the plot is similar to that of CCFR, but the length
separation is made energy depedent.}
\label{CDHS Ldis}
\end{figure}

Two approaches have been successfully employed in high statistics
measurements: event-by-event CC identification through muon track finding
(CHARM \cite{CHARM sin2thw-A,CHARM sin2thw-B}) and statistical separation
based on the distribution of longitudinal energy deposition in $\nu _\mu N$
interactions [CDHS \cite{CDHS sin2thw-A,CDHS sin2thw-B}, CCFR \cite{CCFR
sin2thw}]. Figure \ref{CCFR Ldis} illustrates the latter idea as implemented
by CCFR in the form of an event length distribution. The essential idea is
that CC events produce penetrating muons that deposit energy over a large
distance in the detector, whereas NC event lengths are characteristic of a
hadron shower. CDHS employed a variant of this technique that allowed the
separation length to vary with energy (Fig. \ref{CDHS Ldis}). Corrections
must be made for the fraction of short events which are actually CC,
typically $5-20\%$ and for the much smaller ``punchthrough'' of NC events
into the long event length category.

The second most important experimental effect is the correction for $\nu _e$
and $\bar{\nu}_e$ in the beam. Both NC and CC electron neutrino interactions
mimic the experimental signature of NC\ $\nu _\mu $ interactions, and their
contribution to the apparent NC sample must therefore be subtracted. The
subtraction is estimated by a detailed Monte Carlo simulation of the
neutrino beam. This simulation can be tuned to describe $\nu _e,\bar{\nu}_e$
produced from charged kaon decay with high accuracy since the $K^{\pm }$
decay contribution is tightly constrained by measurements of the $\nu _\mu ,%
\bar{\nu}_\mu $ flux (Sec. \ref{Flux Section}). The largest uncertainty in
the calculated electron neutrino flux comes from other sources of $\nu _e$
such as beam scraping or $K_L^0$ decay. CCFR has recently directly measured
the $\nu _e$ flux in their experiment using the difference in longitudinal
energy deposition for electrons compared to hadrons \cite{Romosan}; the
result confirms the calculated flux used in electroweak measurements.

CHARM and CDHS attempted to extract values for $R^\nu $ and $R^{\bar{\nu}}$,
and then infer $\sin ^2\theta _W$, whereas CCFR varies the mixing angle in a
parametric Monte Carlo approach to describe the data. Table \ref{Us vs Them}
compares properties of CCFR and CDHS/CHARM. \footnote{%
The latter pair ran in the same beam.}  We focus on these three experiments
in this section because other recent measurements \cite{FMMF sin2thw,Old CCFR
sin2thw} suffer from relatively poor statistical and systematic errors.

\begin{table}[tbp] \centering%
\begin{tabular}{|c|c|c|}
\hline
{\bf Property} & {\bf CDHS/CHARM} & {\bf CCFR} \\ \hline
mean neutrino energy & 70 GeV & 160 GeV \\ 
minimum hadron energy & 10 GeV & 30 GeV \\ 
mean $Q^2$ & 15 GeV$^2$ & 35 GeV$^2$ \\ 
statistics & $2.0\times 10^5$ & $8.1\times 10^5$ \\ \hline
\end{tabular}
\caption{Comparison of properties of CCFR and CDHS experiments.\label{Us vs
Them}}%
\end{table}%

\subsubsection{Standard Model Results}

The most accurate determination of $\sin {}^2\theta _W$ from $\nu N$
scattering is the recently updated final CCFR result \cite{Kevin97}: 
\[
\sin {}^2\theta _W\text{(on-shell)}=0.2236\pm 0.0027_{{\rm expt.}}\pm
0.0030_{{\rm model}}\quad \text{(CCFR),} 
\]
where the first error is the total statistical and systematic experimental
contribution and the second represents uncertainties due to model
corrections. This value was extracted from the ratio of NC to CC events
assuming values of $m_c^{eff}=1.31$ GeV, $M_{top}=175$ GeV, and $M_H=150$
GeV; the explicit dependence on these parameters is 
\begin{eqnarray*}
\sin {}^2\theta _W\text{(on-shell)} &=&0.2236+0.0111\left( m_c^{eff}-1.31%
\text{ GeV}\right) \\
&&+\left( 2.1\times 10^{-5}\right) \left( M_{top}-175\text{ GeV}\right)
-0.0002\ln \left( \frac{M_H}{150\text{ GeV}}\right) \quad \text{(CCFR),}
\end{eqnarray*}
where the masses are in units of GeV. The significant dependence of the
result on the effective charm quark mass and the weak dependence on
radiative corrections in translating to the on-shell mixing angle are
evident. A detailed breakdown of systematic errors is provided in Table \ref
{CCFR errors}. The dominant contributions are from statistics, uncertainties
in $\nu _e$ production (mainly from neutral kaon sources), and charm quark
production.

\begin{table}[tbp] \centering%
\begin{tabular}{|l|l|}
\hline
{\bf SOURCE OF UNCERTAINTY} & ${\bf \delta }\sin ^2\theta _W$ \\ \hline
Data Statistics & {\em 0.0019} \\ 
Monte Carlo Statistics & 0.0004 \\ \hline
TOTAL STATISTICS & 0.0019 \\ \hline\hline
Charm Production ($m_c=1.31\pm 0.24$GeV) & {\em 0.0027} \\ 
Charm Sea & 0.0006 \\ 
Longitudinal Cross Section & 0.0008 \\ 
Higher Twist & 0.0010 \\ 
Non-Isoscalar Target & 0.0004 \\ 
Strange Sea & 0.0003 \\ 
Structure Functions & 0.0002 \\ 
Radiative Corrections & 0.0001 \\ \hline
TOTAL PHYSICS\ MODEL & 0.0030 \\ \hline\hline
$\nu _e$ Flux & {\em 0.0015} \\ 
Transverse Vertex & 0.0004 \\ 
Energy Measurement &  \\ 
\hspace{0.5in}Muon Energy Loss in Shower & 0.0003 \\ 
\hspace{0.5in}Muon Energy Scale ($\pm 1\%$) & 0.0004 \\ 
\hspace{0.5in}Hadron Energy Scale ($\pm 1\%$) & 0.0004 \\ 
Event Length &  \\ 
\hspace{0.5in}Hadron Shower Length & 0.0007 \\ 
\hspace{0.5in}Vertex Determination & 0.0003 \\ 
\hspace{0.5in}Detector Noise, Efficiency & 0.0006 \\ 
\hspace{0.5in}Dimuon Production & 0.0003 \\ \hline
TOTAL EXPERIMENTAL SYSTEMATIC & 0.0027 \\ \hline\hline
\multicolumn{1}{||l|}{TOTAL\ UNCERTAINTY} & \multicolumn{1}{||l||}{0.0041}
\\ \hline\hline
\end{tabular}
\caption{Uncertainties from updated CCFR extraction of
$\sin^2\theta_W$\label{CCFR errors}}%
\end{table}%

CDHS and CHARM extracted their values for the weak mixing angle from the LS
formula. CDHS \cite{CDHS sin2thw-B} obtains, using $m_c^{eff}=1.50$ GeV, $%
M_{top}=60$ GeV, and $M_H=100$ GeV, 
\[
\sin {}^2\theta _W\text{(on-shell)}=0.228\pm 0.005_{\text{{\rm expt}}}\pm
0.005_{\text{{\rm model}}}\quad \text{(CDHS),}
\]
and CHARM \cite{CHARM sin2thw-B} reports a very similar quality result 
\[
\sin {}^2\theta _W\text{(on-shell)}=0.236\pm 0.005_{\text{{\rm expt}}}\pm
0.005_{\text{{\rm model}}}\quad \text{(CHARM).}
\]
Unfortunately these results were published at a time when the top mass was
thought to be lighter; hence one must undo some of the radiative corrections
to compare to the more recent CCFR result. If a common set of model
parameters is employed, CCFR, CDHS, and CHARM agree in detail \cite
{Gatlinburg}, with the charm production uncertainty dominating in all cases.
The average value of $\sin ^2\theta _W$ obtained is 
\[
\sin ^2\theta _W\text{(on shell)}=0.2256\pm 0.0035\quad \text{(average }\nu N%
\text{).}
\]
This value is obtained by fitting the CHARM, CDHS, and CCFR mixing angles
and the CDHS and CCFR charm mass measurements to a simple model that allows $%
\sin ^2\theta _W$ and $m_c^{eff}$ to vary. The fit has a $\chi ^2$ of 2.9
for 3 degrees of freedom and also yields an estimate of $m_c^{eff}=1.37\pm
0.19$ GeV. The contribution to the $\sin ^2\theta _W$ error from charm
production is $\pm 0.0023$.

Within the SM, the $\nu N$ $\sin ^2\theta _W$ measurement implies 
\[
M_W=80.25\pm 0.18{\rm ~{GeV}}\quad \text{(average }\nu N\text{).} 
\]
The latter agrees well with direct determinations at the Tevatron and LEP II%
 \citeaffixed{Direct W-mass-D0,Direct W-mass-CDF,Direct W-mass-OPAL}{The most
recent high accuracy measurements are:} $M_W=80.33\pm 0.15$ GeV, and the
indirect determination from global fits to $e^{+}e^{-}$ data at the $Z^0$, %
 \citeaffixed{LEP-SLD W-mass}{See, for example,} $M_W=80.359\pm 0.056$ GeV. 
\footnote{%
For a more detailed summary, see  \citeasnoun{PDG}.}

\subsubsection{Model Independent Results}

The three collaborations have also attempted to present their results in a
more model independent form, and we summarize these results here. Where
possible, the explicit dependence of a parameter on the effective charm
quark mass $m_c^{eff}$ is given since correcting for CC charm production
produces the largest systematic uncertainty in all results. In these cases,
the charm mass contribution to the parameter is not included in the quoted
error, but it can be easily calculated from $m_c^{eff}=1.5\pm 0.3$ GeV used
by CHARM/CDHS and $m_c^{eff}=1.31\pm 0.24$ GeV\ used by CCFR.

CHARM and CDHS quote values of $R^\nu $ and $R^{\bar{\nu}}$ (Fig. \ref{CDHS
Rnu}), as well as separate extractions of $\Delta \rho $ and $\sin ^2\theta
_W$. CHARM obtains 
\begin{eqnarray*}
R^\nu &=&0.3093\pm 0.0031 \quad \text{(CHARM)}, \\
R^{\bar{\nu}} &=&0.390\pm 0.014 \quad \text{(CHARM)},
\end{eqnarray*}
using $r=0.456\pm 0.011$ (CHARM); whereas CDHS extracts 
\begin{eqnarray*}
R^\nu &=&0.3135\pm 0.0033 \quad \text{(CDHS),} \\
R^{\bar{\nu}} &=&0.376\pm 0.0016 \quad \text{(CDHS)},
\end{eqnarray*}
using $r=0.409\pm 0.014$(CDHS). These cross section ratios represent the
most model independent expression of $\nu N$ electroweak results; only
corrections for experimental effects and the non-isoscalar nuclear targets
are applied. As such, they cannot be directly used for electroweak tests
without further model corrections. CDHS has applied these corrections to get 
\begin{eqnarray*}
R^{\nu 0} &=&0.3122\pm 0.0034_{{\rm expt}}-0.009(m_c^{eff}-1.5\text{ GeV})
\quad \text{(CDHS),} \\
R^{\bar{\nu}0} &=&0.378\pm 0.014_{{\rm expt}}-0.019(m_c^{eff}-1.5\text{ GeV}%
) \quad \text{(CDHS).}
\end{eqnarray*}
The ``$0$'' superscript denotes that these quantities have effects due to
heavy quarks, radiative corrections, and other factors removed from the
physical cross section ratios. Application of these corrections requires
slightly different $\nu _\mu $ to $\bar{\nu}_\mu $ CC ratios to be used in
the LS formulas, $r^0=0.383\pm 0.014_{{\rm expt}}+0.004(m_c^{eff}-1.5$ GeV$)$
for neutrinos and $\bar{r}^0=0.371\pm 0.014_{{\rm expt}}+0.004(m_c^{eff}-1.5$
GeV$)$ for antineutrinos.

\begin{figure}[pthb]
\psfig{file=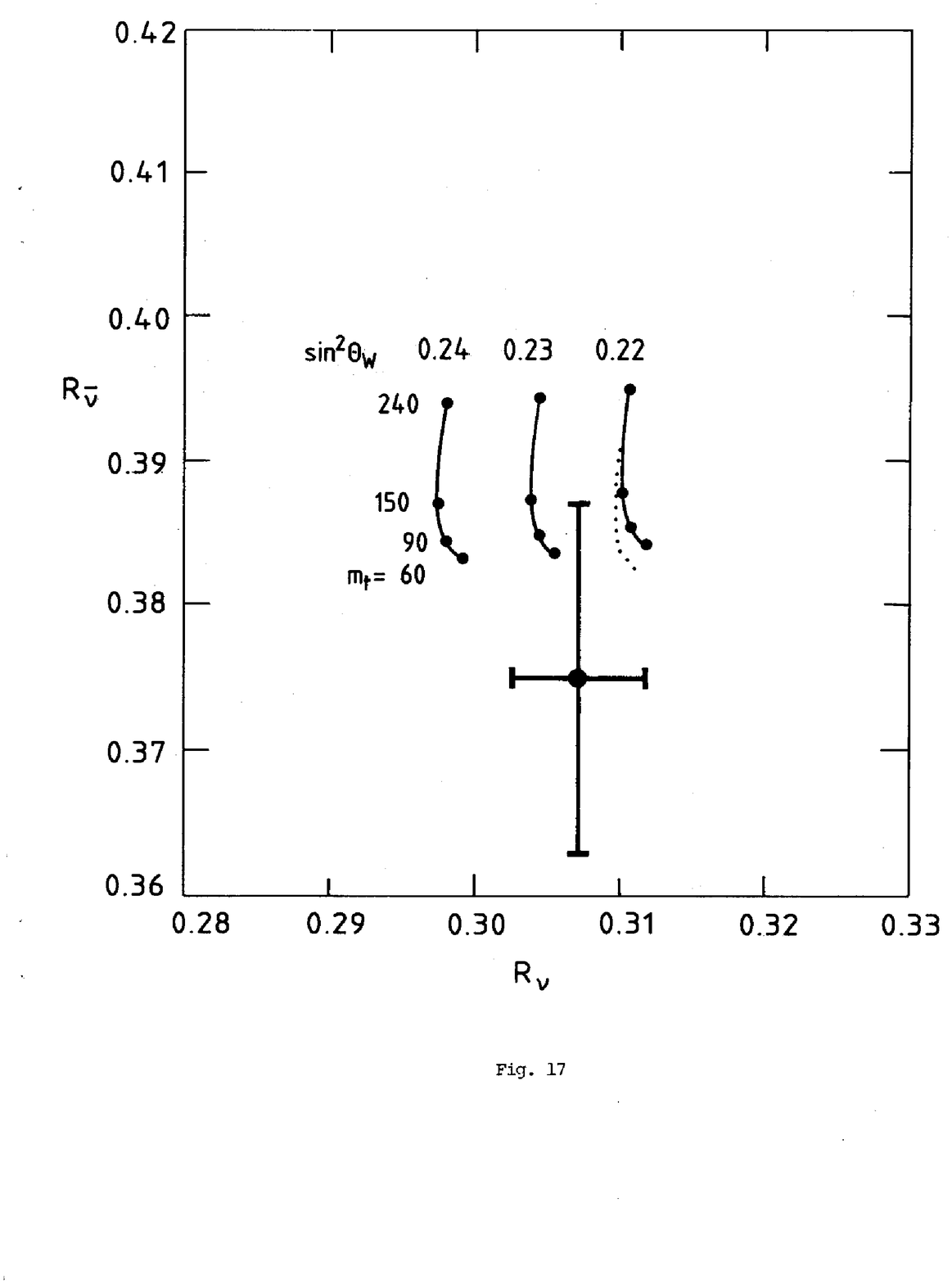,clip=,height=3.5in}
\caption{$R_\nu$ vs $R_{\bar\nu}$ from CDHS. This measurement, along with a
comparable result from CHARM imposed stringent constraints on the top mass
as soon as the $Z$ boson mass was measured precisely.}
\label{CDHS Rnu}
\end{figure}

Because CDHS and CHARM took data in sign-selected beams, they could extract
two electroweak parameters from their data without assuming the SM.
Translated to the notation of Eq.'s \ref{KappaRadCor}-\ref{RhoRadCor}, CDHS
extracted values for these parameters (originally expressed as ``$\rho $''
and ``$\sin ^2\theta _W$''), using only their own data, of 
\begin{eqnarray*}
\Delta \rho &=&0.009\pm 0.020_{{\rm expt}}-0.023(m_c^{eff}-1.5\text{ GeV}%
)\quad \text{(CDHS),} \\
\kappa \sin ^2\theta _W &=&0.218\pm 0.021_{\text{{\rm expt}}%
}-0.011(m_c^{eff}-1.5\text{ GeV})\quad \text{(CDHS).}
\end{eqnarray*}
The larger experimental error on the mixing angle simply reflects the weaker
constraints of a two-parameter fit. The strong top mass dependence of $%
\Delta \rho $ and $\kappa $ in the SM allowed this measurement to be used to
constrain $M_t<200$ GeV well before the direct observation of the top quark.
CDHS and CHARM weak mixing angle measurements also decisively ruled out the
once promising grand unified theory based on minimal SU(5) (which predicted $%
\sin ^2\theta _W=0.218\pm 0.004$), corroborating results from proton decay
experiments \cite{SirlinAndMarciano}.

CCFR quotes an experimental value \cite{Kevin97} for a sum of couplings that
is closest to what it actually measured in its mixed $\nu _\mu /\bar{\nu}%
_\mu $ beam: 
\begin{eqnarray}
\hat{\kappa} &\equiv &1.7897\left( \ell _u^2+\ell _d^2\right) +1.1479\left(
r_u^2+r_d^2\right) \\
&&-0.0916\left( \ell _u^2-\ell _d^2\right) -0.0782\left( r_u^2-r_d^2\right) ,
\nonumber \\
&=&0.5820\pm 0.0031-0.0111(m_c^{eff}-1.31\text{ GeV})\text{ }\quad \text{%
(CCFR),}  \nonumber
\end{eqnarray}
This value can be used to constrain values of left- and right-handed
couplings to quarks as shown in Fig. \ref{CCFR couplings}. Using this
measurement and the SM prediction of $\hat{\kappa}_{{\rm SM}}=0.5817\pm
0.0013$, with SM parameters fixed from other precision electroweak
measurements of $M_Z$, $M_W$, and $M_{{\rm top}}$, CCFR sets limits on
several new physics processes:

\begin{figure}[pthb]
\psfig{file=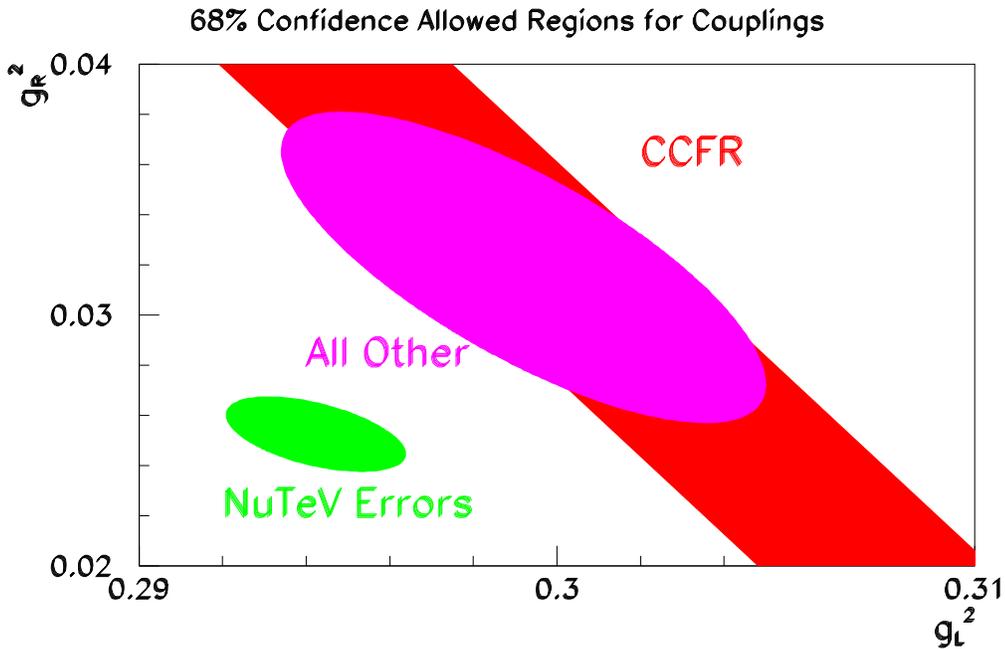,clip=,height=3.5in}
\caption{ Constraints on $g_L^2=\ell^2_u+\ell^2_d$ and $g_R^2=r^2_u+r^2_d$
from CCFR model independent result and all neutrino data combined. Also
shown are the projected constraints on these couplings from the NuTeV
experiment. }
\label{CCFR couplings}
\end{figure}

\begin{itemize}
\item  For ``left-left'' four-fermion contact operator described by a
Lagrangian ${-{\cal L}}=\pm (4\pi /\Lambda _L^{\pm })\bar{l}_L\gamma ^\nu l_L%
\bar{q}_L\gamma _\nu q_L,$%
\begin{eqnarray*}
\Lambda _{LL}^{+} &>&4.7{\rm ~{TeV~at~95\%~C.L.,}} \\
\Lambda _{LL}^{-} &>&5.1{\rm ~{TeV~at~95\%~C.L.}}
\end{eqnarray*}
The approximate equality of these limits reflects the equal sensitivity to
destructive and constructive interference effects in $\nu N\;$scattering
mentioned earlier and compares favorably with results set by CDF \cite{CDF
Composite}. Limits on a set of other composite operators are also set.

\item  For SU(5) Leptoquarks that do no induce flavor-changing neutral
currents and which have dominant left-handed couplings, 
\[
M_{LQ}\left| \eta _L\right| ^{-1}>0.8{\rm ~{TeV~at~95\%C.L.}}
\]
where $M_{LQ}$ is the leptoquark mass and $\eta _L$ the left-handed coupling
parameter.
\end{itemize}

The CCFR result can be input into the formalism of  \citeasnoun{Langacker} to
obtain parametric limits on a variety of new physics models. These results
are summarized in Table \ref{New Physics}. \footnote{%
The values in Table \ref{New Physics} are inferred from the CCFR measurement
by the authors of this Review; the experiment may publish more extended
results that differ slightly from those presented here.}

\begin{table}[tbp] \centering%
\begin{tabular}{|l|l|l|}
\hline
{\bf Model} & {\bf 95\% C.L. Limit} & {\bf comment} \\ \hline
$M_{Z(\chi )},C=\sqrt{\frac 25}$ & $M_{Z^{\prime }}>691$ GeV & $Z(\chi )$
with $\left( \frac{g_2^{\prime }}{g_2}\right) ^2=\frac 53\sin ^2\theta _W$
\\ \hline
$M_{Z(\chi )},C=0$ & $M_{Z^{\prime }}>215$ GeV & $Z(\chi )$ with $\left( 
\frac{g_2^{\prime }}{g_2}\right) ^2=\frac 53\sin ^2\theta _W$ \\ \hline
$M_{Z(\psi )},C=\sqrt{\frac 23}$ & $M_{Z^{\prime }}>843$ GeV & $Z(\psi )$
with $\left( \frac{g_2^{\prime }}{g_2}\right) ^2=\frac 53\sin ^2\theta _W$
\\ \hline
$M_{Z(\psi )},C=-\sqrt{\frac 23}$ & $M_{Z^{\prime }}>513$ GeV & $Z(\psi )$
with $\left( \frac{g_2^{\prime }}{g_2}\right) ^2=\frac 53\sin ^2\theta _W$
\\ \hline
$M_{Z(\psi )},C=0$ & $M_{Z^{\prime }}>54$ GeV & $Z(\psi )$ with $\left( 
\frac{g_2^{\prime }}{g_2}\right) ^2=\frac 53\sin ^2\theta _W$ \\ \hline
$M_{Z(\eta )},C=-\sqrt{\frac 1{15}}$ & $M_{Z^{\prime }}>101$ GeV & $Z(\eta )$
with $\left( \frac{g_2^{\prime }}{g_2}\right) ^2=\frac 53\sin ^2\theta _W$
\\ \hline
$M_{Z(\eta )},C=\sqrt{\frac{16}{15}}$ & $M_{Z^{\prime }}>879$ GeV & $Z(\eta
) $ with $\left( \frac{g_2^{\prime }}{g_2}\right) ^2=\frac 53\sin ^2\theta _W
$ \\ \hline
$M_{Z(\eta )},C=0$ & $M_{Z^{\prime }}>87$ GeV & $Z(\eta )$ with $\left( 
\frac{g_2^{\prime }}{g_2}\right) ^2=\frac 53\sin ^2\theta _W$ \\ \hline
$M_{Z(3R)},C=\sqrt{\frac 35}\alpha $ & $M_{Z^{\prime }}>988$ GeV & $Z(3R)$
with $\left( \frac{g_2^{\prime }}{g_2}\right) ^2=\frac 53\sin ^2\theta _W$
\\ \hline
$M_{Z(3R)},C=-\sqrt{\frac 35}\alpha $ & $M_{Z^{\prime }}>189$ GeV & $Z(3R)$
with $\left( \frac{g_2^{\prime }}{g_2}\right) ^2=\frac 53\sin ^2\theta _W$
\\ \hline
$M_{Z(3R)},C=0$ & $M_{Z^{\prime }}>244$ GeV & $Z(3R)$ with $\left( \frac{%
g_2^{\prime }}{g_2}\right) ^2=\frac 53\sin ^2\theta _W$ \\ \hline
Non-SM Higgs & $\Delta \rho <0.0049$ & rad. corr. parameter \\ \hline
SU(5) Leptoquark & $\left| M_{LQ}/\eta _L\right| >887$ GeV & RH coupling $%
\eta _R\simeq 0$ \\ \hline
extra $u_L$ fermion & $\sin ^2\theta _L^u<0.011$ & mixing with ordinary $u_L$
\\ \hline
extra $u_R$ fermion & $\sin ^2\theta _R^u<0.037$ & mixing with ordinary $u_R$
\\ \hline
extra $d_L$ fermion & $\sin ^2\theta _L^d<0.009$ & mixing with ordinary $d_L$
\\ \hline
extra $d_R$ fermion & $\sin ^2\theta _R^d<0.074$ & mixing with ordinary $d_R$
\\ \hline
extra $e_L$ fermion & $\sin ^2\theta _L^e<0.043$ & mixing with ordinary $e_L$
\\ \hline
extra $\nu _{eL}$ fermion & $\sin ^2\theta _L^{\nu _e}<0.043$ & mixing with
ordinary $\nu _{eL}$ \\ \hline
extra $\mu _L$ fermion & $\sin ^2\theta _L^\mu <0.006$ & mixing with
ordinary $\mu _L$ \\ \hline
extra $\nu _{\mu L}$ fermion & $\sin ^2\theta _L^{\nu _\mu }<0.043$ & mixing
with ordinary $\nu _{\mu L}$ \\ \hline
4-Fermi plus & $\Lambda _{+}>4.1$ TeV & compositeness scale \\ \hline
4-Fermi minus & $\Lambda _{-}>4.0$ TeV & compositeness scale \\ \hline
\end{tabular}
\caption{Summary of limits on parameters for models of 
new physics beyond the SM as
described by Langacker {\it{et al.}} based on CCFR preliminary 
model independent result.\label{New Physics}}%
\end{table}%

\section{Summary and Outlook}

High-energy, high-precision neutrino measurements 
have become an important
tool for measuring the parameters of the Standard Model and for testing our
understanding of the electroweak and strong interactions. This Review has
presented the current status of these measurements and has pointed out some
of the open questions. These topics are summarized below along with comments
on future measurements, now underway or being planned, which will have even
higher precision and be able to address some of these questions.

\subsection{Structure Functions and QCD Tests}

Precision neutrino experiments have made important contributions to our
knowledge of nucleon structure and the strong interaction through the
measurements of the structure functions $F_2$ and $xF_3$. Data have been
taken over a wide range of targets and energies. The total neutrino cross
section divided by energy has been found to be constant with energy,
regardless of target. 
The structure function measurements have different assumptions concerning $%
R_L$ and radiative corrections, and are made on different types of target
nuclei, making cross experiment comparisons difficult. Iron data between
CCFR and CDHSW can be directly compared and a discrepancy exists,
particularly in the low-$x$ region and as a function of $Q^2$.

Neutrino scattering studies yield important determinations of the parton
distribution functions. These results are important inputs to the global
fits due to the unique flavor differentiation of the neutrino measurements
combined with the good precision. The $xF_3$ structure function is
particularly important due to its close connection to the valence quark
distribution.

The recent neutrino measurements have comparable statistics to the
charged-lepton structure function measurements. Comparisons of the neutrino $%
F_2$ measurements to charged-lepton experiments allows an important test of
the universality of the parton distributions measured in different
processes. The comparison indicates general agreement over most of the
kinematic region but a small discrepancy at low $x$ after quark charge and
nuclear corrections are applied. At present, this differences is not
understood. The discrepancy may be related to assumptions concerning nuclear
corrections or to systematic errors in one or more of the experiments at low 
$x$. The difference is too large to be explained by uncertainties in the
strange sea which has been measured from neutrino charm production.

Precise tests of QCD are now possible with neutrino scattering data. The
CCFR E744/E770 data is the most precise and the 
$Q^2 $ evolution of the structure functions yields $\Lambda _{\overline{MS}%
}^{NLO-4f}=337\pm 31(exp)$~MeV. This is equivalent to 
$\alpha_s(M_Z^2)=0.119\pm 0.002(exp) \pm 0.004(scale)$. 
This is the most precise
measurement of $\alpha _s$ at low and moderate $Q^2$ values. The agreement
of the CCFR data with the QCD expectation is quite good. The GLS sum rule
results have sufficient accuracy to test the predicted QCD corrections.
Evaluation of this sum rule yields another measurement of the strong
coupling constant which is consistent with the CCFR $\alpha _s$ result from
QCD evolution. Because of the inclusive nature of the GLS and evolution
measurements, these results have small theoretical errors and have been
important in the past in establishing the QCD theory and more recently in
making precise measurements of the fundamental parameters.

The NuTeV experiment \cite{NuTeV} at Fermilab, which uses the upgraded CCFR
detector and a new sign-selected beam \cite{NuTeV Beam}, 
began taking data in May, 1996. This experiment will address many of the
issues concerning structure functions raised in this review. Among the NuTeV
goals are more precise measurements of $\alpha _s$ and $R_L$. In order to
investigate fully the discrepancy in the $F_2$ measurements, more data in
the low-$x$ region with improved systematic understanding are required. The
NuTeV experiment is expected to obtain approximately equal statistical
errors on $F_2$ with substantially reduced systematics. The most important
systematic uncertainties are related to the absolute energy calibration of
the detector. For the NuTeV experiment, precise hadron and muon energy
calibrations will be obtained over a wide range of energies using precision
test beam measurements throughout the run. 
With these improvements, the experiment should be able to measure the QCD
scale parameter, $\Lambda_{QCD}$ to better than 25 MeV and improve
significantly the uncertainties in the $F_2$ structure function measurements
at low-$x$.

\subsection{Measurements of the Strange Sea, Charm Sea, and $|V_{cd}|$}

Neutrino charm production has been used to isolate the nucleon strange quark
distributions, $xs(x)$ and $x\overline{s}(x)$, and study the transition to
the heavy charm quark. The CCFR NLO analysis agrees well with their dimuon
data over a broad range of energies with a charm quark mass, $m_c = 1.70 \pm
0.19 $ GeV. At NLO, the strange quark distribution is found to have the same
shape as the non-stange sea but with a magnitude that is $(48 \pm 5)\%$ of a
SU(3) symmetric sea. These measurements imply that an increased strange sea
cannot be the explanation for the discrepancy between the neutrino and
charged-lepton measurements of $F_2$ at low-$x$. Investigations by the CCFR
collaboration, also, indicate that the $s(x)$ and $\overline{s}(x)$
distributions are similar as expected from a perturbative QCD source but
inconsistent with non-perturbative models of intrinsic strangeness in the
nucleon.

The CKM matrix elements $|V_{cd}|$ and $|V_{cs}|$ are determined from the
neutrino dimuon data using an effective charm semileptonic branching ratio
found from $\nu $-emulsion and $e^{+}e^{-}$ data. The combined CCFR/CDHS
data yields a value of $|V_{cd}|=0.220\;_{-\;0.021}^{+\;0.018}$ which is in
excellent agreement with the value of $|V_{cd}|=0.221\pm 0.003$ found from
other measurements combined with the unitarity of the CKM matrix. With the
assumption that the strange sea must be smaller than the non-strange sea $%
(\kappa \le 1)$, $|V_{cs}|$ is limited by the dimuon data to be $>0.69\;%
\text{at 90\% C.L.}$. The parameter $\kappa $ could possibly be determined
in the future from a measurement of $xF_3^\nu (x,Q^2)-xF_3^{\overline{\nu }%
}(x,Q^2)=4x\left[ s(x,Q^2)-c(x,Q^2)\right] $.  
A $20\%$ measurement of $\kappa $ will provide a $10\%$ measurement of $%
|V_{cd}|/|V_{cs}|$ .

Present and future high-energy neutrino experiments should be able to
significantly improve these measurements. The NuTeV experiment 
at Fermilab is running with a sign-selected beam which will eliminate the $%
\nu /\overline{\nu }$ confusion in the dimuon channel at low $x$ with
statistics similar to the CCFR (E744/770) sample. NOMAD \cite{NOMAD} and
CHORUS \cite{CHORUS-A,CHORUS-B} at CERN are presently running with a
high-energy, horn-focused beam searching for neutrino oscillations. These
experiments will also record a sizeable number of charm production events
(about 20,000 dileptons for NOMAD and 20,000 reconstructed charm particles
for CHORUS.) As in the E531 experiment, CHORUS will be able to reconstruct
exclusive channels allowing precise tests of the theory. The ultimate
neutrino charm production experiment is COSMOS(E803 at Fermilab)  \cite
{COSMOS}. Like CHORUS, the experiment is a hybrid emulsion experiment that
can reconstruct exclusive charm final states. The estimated data sample is
200,000 reconstructed charm events which might allow a determination of $%
|V_{cd}|$ to $\sim 2\%$ \cite{Bolton:CKMa,Bolton:CKMb} .

WSM production has also been used to probe the charm content in the nucleon
through the neutral current scattering of a $\nu _\mu $ off a charm quark
which subsequently decays into a $\mu ^{+}$ . From their WSM data with $%
E_{vis}>100$ GeV, the CCFR collaboration has set a $90\%$ confidence level
(CL) upper limit on the WSM rate of $4.3\times 10^{-4}$ which is much larger
than the estimate of $1.4\times 10^{-5}$ for NC charm scattering.

The currently running NuTeV experiment will have a 
substantially reduced $\overline{\nu }_\mu $ contamination in their new SSQT
beam which corresponds to a $2\times 10^{-5}$ WSM rate for $E_{vis}>100$
GeV. If additional cuts can be used to minimize the other backgrounds, the
NuTeV experiment may be able to isolate a NC charm scattering signal and
make the first direct estimate of the charm sea in the nucleon.

\subsection{Electroweak Measurements with Neutrinos}

Experimental results on $\nu _\mu $ scattering on electrons and nucleons
confirm predictions of the SM at low to moderate space-like momentum
transfer, and in some cases, provides interesting limits on possible new
physics beyond the SM.

CHARM II, and earlier experiments, have demonstrated from NC measurements
with electron targets that $Z^0$ couplings to neutrinos are independent of
family number and that the $Z^0$ couples to electrons in processes with very
small space-like momentum transfer with precisely the strength specified by
the SM and precision measurements at very high energy collider experiments.
CCFR and CHARM II\ measurements of CC scattering of neutrinos from electrons
demonstrate conclusively that the CC process is uniquely V-A and set tight
constraints on possible exotic couplings beyond the SM.

Precision NC measurements with nucleon targets by CDHS, CHARM, and CCFR
allow an independent and competitive ($\Delta M_W=\pm 190$ MeV) prediction
of the $W$-boson mass within the SM in a very different kinematic regime
than direct and indirect collider determinations. The consistency of
neutrino measurements with SM predictions allows limits to be placed on a
variety of possible new physics parameters, especially those involving light
quarks.

The only way to probe for the expected new physics beyond the SM is
currently through precision electroweak tests, and this situation will
likely persist at least until the LHC turns on and begins taking data.
Because effects of new physics at low energies will be small and subtle, if
they are there at all, it is desirable to perform as many cross checks as
possible. Neutrino-nucleon scattering experiments are perhaps the only
non-collider measurements that have the required sensitivity to provide
these checks. And indeed, there are prospects for further gains from two
near-term efforts that should improve electroweak measurements in the
neutrino sector.

The NuTeV experiment is currently taking data with high intensity
sign-selected beams 
that will permit the first significant exploitation of the PW relationship,
which will reduce the error due to heavy flavor production by a factor of
4-5. The new beam design also eliminates much of the uncertainty attributed
to $K_L^0$ production of $\nu _e$. Depending in the total number of protons
ultimately delivered, NuTeV should measure $\sin ^2\theta _W$ with an error
of $0.002-0.003$, which is equivalent to a $W-$mass measurement of 100-150
MeV.

High sensitivity searches for $\nu _\mu \rightarrow \nu _\tau $ oscillations
are now underway with CHORUS and NOMAD at CERN soon to be joined by COSMOS
and MINOS  \cite{MINOS} at 
Fermilab. While these experiments probably will not directly measure
electroweak parameters to interesting precision, they should greatly improve
our understanding of neutrino induced charm production. It is possible that
the error on the effective charm mass will be reduced by an
order-of-magnitude or more, particularly by CHORUS and COSMOS, which will
have unique abilities to measure inclusive charm production in their
emulsion targets. With the largest systematic uncertainty in $\nu _\mu N$
perhaps nearly eliminated, one might envision a successor to the NuTeV
experiment running at Fermilab with the higher intensity of the Fermilab
Main Injector, or possibly even a new neutrino experiment at LHC or a muon
collider.

\acknowledgments
We would like to thank the members of the CCFR and NuTeV collaborations for
many valuable discussions and for making their results available for this
review. This work was supported by the Department of Energy contract
DE-FG02-94ER40814 and the National Science Foundation grant PHY-95-12810.

\end{document}